\pdfoutput=1
\documentclass[]{amsart}

\usepackage{amssymb,latexsym}
\usepackage{graphicx}
\usepackage{subfig}
\usepackage{xcolor}

\usepackage{array}
\usepackage[]{natbib}
\usepackage[]{amsaddr}

\usepackage{savesym}
\usepackage{caption3}
\savesymbol{captionbox}
\usepackage{caption}
\restoresymbol{CAP3}{captionbox}

\begin{document}

\title{Planetary Embryo Bow Shocks as a Mechanism for Chondrule Formation}
\author{Christopher R. Mann and Aaron C. Boley}
\address{
	Department of Physics and Astronomy\\
         University of British Columbia\\
         Vancouver, BC V6T 1Z1\\
         Canada} 
	
\author{Melissa A. Morris}
\address{
	Physics Department\\
	State University of New York at Cortland\\
	Cortland, NY 13045\\
	USA}

\begin{abstract}
We use radiation hydrodynamics with direct particle integration to explore the feasibility of chondrule formation in planetary embryo bow shocks. 
The calculations presented here are used to explore the consequences of a Mars-size planetary embryo traveling on a moderately excited orbit through the dusty, early environment of the solar system.
The embryo's eccentric orbit produces a range of supersonic relative velocities between the embryo and the circularly orbiting gas and dust, prompting the formation of bow shocks.
 Temporary atmospheres around these embryos, which can be created via volatile outgassing and gas capture from the surrounding nebula, can non-trivially affect thermal profiles of solids entering the shock. 
We explore the thermal environment of solids that traverse the bow shock at different impact radii, the effects that planetoid atmospheres have on shock morphologies, and the stripping efficiency of planetoidal atmospheres in the presence of high relative winds. 
Simulations are run using adiabatic and radiative conditions, with multiple treatments for the local opacities. 
Shock speeds of 5, 6, and 7 km s$^{-1}$ are explored.  We find that a high-mass atmosphere and inefficient radiative conditions can produce peak temperatures and cooling rates that are consistent with the constraints set by chondrule furnace studies.  For most conditions, the derived cooling rates are potentially too high to be consistent with chondrule formation. 
 
\end{abstract}

\maketitle


\section{Introduction}\label{S:intro}

 Chondrules are 0.1 to 1 mm igneous spherules found in abundance in the majority of stony meteorites (chondrites).
They were processed by transient heating events that took place during the formation of the solar system, many of which have not been substantially altered since their incorporation into undifferentiated parent bodies.
The mass in chondrules is estimated to be comparable to the mass of the present-day asteroid belt itself \citep{grossman_book_1988}, suggesting that chondrule formation was common. 
In addition to being ubiquitous, chondrules formed for several million years after the formation of calcium-aluminum-rich inclusions (CAIs)\citep{villeneuve_etal_sci_2009}, a duration that is similar to the  inferred lifetimes of gaseous disks \citep{mamajek_aipcs_2009}.
Because CAIs are the oldest known processed solids in the solar system, with ages of 4568.2~Myr \citep{bouvier_wadhwa_2010}, chondrules are presumed to have recorded the conditions that prevailed during major periods of planet building.

Despite their importance, the formation mechanism(s) of chondrules has yet to be  identified conclusively, which frustrates attempts to interpret and use the meteoritic record for understanding planet formation.
Suggested models include, e.g., winds \citep{shu_etal_apj_2001}, electrical shorts \citep{mcnally_etal_apjl_2013}, asteroid collisions \citep{sanders_scott_mps_2012}, and nebular shocks \citep{morris_desch_2010}. 
Here, we explore the hypothesis that planetesimals and/or planetoids themselves are responsible for producing at least some and quite possibly the majority of chondrules \citep[e.g.,][]{hood_mps_1998,ciesla_etal_mps_2004,hood_weidenschilling_mps_2012}.
In particular, we focus on the high-temperature processing of solids in planetoidal bow shocks, building on the results of \cite{morris_etal_apj_2012} and \cite{boley_etal_apj_2013}. 
While using planetesimals and planetoids to form chondrules may seem to be contradictory, as a significant fraction of the asteroid's mass may be chondritic, we note that there is evidence that significant planet building was already underway before the peak of known chondrule formation.  For example, 
iron meteorite parent bodies are thought to have formed $\le$ 1.5 Myr after CAIs \citep{schersten_etal_epsl_2006}, and Mars may have accreted $\sim50$\% of its mass within 0.8--2.7 Myr after CAIs \citep{dauphas_etal_nat_2011}.  
This can be compared with the age range of the majority of known chondrules, which is between 1.7--3.0 Myr after CAI formation \citep{kurahashi_etal_geoat_2008}.
In this view, chondrules are byproducts of the planet formation process.

Many of the basic structures of bow shocks have been characterized in previous studies \citep[e.g.,][]{ciesla_etal_mps_2004,morris_etal_apj_2012,boley_etal_apj_2013}, suggesting that larger objects, such as planetoids, are preferred over planetesimals. 
This is for a number of reasons. For example,
the standoff distance between the shock front and the surface of the obstacle (planetesimal or planetoid) is typically between 10--20\% of the obstacle's radius, although this depends on the details of the shock.   
Chondrule-size solids have stopping times $\sim 1$ min for post-shock gas densities $\sim 10^{-8}$ g cm$^{-3}$, so if chondrule precursors have a relative speed $\sim 6$ km s$^{-1}$, the required stopping distance is several hundred kilometers.
Precursors that traverse the bow shock at impact radii less than the size of the planetesimal, and even small planetoids, are not expected to re-couple to the gas flow before striking the obstacle's surface. 
The shock strength decreases rapidly with impact radius, limiting the amount of material that can potentially be melted at distances larger than the size of the planetesimal/planetoid.
The size of the planetesimal/planetoid will also affect the thickness of the post-shock region, which affects the time during which a potential melt will be exposed to a high-temperature, high-density environment as it is deflected around the obstacle.
This thickness of the shock front also affects the radiative optical depth, $\tau$, which will be much smaller than unity for shock widths around planetesimals.
This potentially allows any chondrule melt to radiate efficiently into space, losing energy faster than what heat can be supplied by thermal contact with the hot gas reservoir.
While this is a potential problem for shocks around planetoids as well, it is not as severe.  
We will return to this later in the manuscript.
Regardless, even in the adiabatic limit, the gas behind the bow shock cools rapidly due to expansion, effectively removing the thermal reservoir needed to keep chondrules hot after shock passage.
Only when considering planetary embryos (large planetoids) can bow shocks approach conditions required to meet laboratory constraints on chondrule formation.

However, if we consider obstacles that are planetoids, then we need to also consider whether there is a primitive, molecular hydrogen atmosphere that will affect the bow shock structure. 
\citet{morris_etal_apj_2012} included such an atmosphere in their planetoidal bow shock simulation, noting that the atmosphere could affect the shock structure and might be retained through outgassing.  
Such an atmosphere could be volatile-rich, which would help to explain some of the cosmochemistry results, which suggest that chondrules need to be formed in regions of very high Na abundances \citep{alexander_etal_sci_2008}.
Their bow shock calculations were nonetheless only adiabatic and assumed a perfect gas (fixed adiabatic index).
They also did not explicitly track atmosphere loss, and particle tracing was conducted as a post-processing analysis instead of a self-consistent tracking during simulation. 
They also used the post-processing to inform detailed 1D shock models.
Based on their analysis, they found that atmosphere stripping rates were $\sim10^{12}$ g s$^{-1}$, suggesting that the atmospheres would be long-lived. 
\cite{boley_etal_apj_2013} extended these calculations by running 3D simulations using adiabatic and radiative gas dynamics, integrating trace particles along with the gas flow, and including a non-trivial equation of state for the gas.
They did not, however, run simulations with atmospheres around the planetoids.
In this manuscript, we examine whether atmospheres can be retained around planetoids during bow shock processing and explore how the atmosphere itself can affect the shock and post-shock structures.

This manuscript is organized as follows.  
In section \ref{S:shock_model}, we summarize several constraints on the thermal histories of chondrules, features of the general shock model, and details of the bow shock model.
Our numerical methods and setup are described in section \ref{S:num_experiments}, with the results in section \ref{S:results}.  
Further discussion is presented in section \ref{S:discussion}  with the conclusions in section \ref{S:conclusions}.


\section{Shock Model\label{S:shock_model}}

\subsection{Thermal History Constraints}\label{SS:thermal_hist_constraints}

Petrological laboratory studies of chondrule analogues have revealed some of the basic conditions that are consistent with chondrule formation
\citep{hewins_1997,jones_ppvi_2000,jones_chondrules_2005,connolly_desch_2004,beckett_etal_2006}.
Chondrule textures are best explained if their precursors were heated to above their melting temperature within minutes and then independently cooled at protracted rates while orbiting freely within the nebula. 
Specific peak temperatures and detailed cooling times vary among types of chondrules.  
Briefly, there are three commonly observed chondrule textures:  porphyritic, barred, and radial.
Porphyritic textures are the most common type, making up around 80\% of all chondrules.
Producing them requires retaining numerous nucleation sites after the chondrule is melted.  
This suggests peak temperature only slightly (80-120~K) above the liquidus temperature, or in the range of 1750--2090~K  \citep[see][and references therein for a summary of cooling rates]{desch_etal_mps_2012}. 
Porphyritic olivine chondrules are constrained to a cooling rate of lower than 1000~K hr$^{-1}$, whereas porphyritic pyroxene chondrules have been found to require a 2--2000~K hr$^{-1}$ cooling rate range.

In contrast, radial textures require the complete destruction of all nucleation sites. 
This necessitates a higher peak temperature, somewhere within the range of 150-400 K above liquidus, or 1820-2370 K.
Melts must then cool well into their crystallization temperature before any crystallization begins. 
This creates a supercooled chondrule precursor, which allows rapid crystal formation when a nucleation site is injected or forms spontaneously.

Intermediate between radial and porphyritic, barred textures require the retention of only a small number of nucleation sites.  
The required peak temperatures are similar to the radial textures (1820--2370 K).
Cooling rates are in the range of a hundred to about 3000 K hr$^{-1}$.
Less common textures such as skeletal chondrules can be formed when cooling rates exceed those mentioned above \citep{hewins_etal_2005}.

In addition to these cooling rates, chondrule formation mechanisms must also match, for example, the frequency of compound chondrules, the suppression of isotopic fractionation, and the cosmochemical complementarity of chondules with their matrix. As in \cite{boley_etal_apj_2013}, we focus on the cooling rate constraints, as this alone can potentially rule out several models~\citep{desch_etal_mps_2012}.


\subsection{The Shock Model in Brief\label{sec:shocks_brief}}

The shock model is able to accommodate many of the chondrule formation constraints discussed above, assuming a 1D geometry.
As a chondrule precursor enters a shock, it is rapidly heated by gas-drag friction, radiation, and thermal contact with the gas in the high-density, high-temperature post-shock region.
Thermal buffering by H$_2$ dissociation regulates the post-shock temperatures to be around $\sim 2000$ K, except during a short ($\sim 10$ seconds), initial period when the shocked gas is out of chemical equilibrium. 
As long as the optical depth through the shock is large, high-temperatures can prevail downstream.
Under this condition, the solid and gas temperatures will track each other closely, and cooling will be controlled by the rate at which energy can diffuse radiatively from the shock.   
If however the shock is optically thin, then any melt would radiate away its energy in about one second \citep[see, e.g.,][]{hood_horanyi_icarus_1991}, even while in contact with a large heat reservoir.  

To highlight this point, consider the following.
The rate of energy transfer from the gas to the chondrule, assuming the chondrule is cooler than the gas, can be estimated by
\begin{equation}
	\dot{E}_{\rm col} \approx 2 \pi C_K s^2 v_{th}^3 \rho_g,
\end{equation}
where $s$ is the chondrule size, $\rho_g$ is the gas density,  $v_{\rm th}$ is the thermal velocity of gas particles, and $C_K$ is a thermal coupling constant, which we take to be unity for very efficient transfer of energy.
Chondrules will radiate energy at approximately
\begin{equation}
	\dot{E}_{\rm rad} \approx \frac{ \epsilon 4 \pi s^2 \sigma T_{\rm c}^4 }{1+\tau},
\end{equation}
where $\epsilon$ in the material emissivity, $\sigma$ is the Stephan-Boltzmann constant, $T_c$ is the chondrule temperature, and $\tau$ is the optical depth across the shock front.
The energy content of a chondrule is approximately $\frac{4 \pi}{3} \rho_m s^3 c_p T_{\rm c}$, where
$\rho_m\sim 3$ g cm$^{-3}$ is the internal density of the chondrule and $c_p$ is the specific heat (about $10^7$ erg g$^{-1}$ K$^{-1}$).

For $s\sim 0.03$ cm and $T_c\sim$ 2000 K, an isolated chondrule (no thermal reservoir) will radiate away its energy in a few seconds (even including the latent heat of crystallization, $4\times10^9$ erg g$^{-1}$).
If the chondrule is in thermal contact with a gas reservoir, then the equilibrium chondrule temperature will become
\begin{equation}
	T_{c}^4 \approx \frac{ \rho_g v_{th}^3 (1+\tau) }{2 \sigma \epsilon}
\end{equation}
for $T_c< T_{\rm gas}$. 
Taking post-shock gas conditions $\sim 10^{-8}$ g cm$^{-3}$ , $T\sim 2000$ K, and $\epsilon\rightarrow 1$ (perfect radiator), the equilibrium temperature for the chondrule will be about $T_c\sim1600 K$ for $\tau\sim 0$.  Here, we have taken $v_{\rm th} = (8 \mathcal{R} T / (\pi \mu))^{1/2}\sim 4.3$ km s$^{-1}$ for gas constant $\mathcal{R}$ and $\mu=2.3$ g mol$^{-1}$.
At lower gas densities, the equilibrium chondrule temperature will also be lower. 
High optical depths or very high densities are thus necessary to limit chondrule cooling and to allow chondrules to maintain thermal contact with the gas.
We will return to these points.

We further note that frictional drag heating alone cannot keep chondrules hot enough to prevent rapid cooling, unless the chondrules are in a very low-density gas (giving long stopping times) and encounter a very high shock speed.
Balancing gas-drag heating with radiative losses yields
\begin{equation}
	\frac{1}{8} \rho_g  \pi s^2 |v_{rel}|^3 = \epsilon 4 \pi \sigma s^2 T_{eq}^4, {\rm~and}
\end{equation}
\begin{equation}
	T_{eq}^4 = \frac{ \rho_g |v_{rel}|^3}{32  \sigma \epsilon}.
\end{equation}
For example, a chondrule moving through a $\sim 2000$ K gas with a density $\sim 10^{-10}$ g cm$^{-3}$ will have a stopping time between approximately 1 to 2 hr.  
However, to maintain a temperature of at least 1400 K, the chondrule would need a relative speed with the gas $\sim 40$ km s$^{-1}$.  
Lower densities require even higher speeds.
While such conditions may be attainable in, e.g., X-ray flares \citep{nakamoto_etal_2005}, they are not plausible in known midplane conditions.

The most ideal 1D shocks are those with pre-shock gas densities $\sim 10^{-9}$ g \rm~cm$^{-3}$, pre-shock temperatures less than 650 K, shock speeds $\sim$ 8 km s$^{-1}$, and sufficient dust and/or chondrule precursors present to keep the optical depths such that the radiative cooling rates through the crystallization temperature range are $\lesssim 1000$ K hr$^{-1}$ as in \citet{desch_connolly_mps_2002,morris_desch_2010}.
The main issue for the shock heating hypothesis is the identification of a suitable shock-producing mechanism.


\subsection{Bow Shocks\label{sec:bow_shocks}}

Bow shocks have many similarities to 1D shocks, but there are at least two very important differences.
First, even if the gas has negligible energy loss through radiation, the 3D structure of the shock will allow adiabatic expansion in the post-shock region.
In the adiabatic limit, the gas temperature can still drop at a few thousand Kelvin per hour behind the shock front \citep{boley_etal_apj_2013}.  
Thus, the thermal reservoir required to keep a melt hot can be controlled by non-radiative processes.
Second, because shock widths are comparable to about 10\% of the perturber's radius, the post-shock region can be physically narrow.  
This can lead to low optical depths, allowing radiation to redistribute significant energy from the shocked gas (and heated chondrule precursors) into the post- and pre-shock regions.
Combined with adiabatic expansion, low optical depths can promote very rapid cooling of any melts (a few 10,000 K hr$^{-1}$), even around fairly large bow shocks produced by planetoids. 
This was demonstrated by \cite{boley_etal_apj_2013}, building on previous work \citep[][]{ciesla_hood_2003,morris_etal_apj_2012}, and will be revisited here.

For example, consider a bow shock produced by a $R\sim 3000$ km planetoid with a resulting shock width $\sim 300$ km\footnote{We note that many of the 1D studies consider shock regions that are too wide (by at least an order of magnitude) to be consistent with the high-density, high-temperature region of a bow shock's structure \citep[e.g.,][]{hood_etal_2010,morris_desch_2010}.
The results of \citet{morris_desch_2010} appear to be most applicable to disk-wide, large-scale shocks, such as those driven by gravitational instabilities.}.
If all of the small solid mass is in chondrules of size $s\sim0.03$ cm with internal density $\rho_m$, then the opacity (per gram of {\it chondritic material}) is roughly $\kappa_c=\frac{3}{4 \rho_m s}\sim 10$ cm$^{2}$ g$^{-1}$ for $\rho_m\sim2$-3 g cm$^{-3}$.
For a chondritic mass fraction of 0.00375 in a gas density $10^{-8}$ g cm$^{-3}$ (in the shock), the optical depth across the shock is $\tau\sim 0.01$, much too low to prevent rapid cooling and strong coupling with the gas's thermal reservoir.  
Only at chondrule concentrations of $C\sim 100$ will the opacity of chondrules alone limit cooling.  
Furthermore, such a chondrule concentration would require a mass fraction of 0.375, which would change the characteristics of the shock model due to significant momentum feedback by the chondrule precursors on the gas. 

However, if the chondrule precursors were embedded in an environment with small-grain dust of, say, size $s\sim\mu m$, the opacity (per gram of {\it gas}), would be about $\kappa_g\sim10$ cm$^2$ g$^{-1}$ $f_{\rm rad}$,  assuming a well-mixed dust-gas mass ratio of 0.00375. 
The factor $f_{\rm rad} \approx s T_{\rm rad}/(0.3~{\rm cm~K})$ if the result is $<1$, and 1 otherwise.
This roughly approximates the effect of the radiation temperature $T_{\rm rad}$ (in a grey sense) on the dust opacity.
For radiation in the shock ($\sim 2000$ K), the small-grain dust opacity is about $7$ cm$^{2}$ g$^{-1}$ under the given approximations, which yields an optical depth through the shock front of $\tau\sim2$. 
An increase in the fine-grain dust by a factor of ten would create an opaque environment without requiring so much mass that the shock will obviously be altered by solid-gas feedback.
 We are thus led to the conclusion that if bow shocks are able to produce chondrules, then the shocked environment must contain a large abundance of fine-grained dust in addition to chondrule precursors.

\begin{figure}
	\centering
	\includegraphics[width=0.7\textwidth]
		{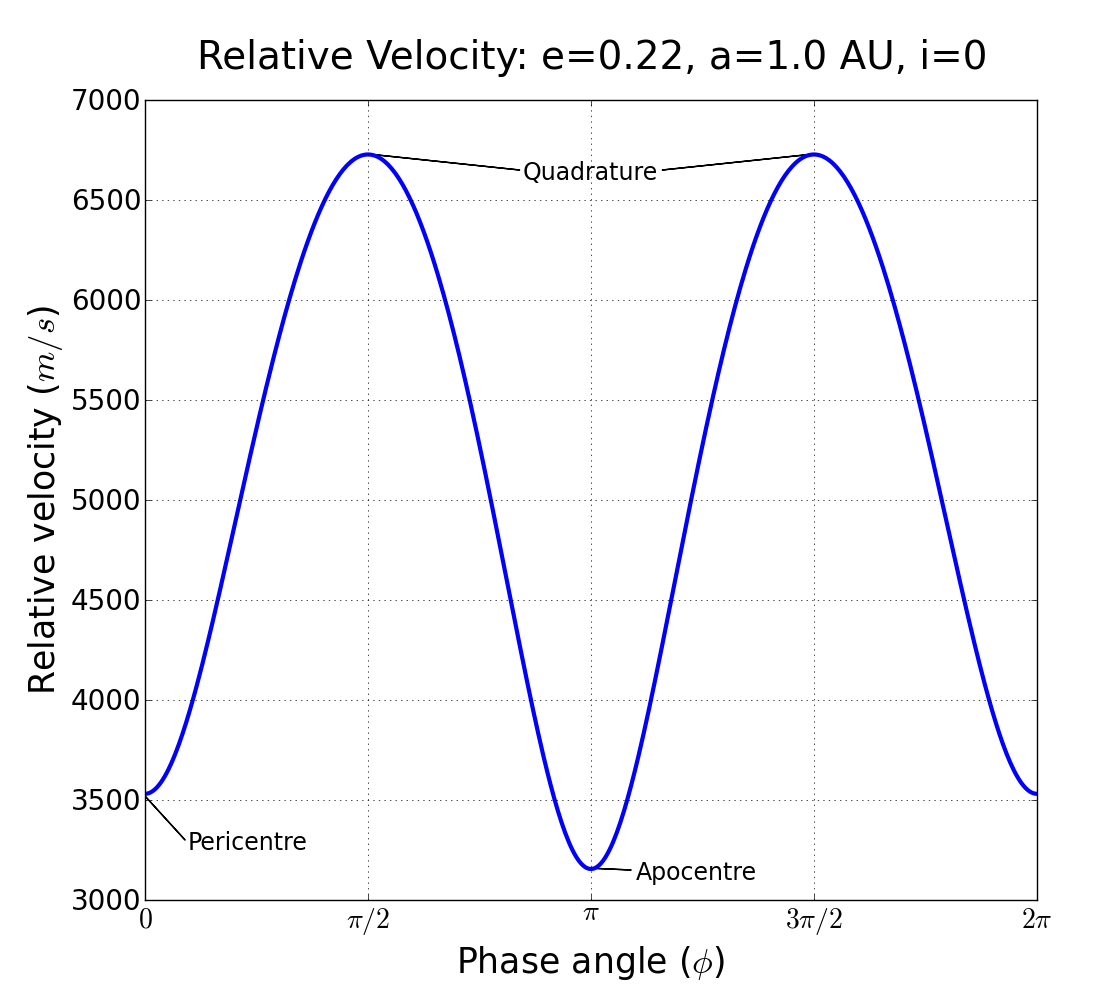}
	\caption{Relative velocity between embryo and the disk's gas throughout the embryo's eccentric orbit. Phase angle of $\phi=0$ corresponds to pericentre.  Fairly large eccentricities are required to reach viable wind speeds for the bow shock model, which has implications for the formation environment of planets if this model is correct.  Only inclination angle $i=0$ is shown here, but an $i>0$ would increase the relative velocity during midplane passage. Depending on the given orbit relative to the gaseous disk, a high inclination would increase the peaks of $V_{rel}$ if midplane passage occurs at quadrature or bring up the minima if it occurs at apo- and pericentre.}
	\label{F:v_rel}
\end{figure}

For strong bow shocks, there must be a source for producing large relative speeds between planetesimals/planetoids and the gaseous, dusty disk.
If we accept that some planetoids form shortly after CAI formation, then a proto-Jupiter could provide a mechanism for exciting the orbits of planetoids \citep{weidenschilling_etal_sci_1998}.
Indeed, simulations by \citet{hood_weidenschilling_mps_2012} showed that resonant interactions with Jupiter can excite large bodies (Moon- to Mars-sized) to eccentricities above 0.33 in $\sim10^5$ years.   
Figure~\ref{F:v_rel} shows the profile of relative velocities $V_{rel}$ for different phases of an eccentric, non-inclined orbit.  At perihelion, the high azimuthal speed of the embryo is offset by the increased gas speed at small $r$.  
Likewise, at aphelion, both the embryo and the gas are moving more slowly.
The largest wind speeds are encountered near quadrature, resulting from the radial motion of the embyro.   
An inclined orbit also affects $V_{rel}$, depending on where the orbit intersects the disk.  
The points of intersection introduce the largest vertical velocity component, which would steepen the gradient should they occur near the high velocity quadrature, or flatten should they occur at the low velocity apsides.  
The orbital relative velocity motions can also impact any primitive atmospheres around planetoids, as will be discussed next.

\subsection{Atmosphere Accretion/Stripping}\label{SS:mass_rates} 

If planetoids are present in a gaseous disk, then they may have substantial atmospheres resulting from a combination of accretion and volatile outgassing.  
Such atmospheres, if not immediately destroyed through ram-pressure stripping, will increase the shock size and can be a source of high volatile partial pressure \citep{morris_etal_apj_2012}.
However, if the atmospheres are stripped rapidly, then their role in chondrule formation is likely limited.  

Consider an embryo of radius $R_e=3000$ km and mass $M_e=3.4\times 10^{26}$ g.  
While the planetoid is stationary with respect to the surrounding gas, we can estimate the atmosphere capture rate using Bondi accretion:
$ \dot{M}_a \approx \frac{4 \pi \rho_g G^2 M^2_e}{c^3_s} \sim 2 \times 10^{15}$ g s$^{-1}$, where $\rho_g$ is taken as $10^{-9} \rm ~g~cm^{-3}$  and the adiabatic sound speed of the gas $c_s\approx 1.2$ km s$^{-1}$ for molecular weight $\mu=2.3$ g mol$^{-1}$ and adiabatic index $7/5$.  
The rate of any volatile outgassing is much harder to constrain, and is not explored here.

 Accretion must ultimately compete with mass stripping. 
Throughout an eccentric orbit, the embryo will go through phases of high and relatively low ram pressure.  
The resulting mass loss can be roughly estimated by comparing the kinetic energy of the gas that is hitting the planetoid with kinetic energy required for gas to escape:
\begin{equation}
	\dot{E}_{in} = \frac{C_w}{2} A_{\rm eff}\rho_g v_{\rm rel}^3\rm .
\end{equation}
Here, $A_{\rm eff} \approx \pi(2R_e)^2 = 4\pi R^2_e$, which includes the envisaged atmosphere, and $C_w$ is a coefficient that accounts for the fraction of energy, on average, that is transferred from the wind to the atmosphere as the ambient gas is deflected around embryo.
Based on analogy to gas drag heating, we expect $C_w\sim 0.1$ to 1. 
Cooling by atmosphere escape balances this heating, i.e.,
\begin{equation}
	\dot{E}_{out} =-\frac{1}{2} \dot{M}_a v^2_{\rm esc} ,
\end{equation}
where the escape velocity is $v_{esc} = \sqrt{ 2GM/r } \approx 3\rm ~km~s^{-1}$ near the top of the atmosphere.  

 Thus, for this simplified model, the stripping rate for a $7~\rm km~s^{-1}$ relative velocity is 
\begin{equation}
	\dot{M}_a = \frac{-4 \pi C_w  R^2_e\rho_g v^3_{rel}}{v^2_{\rm esc}} \sim -4\times10^{15} C_w~\rm g~s^{-1}.
\end{equation}
Assuming an atmosphere  $M_a \approx 2.5\times10^{20}$ g (about 10 Martian atmospheres), complete stripping will occur on a timescale of $\sim1 C_w^{-1}$ day.  
In this work, we will verify this stripping rate in the limit of a predominately molecular hydrogen atmosphere.
If the atmosphere is volatile-rich, e.g., water vapor, then the atmosphere scale height can be reduced substantially.  Nevertheless, the cross section is limited by the planetoid's size, and as such, stripping rates would be expected to be $>10^{14}$ g s$^{-1}$ for the given conditions.  

While informative, the above model is potentially overly simplistic.  For example, the embryo and its atmosphere will sweep through mass at a rate
$\dot{M}_{\rm sweep} = C_{m}\rho_g A_{\rm eff} v_{rel}$,
where $C_{m}<1$ is a coefficient that takes into account the efficiency by which mass can be temporarily accreted by the atmosphere, rather than diverted around it.
For the same conditions noted above, $\dot{M}_{\rm sweep}\approx C_{m} 8\times10^{14} \rm~g~s^{-1}$.
If $C_w$ and $C_m$ are comparable, then the swept-up material can help to reduce some of the mass loss caused by wind stripping.  

Ultimately, the stripping rate needs to be measured directly from simulations (as discussed in section \ref{S:discussion}), particularly as previous estimates suggest that atmospheres should stripped at a rate $\sim 10^{12}$ g s$^{-1}$ \citep{morris_etal_apj_2012}. 
These much lower rates were based on post-processing analyses of simulation results, and may not properly capture the efficacy of ram-pressure stripping from the upper atmosphere.
If stripping is as efficient as our calculation suggests, we do not expect planetoidal atmospheres to be retained throughout all phases of an embryo's orbit.  
  Nonetheless, because the estimates for atmosphere accretion and stripping are of comparable magnitude, we can expect the mass of the atmosphere to vary over time as the embryo enters different windspeed conditions at different orbital phases.  Embryos could thus have periods during which their corresponding bow shock structure are altered by atmospheres and times during which they are not.

It should be noted that ram pressure ultimately removes mass from the low pressure region of the atmosphere at high planetoid altitudes.  
A high surface pressure means that ram pressure cannot remove most of the mass during a single wind crossing time, not that mass cannot be removed.  
Furthermore, while the process can be very slow compared with wind and sound crossing times, it can be very fast compared with orbital times.


\section{Numerical Experiments\label{S:num_experiments}}

We present a series of numerical experiments to study the structure of bow shocks around planetoids and to examine how primitive planetoidal bow shocks interact with the resulting shocked region.
These experiments include a variety of simulations that explore different physics, such a range of shock speeds in the adiabatic limit; different atmosphere masses; and radiative shocks with different opacity assumptions. 
A major objective of this work is to determine whether particles that pass through shocks in the following numerical experiments yield heating/cooling curves that are similar to the constraints set by furnace experiments.  
However, we also pose the following questions: What types of materials would be produced by the heating/cooling curves that \emph{do not} match laboratory constraints? 
Is there evidence of these types of materials in the meteoric record?   
Are there broader implications for primitive atmosphere stripping of planetoids?


\subsection{Parameter Space\label{sec:parameter_space}}

Table~\ref{T:sims} describes the suite of numerical simulations that are run as a part of this study.
Basic parameters such as the simulation space, planetoid position, and resolution are common to every simulation.  
The upper box highlights the different types of adiabatic simulations that are run.  In these simulations, only the wind speed and mass of the planetoid's atmosphere are varied.
Because the runs are adiabatic, gas can only cool via expansion, providing a limiting case for the cooling rates for bow shocks (see section~\ref{S:results}).
The radiative hydrodynamics simulations are shown in the lower box.  
Only wind speeds of 7 km s$^{-1}$ are explored, but the gas opacity is varied in two ways: a fine-grained ``dust'' opacity and a chondrule opacity. 
The f($\kappa_d$) column refers to the scaling factor applied to the fine-grained dust opacity, which is modeled here using  Rosseland grey opacities \citep{pollack_etal_1994}.  
The scaling factor represents the total amount of dust relative to the standard (solar) mixture, and is a convenient way to explore the effects of different concentrations of small-grain dust.
 The $\kappa_c$ column displays the chondrule opacity in $\rm cm^2 ~g^{-1}$, which is the total cross section of chondritic material per gram of chondritic mass.
The value for $\kappa_c$ is also varied, as done for ${\rm f}(\kappa_d)$; however, the chondrule-gas coupling is always calculated using a chondrule concentration appropriate for $\kappa_c=10$ cm$^2$ g$^{-1}$.  
While this adds some inconsistency, it still allows us to explore the radiative transfer consequences for different chondrule concentrations.

\begin{table}
	\begin{center}
	\begin{tabular}{ l l l l c l  }
	  \hline          
	  Name	 	&  Type				   & f($\kappa_d$)	& $\kappa_c$		& $V_{rel}$ & Comment\\
	  			& 					   &				& $\rm (cm^2 g^{-1}$)& ($\rm km~s^{-1}$)   \\
	  \hline
	  AdiLo7	 	& Adiabatic     			   & N/A			& N/A			& 7				 & $a$\\
	  AdiHi7	 	& Adiabatic     			   & N/A			& N/A			& 7				  & $b$ \\
	  AdiHi6 	 	& Adiabatic     			   & N/A			& N/A			& 6				 & $b$ \\
	  AdiHi5 		& Adiabatic     			   & N/A			& N/A			& 5				 & $b$ \\
	  \hline
	  Rd1k10		& Radiative Transfer		   & 1				& 10				& 7				 & $b$ \\
	  Rd0.1k10	& Radiative Transfer		   & 0.1			& 10				& 7				 & $b$ \\
	  Rd0.01k10	& Radiative Transfer		   & 0.01			& 10				& 7				 & $b$ \\
	  Rd0.01k100	& Radiative Transfer		   & 0.01			& 100			& 7				 & $b$ \\
	  Rd30k10 	& Radiative Transfer             & $\leq30$                            & 10                           & 7             & $b,c$ \\
	  \hline
	\end{tabular}
	\end{center}
	  \caption{Summary of simulations listing the run names, whether radiative transfer is used, the opacity assumptions (if applicable), and the wind speeds.    
	  		The Rosseland mean dust opacity \citep{pollack_etal_1994}  is scaled by f($\kappa_d$), which simulates different assumptions for the fine-grained solids.  The chondrule opacity, $\kappa_c$, is defined as the cross section of chondritic material per gram of chondritic mass ($\rm cm^2~ g^{-1})$.
			For the fine-grained dust, the local gas density is used to determine the optical depth across a cell (which assumes the dust is well-mixed with the gas). 
			In contrast, the chondrule opacity uses the chondrule precursor mass density on the grid, as determined by the super-particle distribution.  The column $V_{\rm rel}$ is the relative wind speed imposed by the grid boundary conditions. Comments: $a$ -- Low-mass atmosphere.  $b$ -- High-mass atmosphere.  $c$ -- A suite of short simulations that are used to explore  at which f($\kappa_d$) the radiative simulations become adiabatic-like. 
	  \label{T:sims}}
\end{table}


\subsection{Methods}\label{S:methods}

The simulations are run using the same code described in \citep{boley_etal_apj_2013} (BoxzyHydro), which is a finite volume, Cartesian grid code that allows user-defined zero-flow obstructions within the computational domain (e.g., a planetoid).
BoxzyHydro employs an equation of state that takes into account the rotational, vibrational, and dissociation degrees of freedom of $\rm H_2$, as well as a radiative transfer technique that captures  high, low, and transitional optical depth regimes.  
 It also makes use of a particle-in-cell method that allows for gas-solid dynamical coupling and thermal history tracking.  
 Opacities can be assigned based on particle concentration, which is done for the chondrule contribution to the opacity. 

 The basic setup is a sphere in a wind tunnel.  
The simulation domain is set to $x,y,z=$~35000, 15000, 15000~km, achieving 100 km resolution ($\Delta x,\Delta y,\Delta z=100$ km).
To save computational resources, we simulate only one quadrant of the sphere and its corresponding space.  
 Using this setup, we place the sphere's centroid in the $-y,-z$ corner of the computational domain, and offset the $x$ position upwind by $\sim 1$ embryo radii ($R_e$).  
  As in the above estimates, the planetary embryo is given a radius of $R_e =$ 3000~km, and a mass of $M_e = 3.4\times 10^{26}$ g.  
 For context, this is roughly 90\% of Mars's radius and a little over half its mass.  
 The gas wind originates from the $+x$ boundary with a velocity $v_x,v_y,v_z=-7,0,0$~km s$^{-1}$, and a density of $\rho_g=10^{-9}~\rm g~cm^{-3}$, where the boundary $|v_x|=V_{\rm rel}$ is varied for the different simulation cases.  
 This setup models the embryo's supersonic velocity relative to the gaseous disk  during, e.g., orbital quadrature passage.  
 The $-y$ and $-z$ boundaries of the simulation space are planes of symmetry, and are given zero-flux boundary conditions. 
 Radiation is reflected at these boundaries, as well as along the surfaces of obstructions (the planetoid), the latter of which mimics re-radiation. 
Mass and radiation are allowed to flow out through the other simulation boundaries.

An atmosphere is placed around the embryo upon initializing the run.
For determining the structure of the atmosphere, we assume that the gas is non-self-gravitating and polytropic, i.e., $P=K\rho^\gamma$,
which gives a density structure
\begin{equation} \label{E:atmos}
	\rho(r)= \rho_0  \left[ 1 + H\left( \frac{1}{r} - \frac{1}{R_e} \right)   \right]^{ \frac{1}{\gamma-1}   } .
\end{equation}
Here, $P$ is pressure, $K$ is the polytropic constant, $\rho_0$ is the gas density at the surface, $r$ is the distance from the embryo's centre, and $H=\frac{GM\left(\gamma-1\right)}{\gamma K\rho_0^{\gamma-1}}$ is the atmosphere scale length.  
The mass of an atmosphere itself scales as $\rho_0$, so we run a short simulation of an isolated embryo and gas to test for an appropriate $H$ and an effective (structural) adiabatic index, as the actual (EOS) adiabatic index is a function of temperature in our simulations. 
The atmosphere was initialized with $H\approx R_E$ and $\gamma \sim 1.4$.  
The scale length remained approximately the same after isolated evolution, but the effective adiabatic index for the stable structure was closer to $\gamma\sim1.35$ for our nebula mixture (mean molecular weight $\mu=2.33$)\footnote{One should note that this is not the same as the actual adiabatic index that reflects the equation of state.}.  
 Two different values of $\rho_0$ are used to create two atmosphere masses.  
 The low-mass case has an atmosphere with $\sim2.9\times10^{19}$ g, or roughly 1.2 times the mass of the current Martian atmosphere.  
 The high-mass case is $\sim2.7\times10^{20}$ g, or roughly 11 Martian atmospheres.  
 Practically, these masses are chosen to explore mass stripping on short and long timescales.  
  The values are, nonetheless, reasonable for primitive hydrogen atmospheres for the mass within the first several embryo radii.
  For example, we know that the atmosphere solution is bounded by the ambient nebula density $\rho_{\infty}$ at large distances from the embryo, which is ultimately set by the embryo's Hill sphere.  
  This boundary condition requires that $H<R_E$, giving solutions that asymptote to a fixed density.  
  Taking $r\rightarrow\infty$, $\rho_0 = \left(\frac{GM\left(\gamma-1\right)}{\gamma K R_E} + \rho_\infty^{\gamma-1}\right)^{\frac{1}{\gamma-1}}$.
  This evaluates to $\rho_0\approx 10^{-8}\rm ~g~cm^{-3}$ using a $K$ that corresponds to $\rho_\infty=10^{-9}\rm~g~cm^{-3}$ and background nebular temperature $T_{\infty}=300\rm~K$, as well as assuming $\gamma\approx 1.5$.

After the atmosphere is confirmed to be stable after a short integration,  the wind tunnel is initiated, with gas entering the $+x$ boundary at 7 km/s (or other noted rate).  
Again the simulation is allowed to run until the system has settled into a mostly steady state.

The particle-in-cell method is used to track the envisaged chondrule precursors.  
We use $10^6$ super-particles, in which each represents a swarm of chondrule precursors.
All precursors are assumed to have radii $s=0.03$~cm and density $\rho_s=3 \rm ~g~cm^{-3}$.  
This ignores the actual size distribution of chondrules in the meteoritic record, but the approximation is made for simplicity and the spread of chondrule sizes is small. 
To prevent the particles from being artificially initialized and trapped in the atmosphere or wake, they are injected into the simulation with the wind.
Particles are given random $y,z$ coordinates, then launched with the wind from the $+x$ boundary at a rate to ensure that the envisaged solid-to-gas mass ratio is constant in the pre-shock flow. 
This is done only after the bow shock has taken shape.  
We use a solid-to-gas mass ratio of  0.004 for all calculations, even when the opacities are altered.
If a particle comes in contact with the embryo's surface or one of the outflow boundaries, it is  reset with new $y,z$ coordinates at the $+x$ boundary.
The crossing time for a 7 km/s wind is 80 simulation minutes, but many particles take  $\sim 200$ simulation minutes  to interact with the shock and traverse the computational domain.

At any time, 20 of the active particles can be reassigned as tracers.  
These super-particles are spaced along the $y$-axis with increasing impact parameter ($b_i$) in the range $0<b_i<2R_e$.  
The simulation is then allowed to run until the majority of the tracer particles have traversed the entire simulation space.  
The 20 tracer particles' positions are recorded frequently, as well as the local temperature, density, and pressure.  
From these data, we construct the environmental histories experienced by the chondrule precursors.

For the radiative simulations,  the chondrule optical depth through a  given cell is calculated using the assumed chondrule opacity and the actual chondrule mass density in that cell.  For the small-grain dust, the gas and dust is always assumed to be well mixed.


\section{Results\label{S:results}}

\subsection{Adiabatic, Low-mass and High-mass  Atmospheres (7 km s$^{-1}$)}

\begin{figure}[t]
	\centering
	\subfloat{{\includegraphics[width=\textwidth]
		{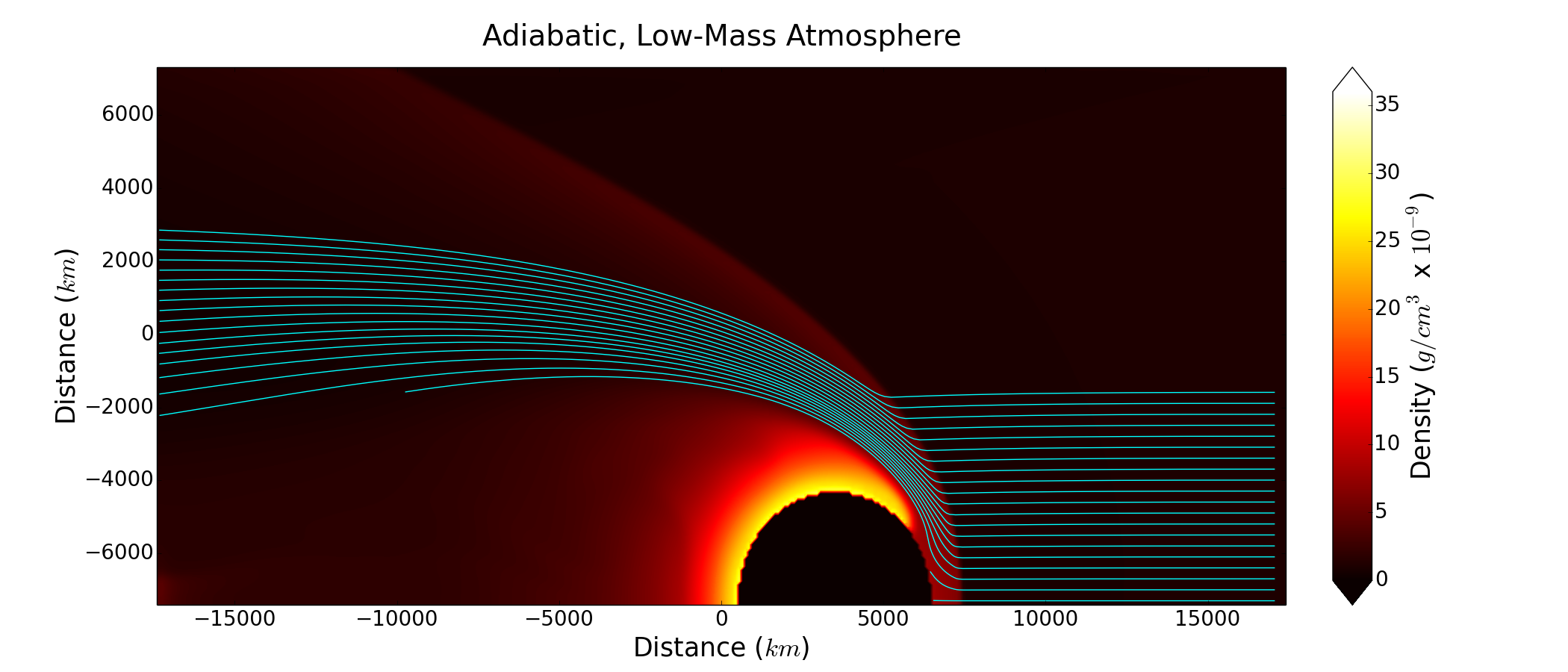} }}\\
	\subfloat{{\includegraphics[width=\textwidth]
		{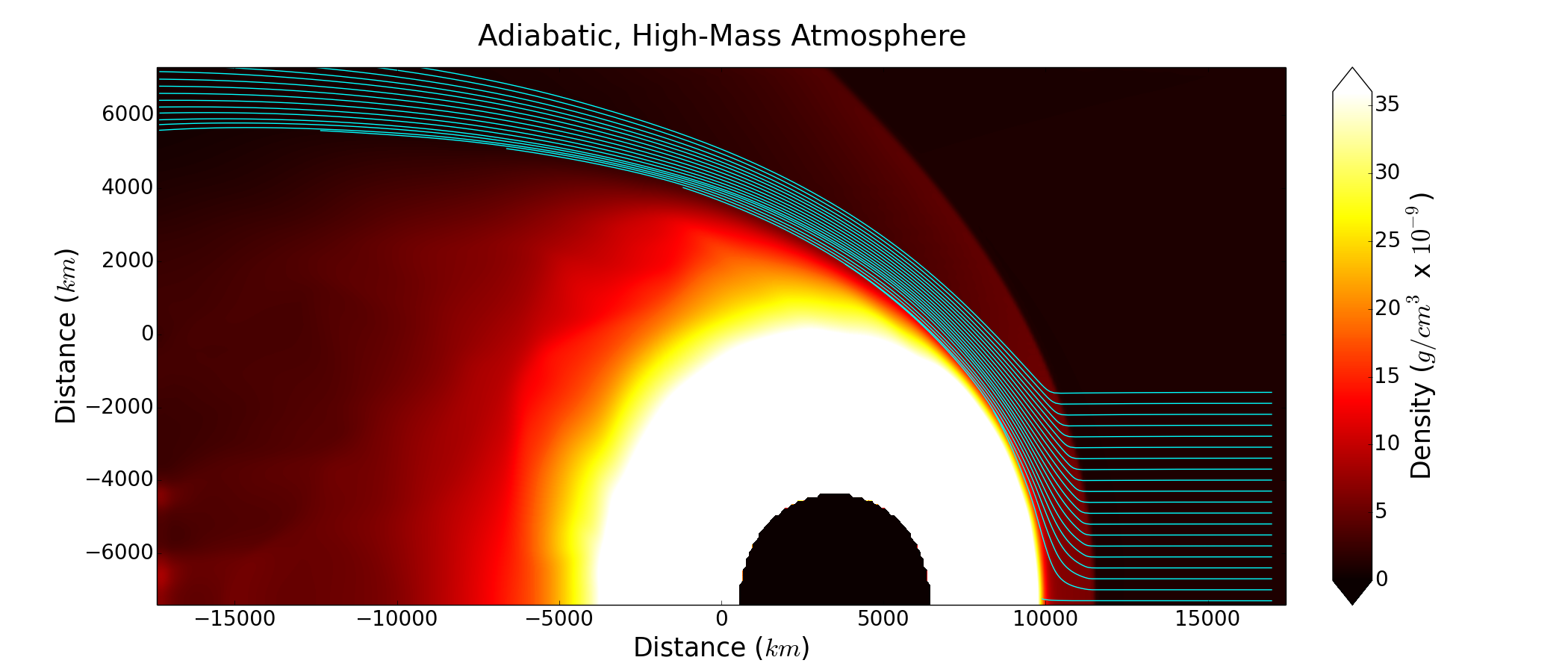} }}
	\caption{The density morphologies of the bow shocks in the AdiLo7 (top) and AdiHi7 (bottom) runs, which include a low-mass ($2.9\times10^{19}$ g) and high-mass ($2.7\times10^{20}$ g)  atmosphere surrounding the embryo.
			The atmosphere increases the effective cross section of the planetoid, creating a larger bow shock.  
			The low-mass atmosphere's surface pressure is too low to prevent the wind from significantly deforming the windward side, while the high-mass atmosphere remains stable for the duration of the simulation. 
			The cyan curves show the trajectories of 20 select super-particles, with impact radii ranging from nearly zero to 2 $R_E$. 
			Particles with small impact radii can directly interact with the embryo's atmosphere. 
			}
	\label{F:adi_traject_dens}
\end{figure}

\begin{figure}[t]
	\centering
	\subfloat{{\includegraphics[width=\textwidth]
		{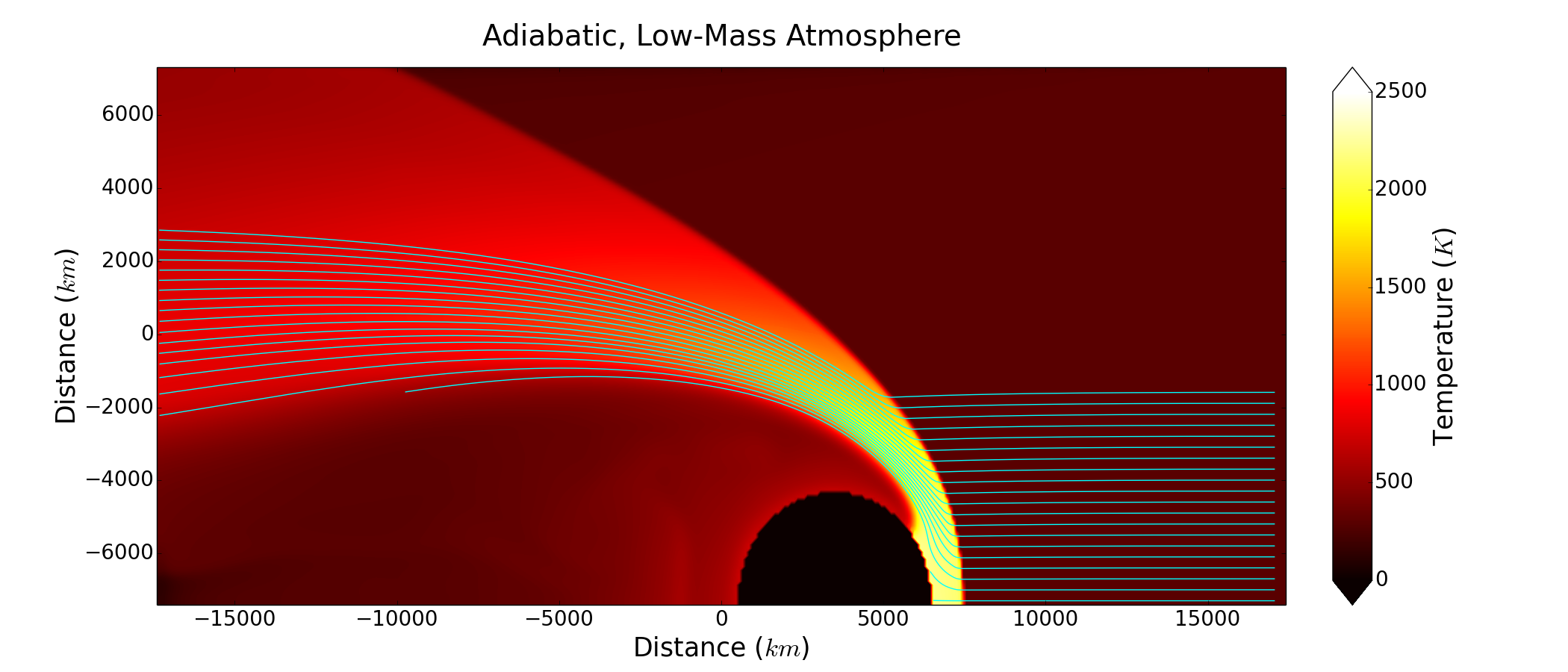} }}\\
	\subfloat{{\includegraphics[width=\textwidth]
		{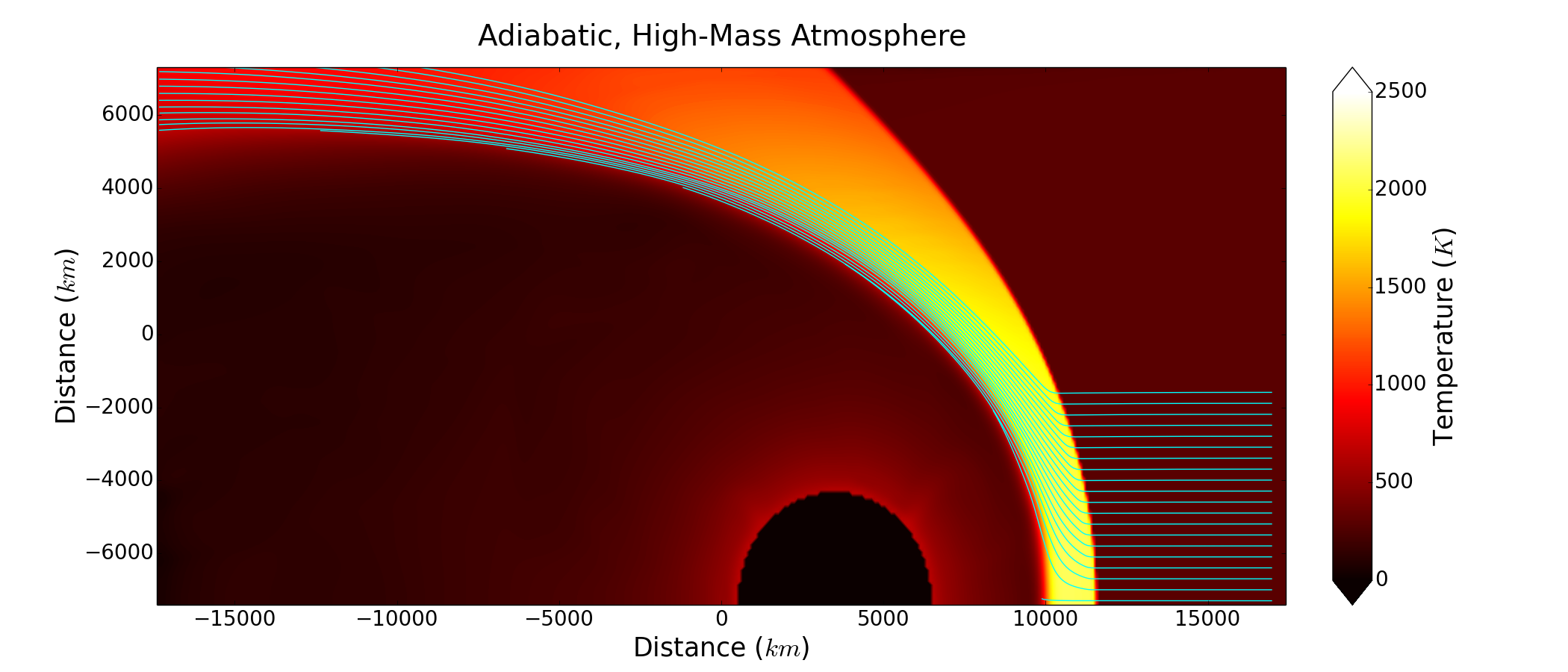} }}
	\caption{Similar to Figure \ref{F:adi_traject_dens}, but for the temperature morphologies.  
			 The high-temperature shock region of the bow reaches the surface of the planetoid for the low-mass atmosphere, but does not have any significant effects on the interior of the high-mass atmosphere. }
	\label{F:adi_traject_temp}
\end{figure}

The density and temperature cross sections for the adiabatic, 7 km s$^{-1}$ simulations are shown in Figures~\ref{F:adi_traject_dens} and \ref{F:adi_traject_temp}, respectively.  
The low-mass atmosphere undergoes significant deformation by the end of the simulation, with material pushed to the leeward side of the planetoid.  
As a result, the hot bow shock reaches the surface of the planetoid.
The pressure at the base of the low-mass atmosphere is about 1.4 mbar (before any stripping occurs). 
The wind creates a ram pressure that is $P_{\rm ram}= \frac{1}{2}\rho_g V_{\rm rel}^2$, which gives  $P_{\rm ram}\approx 0.25$ mbar for the given wind conditions. 
The ram pressure does not immediately remove the atmosphere, but displaces a significant amount of mass in a few wind crossing times.

In contrast, the high mass atmosphere has sufficient surface pressure (about 14 mbar) to prevent significant deformation throughout the simulation, constraining the bow shock to the upper layers of the planetoid's atmosphere. 
Mass is nevertheless removed by the wind, as the atmosphere pressure drops to the ram pressure at the atmosphere's scale length.
We will return to the atmosphere stripping rates in section \ref{S:discussion}.

\begin{figure}[hbt]
	\centering
	\subfloat{{\includegraphics[width=0.48\textwidth]
		{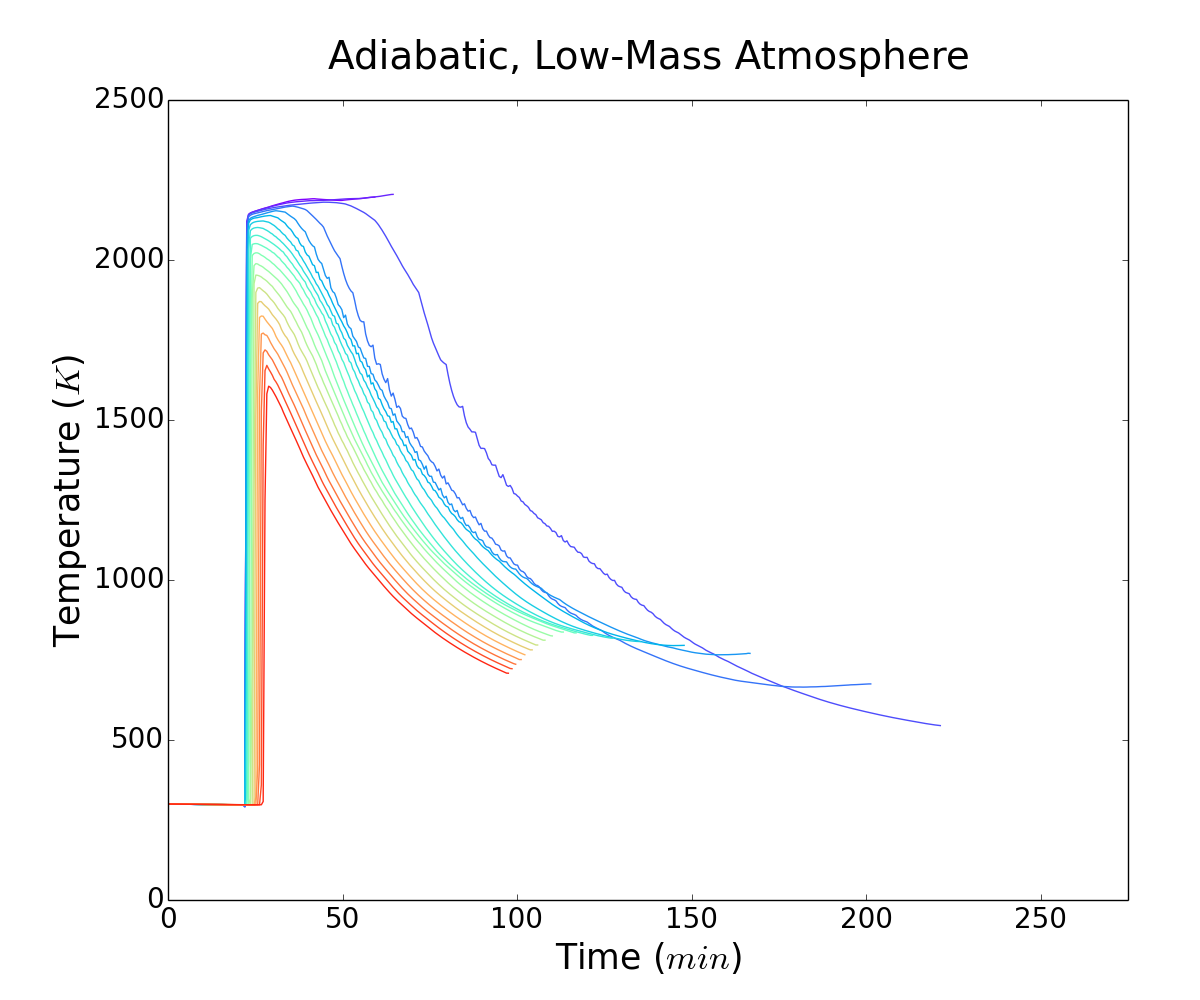} }}
	\subfloat{{\includegraphics[width=0.48\textwidth]
		{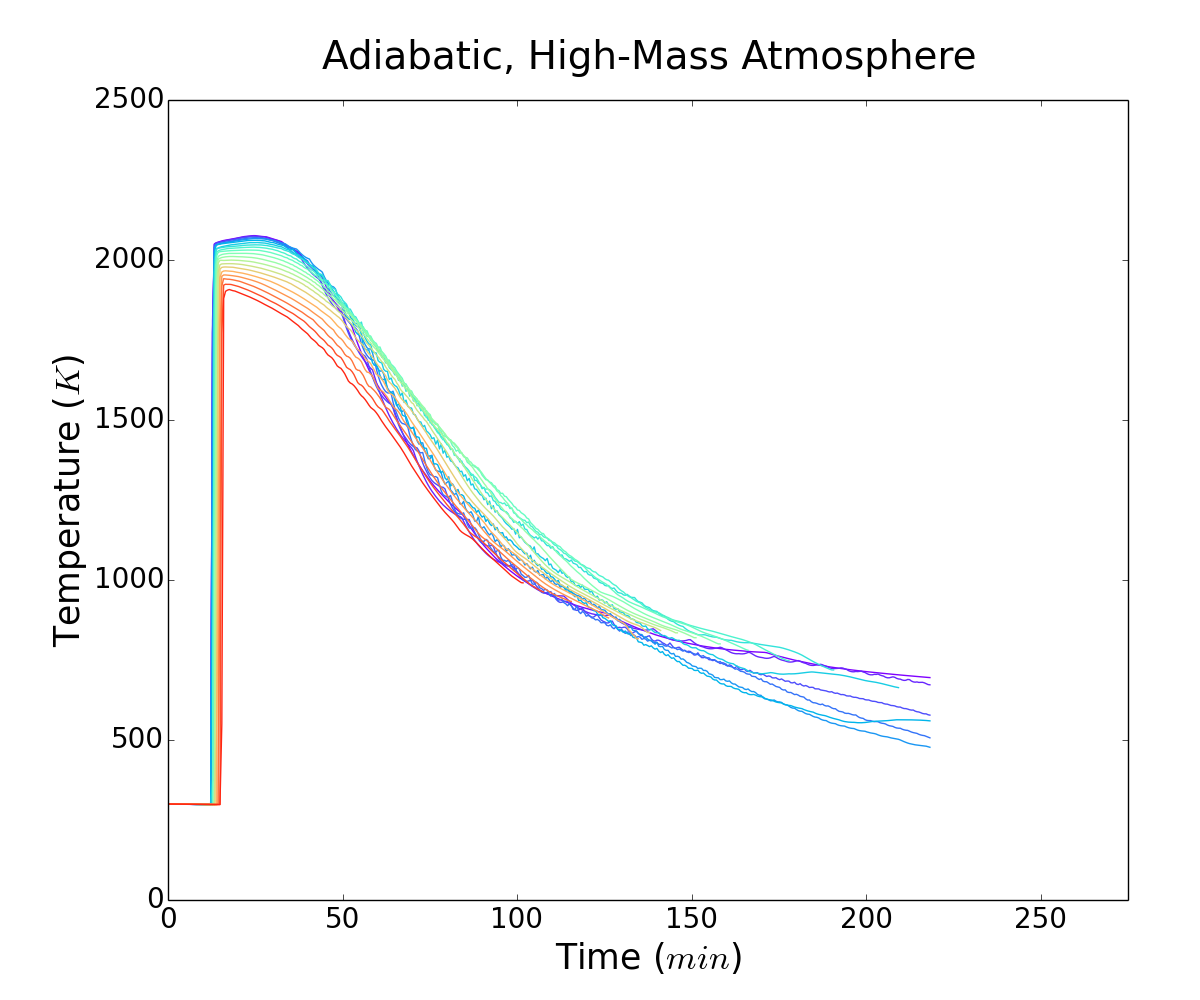} }}\\
	\subfloat{{\includegraphics[width=0.48\textwidth]
		{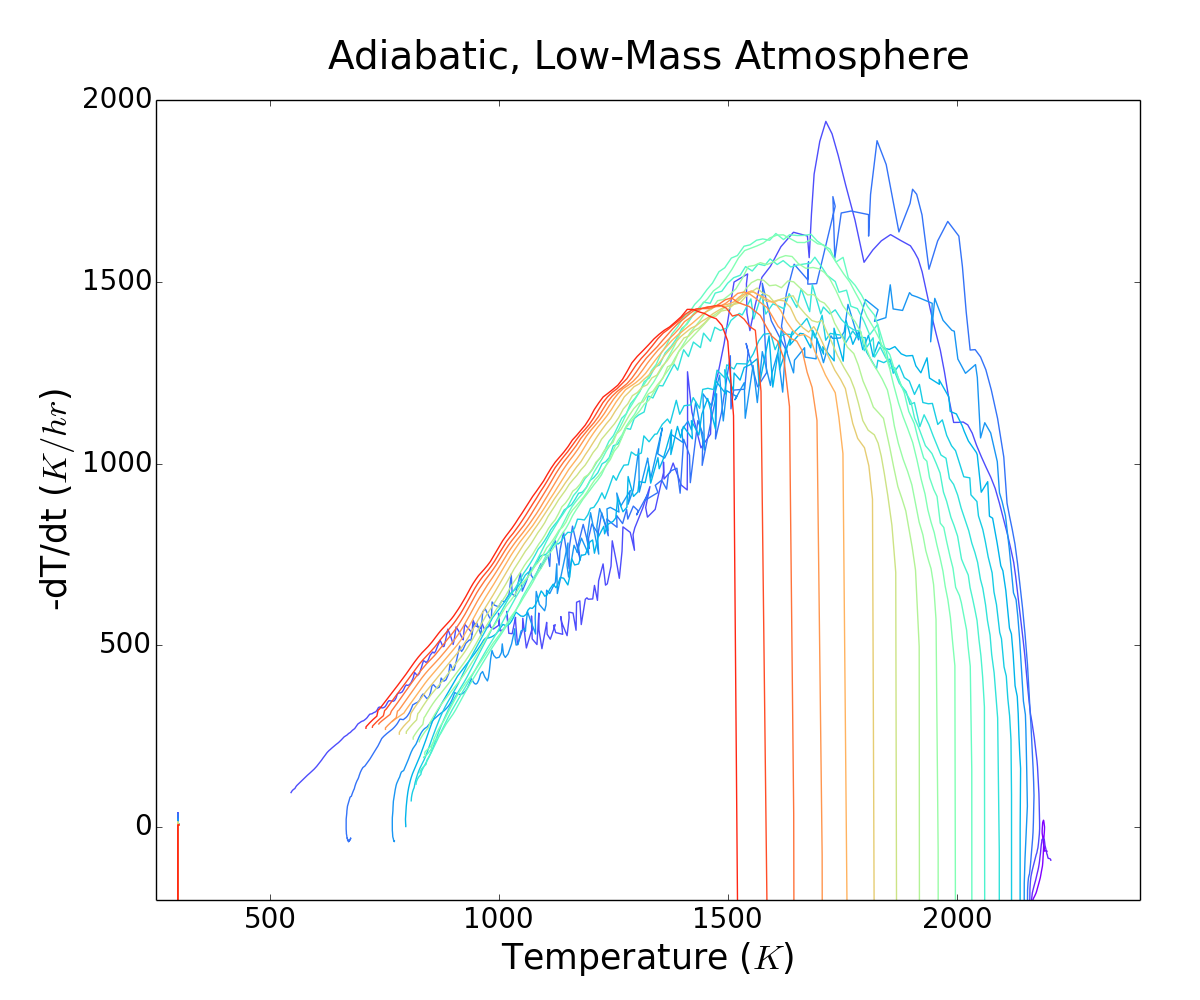} }}
	\subfloat{{\includegraphics[width=0.48\textwidth]
		{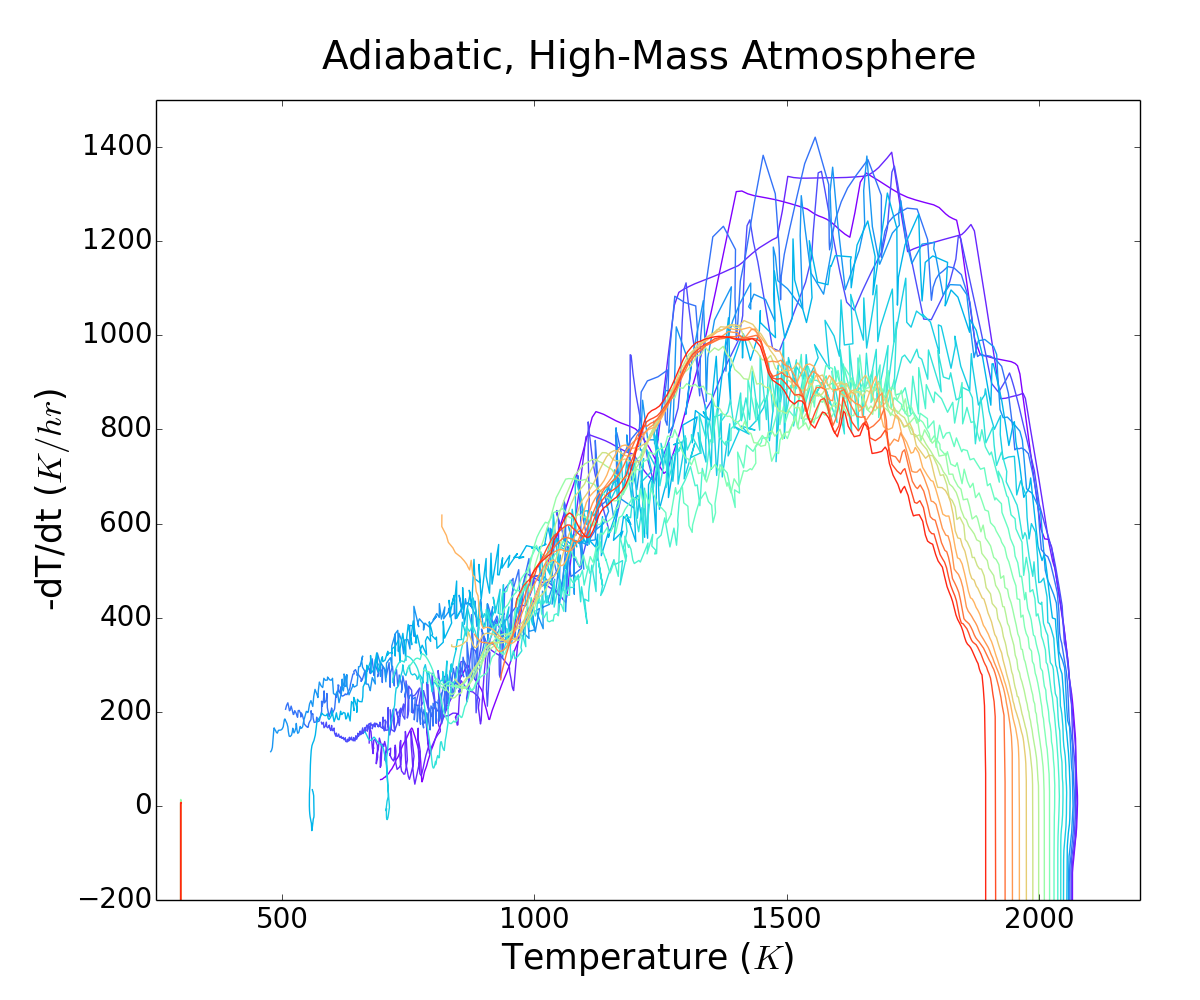} }}
	\caption{ The temperature (top) and cooling (bottom) profiles, showing the environments experienced by the super-particles in the AdiLo7 (left) and AdiHi7 (right) simulations. 
	The low-mass atmosphere experiences faster cooling rates overall, and shows a greater diversity in peak temperatures with impact radius compared with the high-mass atmosphere, although some diversity is still seen in the latter.  
	The low-mass atmosphere has a narrower bow shock, which results in the greater shock diversity across a given cross section. 
			The initial impact radius for each tracer is denoted by color, ranging from $b_i = 0 R_e$ (purple) to 
			$b_i = 2 R_e$ (red).
			Many of the chondrule precursors, if they were to cool with their environment, would experience cooling rates $\sim 1000$ K hr$^{-1}$, consistent with furnace experiments for a range of chondrule textures. }

	\label{F:adi_cooling_temp}
\end{figure}

We trace in detail 20 super-particles as they traverse the shock environment over  a range of initial impact radii $b_i$.
This is done to characterize the shock conditions experienced by solids. 
Figure~\ref{F:adi_cooling_temp} shows the corresponding temperature histories for particles in the low and high-mass atmosphere cases.   
The figure also shows the cooling rates assuming perfect thermal coupling between the solids and the gas. 
Each impact radius is denoted by the color of a given curve, where purple indicates $b_i = 0$.   
Moving through blue, green, orange, and red shows increasing $b_i$ (by $0.1 R_e$), respectively, with red indicating a $b_i = 2 R_e$. 
The  highest peak temperatures of the shocks do not vary significantly between cases ($\sim2200$ K), which is a result of buffering by molecular hydrogen dissociation.   
The low-mass atmosphere results do show slightly higher temperatures ($\sim 100\rm ~K$) due to the loss of the atmosphere on the windward side of the embryo.  
There is also a wider dispersion of shock temperatures with impact radii because the effective cross section of the embryo is smaller with the reduced atmosphere.
This dispersion is a well-known feature of the bow shock model, which can potentially process material in a diverse number of ways in a relatively confined space.
In both the high and low-mass atmosphere cases, solids are diverted around the embryo at a steep angle. 
Shocks in 1D are expected to lead to higher chondrule concentrations simply due to the high-density post-shock conditions.
In the bow shock environment, local concentrations can become even higher as particle trajectories are further concentrated, or even cross, as they traverse the shock. 

The low mass case sees cooling rates of $\sim 1500~\rm~K~hr^{-1}$ through the crystallization temperature range ($\sim1400$--1800 K), 
whereas the high mass case experiences slower rates closer to $\sim 1000~\rm K~hr^{-1}$.  
In comparison, \cite{boley_etal_apj_2013}  found rates of $\sim 4000\rm~K~hr^{-1}$ for their 7 km/s adiabatic simulation with no atmosphere around an identical obstruction. 
From this we infer that the presence of an atmosphere can significantly alter the cooling rates, at least in the adiabatic limit with perfect thermal coupling.
This may be appropriate for a very dusty environment, with the opacity dominated by small grains.  
Introducing efficient radiative cooling will increase cooling rates.

\begin{figure}[t]
	\centering
	\subfloat{{\includegraphics[width=0.48\textwidth]
		{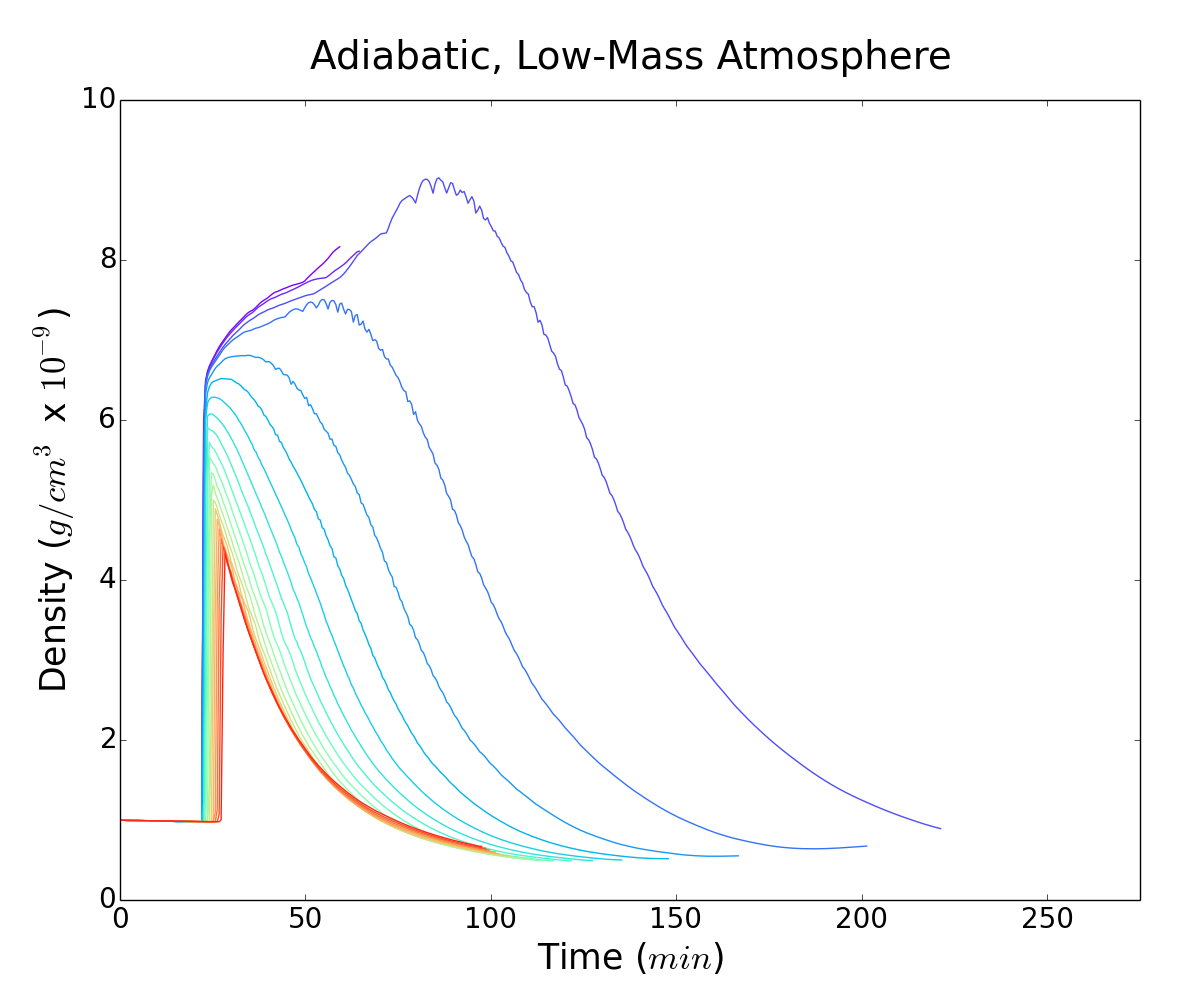} }}
	\subfloat{{\includegraphics[width=0.48\textwidth]
		{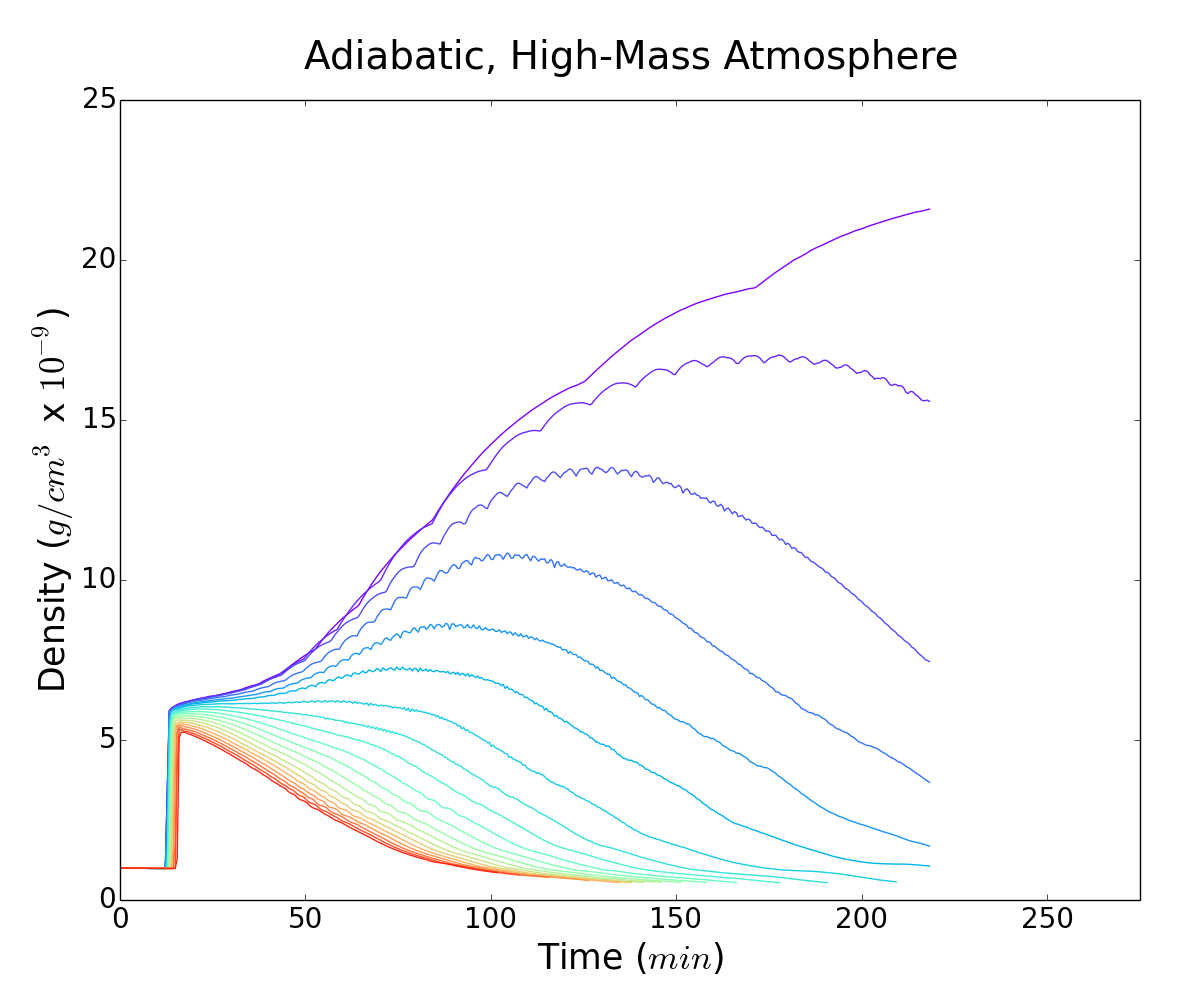} }}\\
	\subfloat{{\includegraphics[width=0.48\textwidth]
		{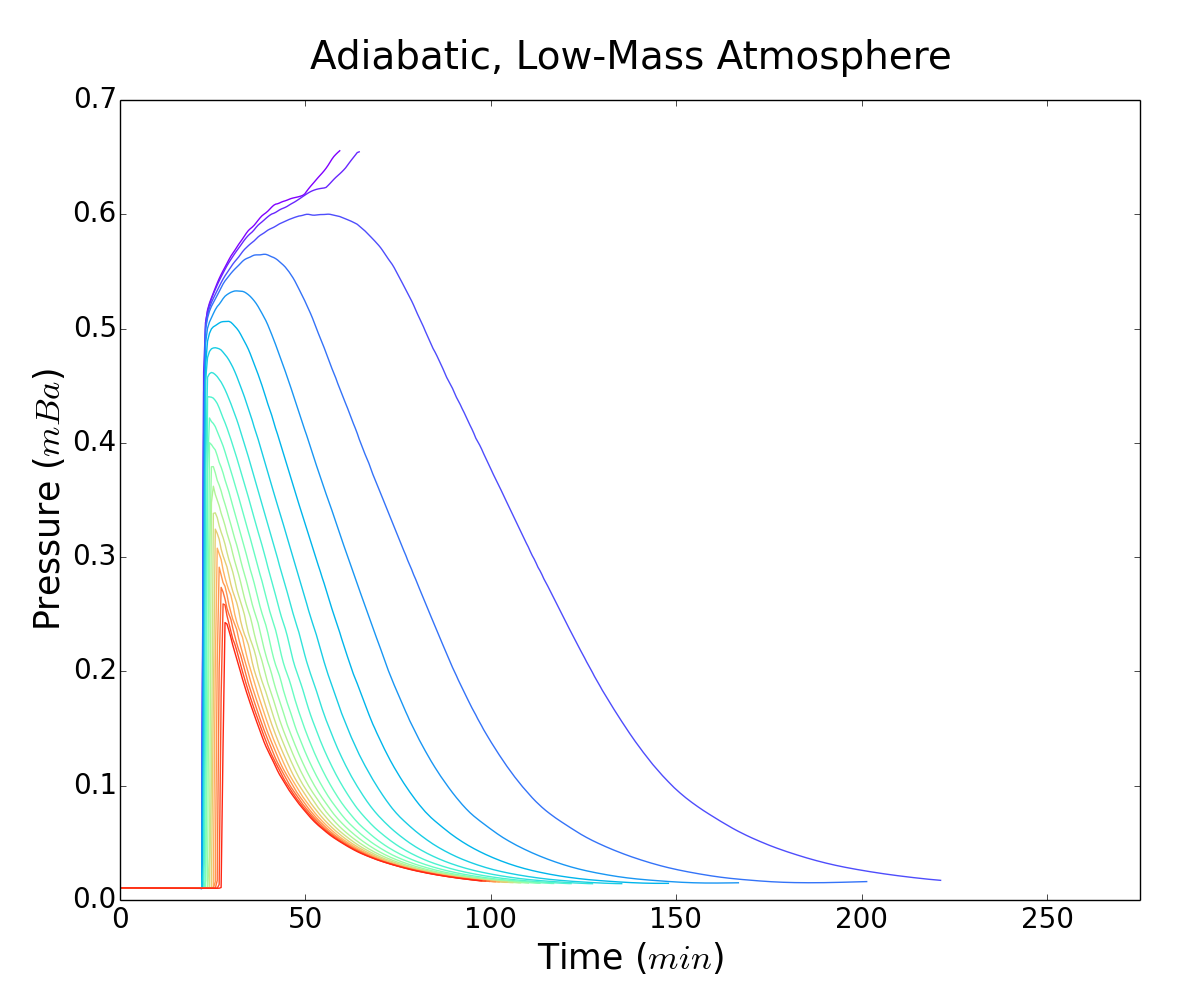} }}
	\subfloat{{\includegraphics[width=0.48\textwidth]
		{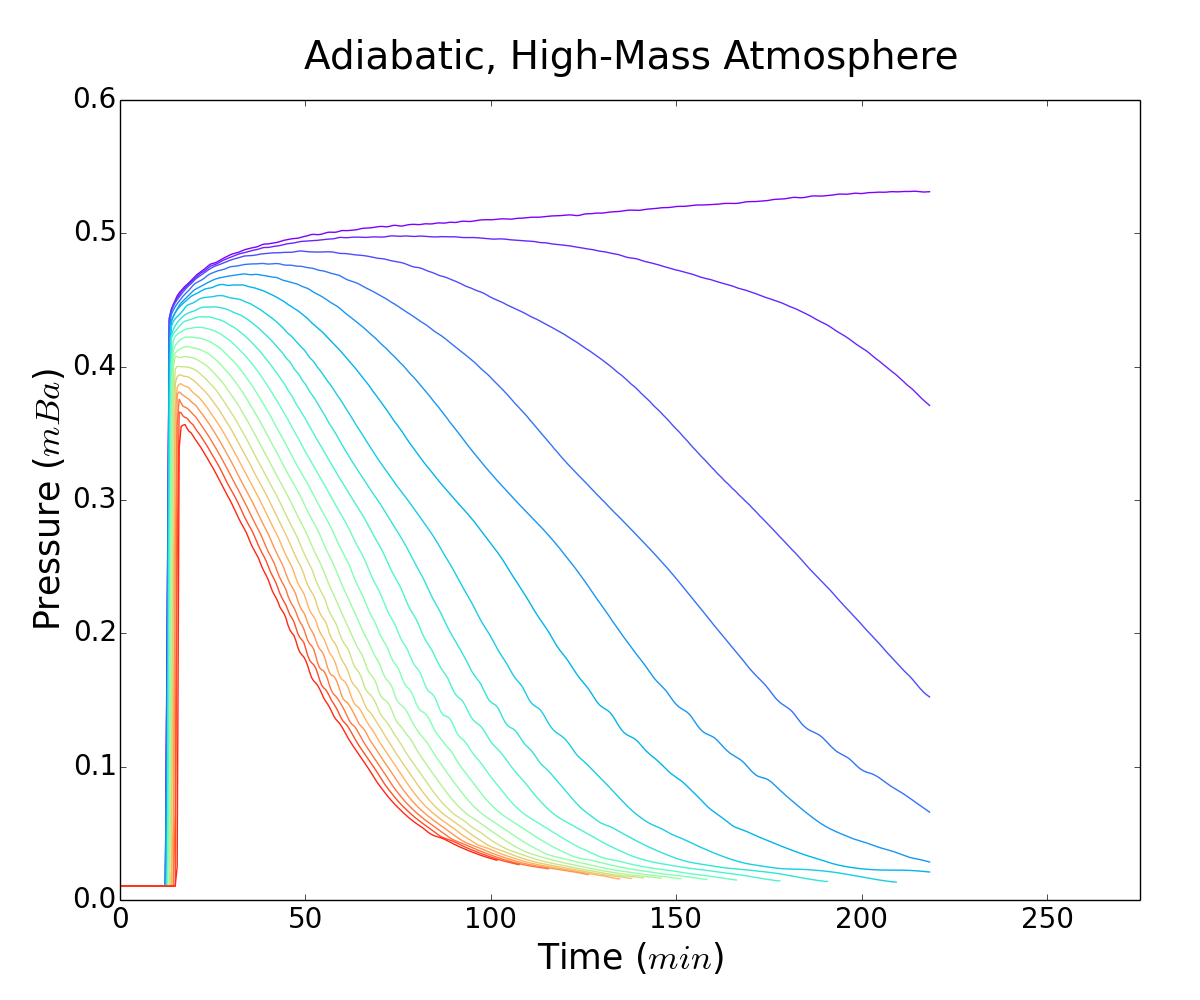} }}\\
	\caption{Similar to Figure~\ref{F:adi_cooling_temp}, but for gas density and pressure.
	The sudden increase is due to the shock. In some cases, the densities continue to increase, which is due to interactions with the embryo's atmosphere. 
	For the low-mass atmosphere (AdiLo7), the two tracers with the smallest impact radii strike the planetoid.  }
	\label{F:adi_dens_press}
\end{figure}

The plots in Figure~\ref{F:adi_dens_press} display the density and pressure histories for the tracer particles.  
The density histories remain very similar until $\sim 50$ minutes, after which the low $b_i$ particles in the high-mass simulation flow along the edge of the atmosphere and its downwind extension. 
For both cases, the density spikes are roughly five-fold upon entering the shocked region from the pre-shock environment.  
Further increases are due to interactions with the atmosphere, reaching $8\times10^{-9} \rm ~g~cm^{-3}$ for low $b_i$ particles in the low-mass case, and~$>15\times10^{-9} \rm ~g~cm^{-3}$ for the high-mass case.

Peak pressures in both simulations are around 4 to 5 mbar. 
The main difference is the longer duration of high pressure exposure for the particles in the high-mass atmosphere case.  
Some of the lowest $b_i$ particles become trapped in the atmosphere, while others are only significantly slowed by it.


\subsection{Radiative cases} 						

The radiative cases use the same setup as the adiabatic simulations, but include radiative transport with several different treatments for the local opacity.  
We focus on only the high-mass atmosphere for these runs. 
As noted in section \ref{sec:parameter_space}, opacity is broken into two components.  A solar-mixture Rosseland grey opacity \citep{pollack_etal_1994} is used to account for the small-grains, assuming they are present.  
This opacity is scaled by a factor, which is denoted in the name of each radiative simulations, e.g., Rad 1, 0.1, and 0.01.  
We refer to these smaller grains as simply dust. 
The other component comes from the chondrule precursors themselves, $\kappa_c$. 
 This value is also varied to reflect an assumed chondrule concentration, which is discussed for each corresponding subsection below. 


\subsubsection{Rad 1, $\kappa_c=$10 (Rd1k10)}  

\begin{figure}[hbt]
	\centering
	\subfloat{{\includegraphics[width=\textwidth]
		{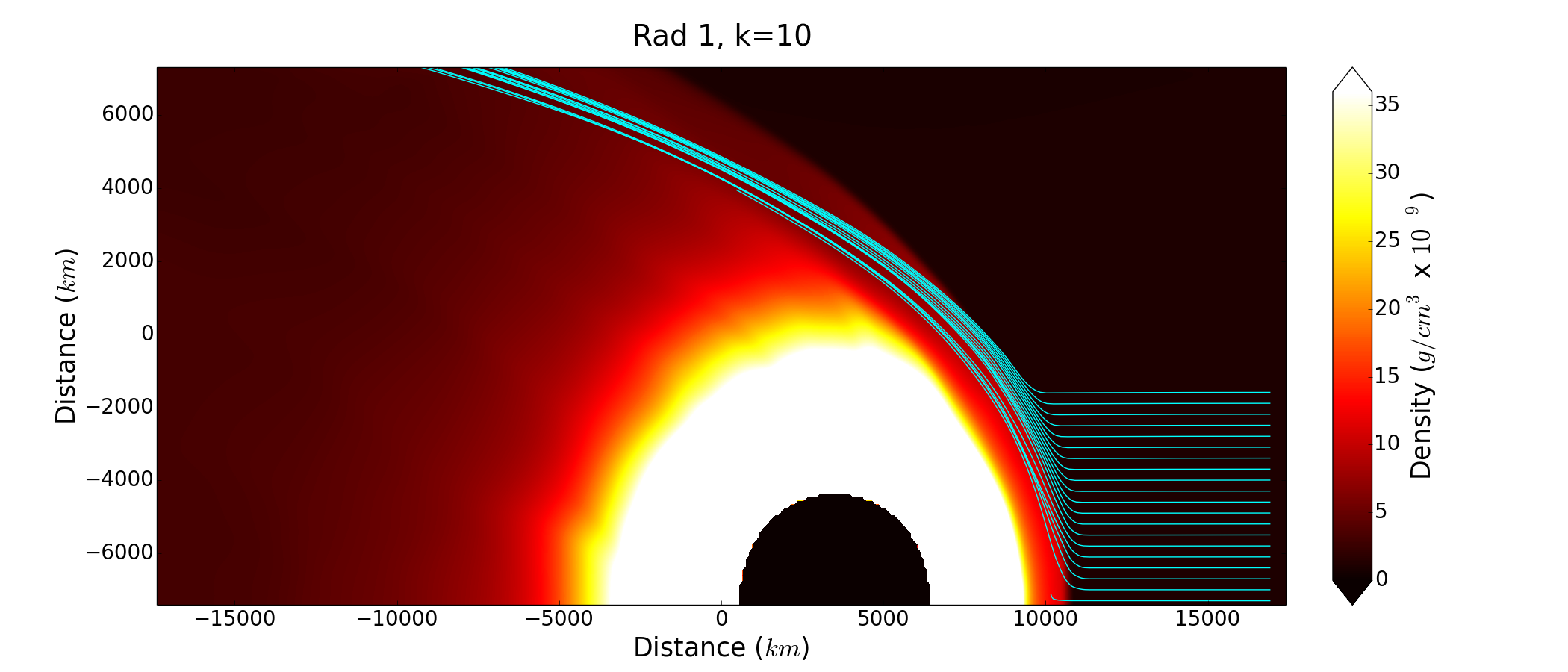} }}\\
	\subfloat{{\includegraphics[width=\textwidth]
		{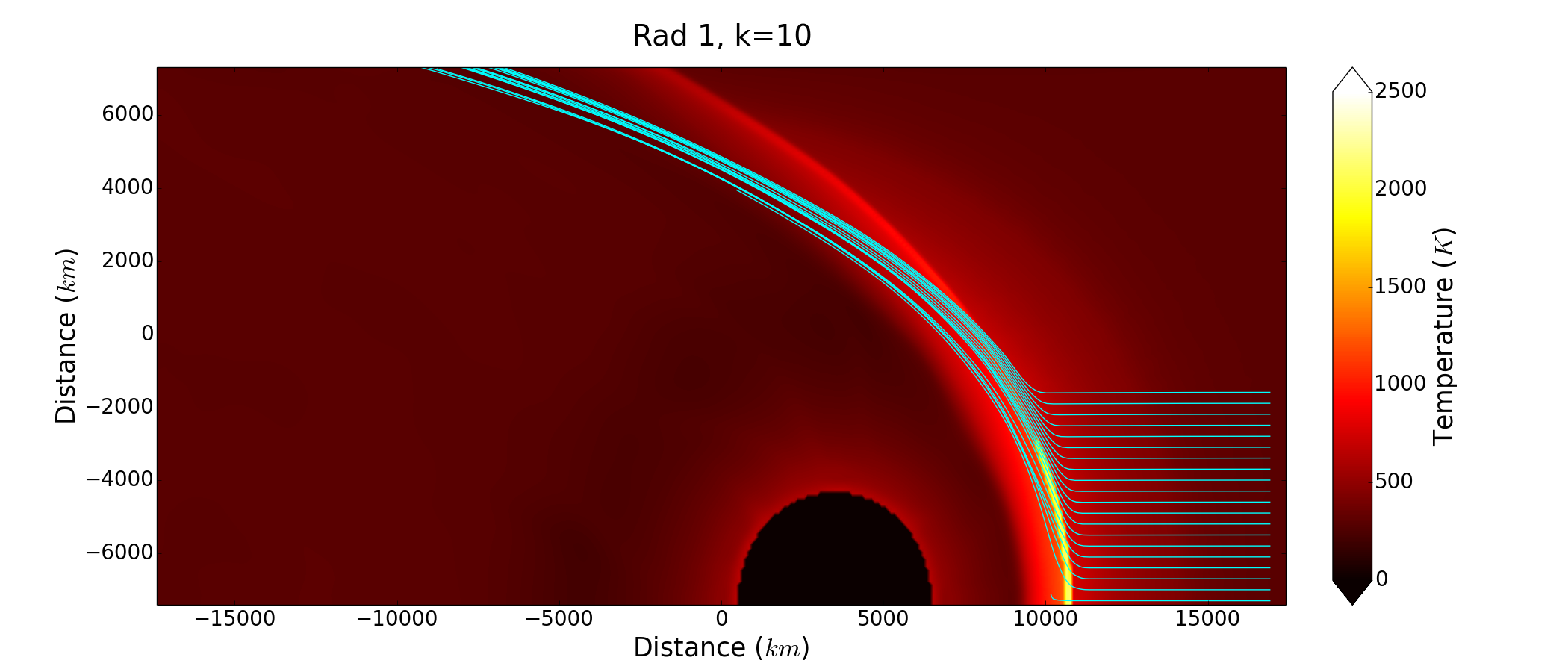} }}\\
	\caption{The density (top) and temperature (bottom) bow shock morphologies for run Rd1k10. 
	                  Particle trajectories (cyan curves) become tightly concentrated as the particles are diverted around the embryo.  
	                  Unlike the adiabatic simulations, the bow shock region above (roughly) 1500 K is limited to a very small region, both in thickness and impact radius. 
			 A radiative precursor is evident, which can heat solids above 500 K before they experience the bow shock. }
	\label{F:rad1_k10_traject}
\end{figure}

This case simulates the standard dust opacity as envisaged for the Solar Nebula \citep{pollack_etal_1994}. 
 The resulting  bow shock structure is shown in Figure \ref{F:rad1_k10_traject}.  
 A radiative shock precursor forms that raises the gas temperature above $\sim 600$~K in the pre-shock region. 
 Only a very thin section of the shock front reaches a 2000 K peak  temperature, dropping by roughly 1000~K in a very short distance. 
 Particle trajectories become concentrated into narrow regions, for which some of the trajectories cross due to the thin shock front and interactions between the wind and the atmosphere.

\begin{figure}[t]
	\centering
	\subfloat{{\includegraphics[width=0.48\textwidth]
		{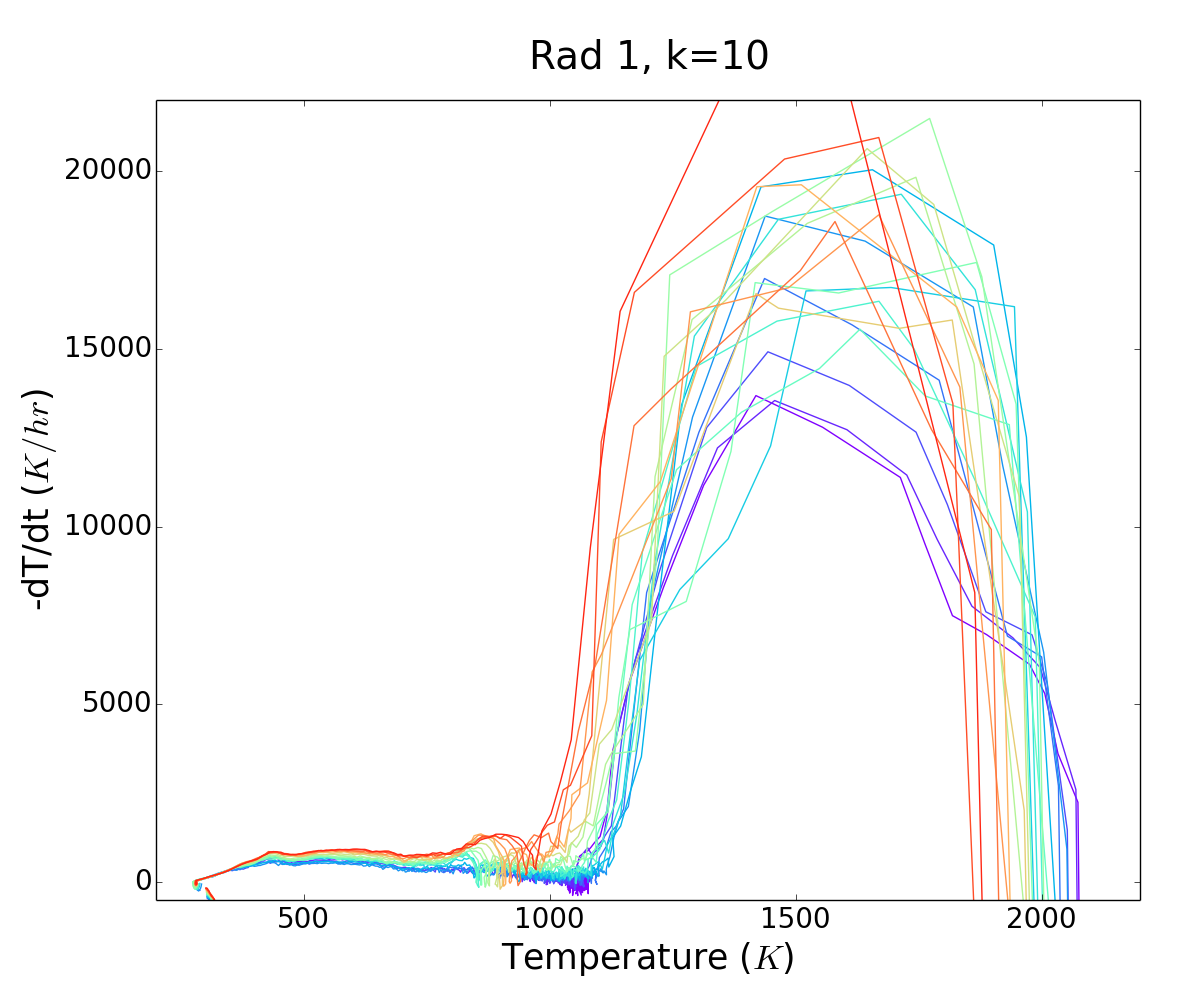} }}
	\subfloat{{\includegraphics[width=0.48\textwidth]
		{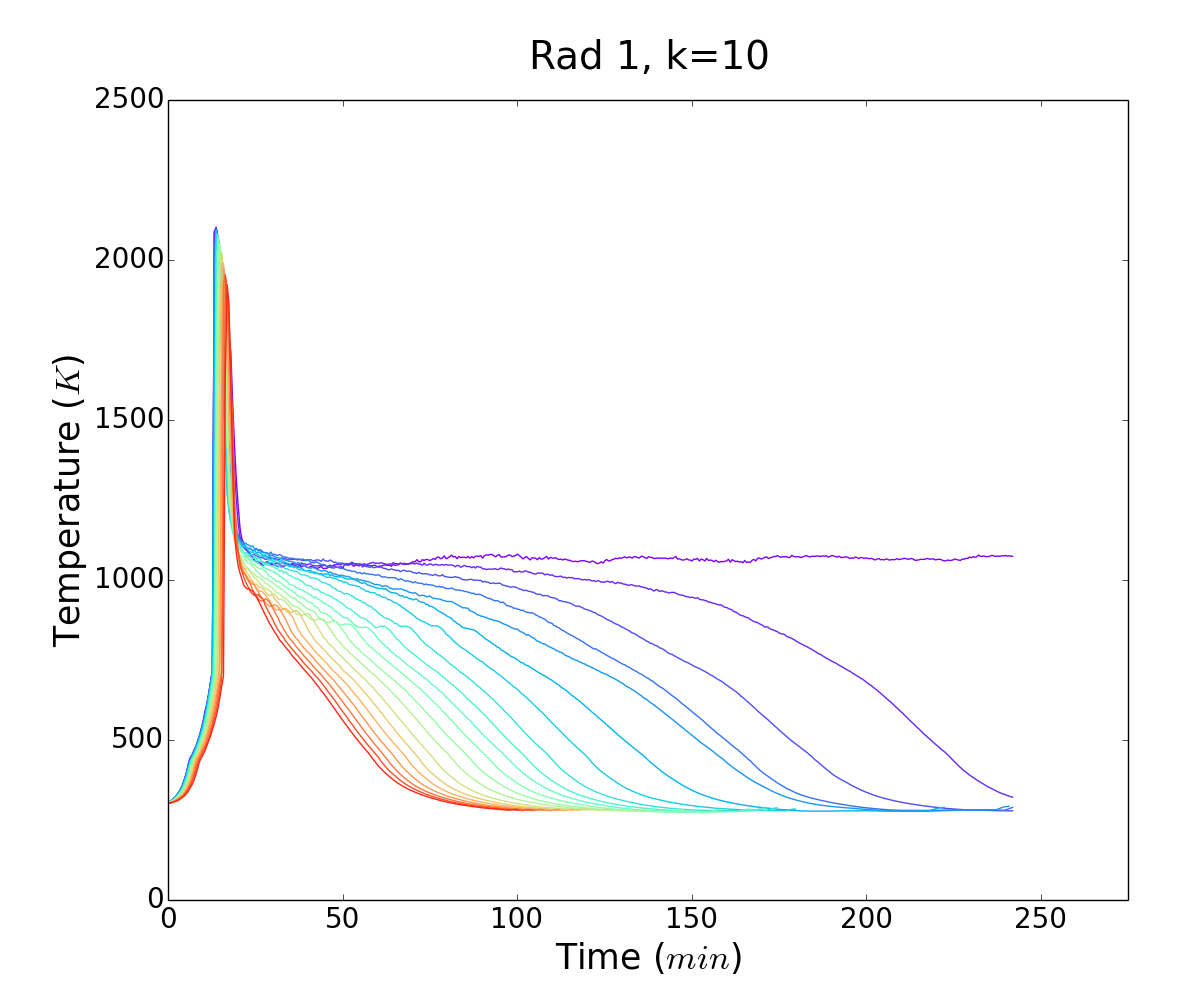} }}\\
	\subfloat{{\includegraphics[width=0.48\textwidth]
		{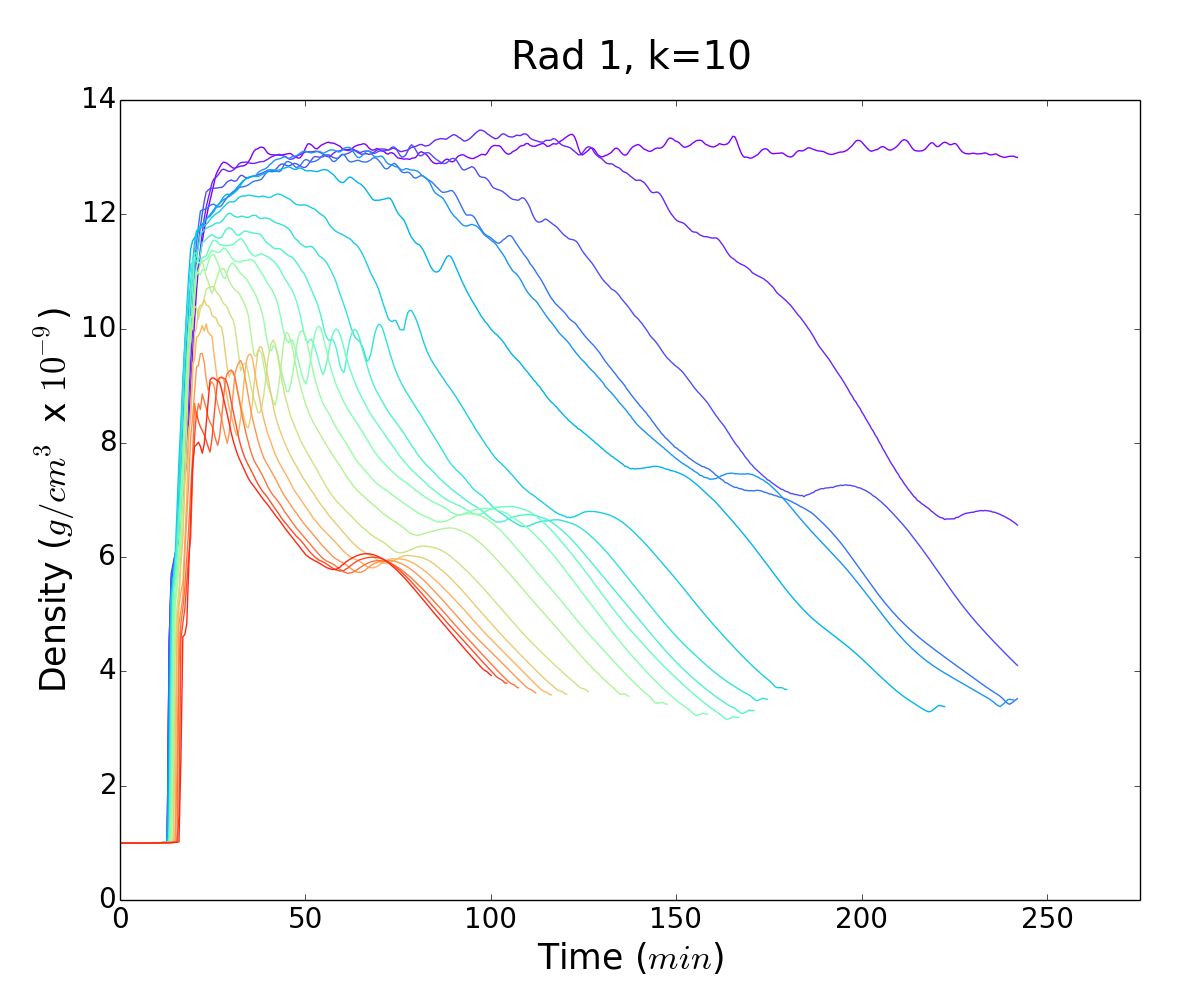} }}
	\subfloat{{\includegraphics[width=0.48\textwidth]
		{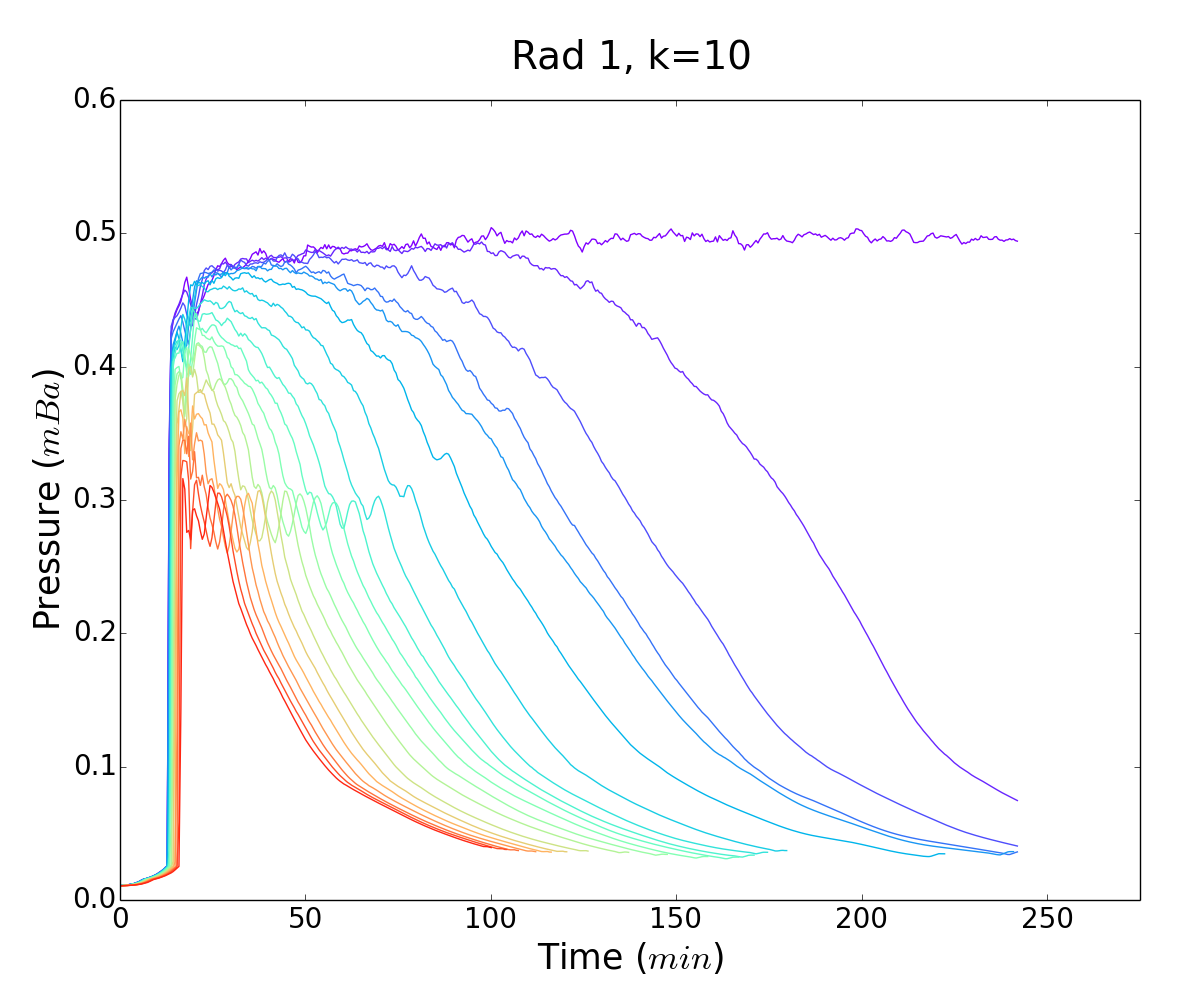} }}\\
	\caption{Similar to Figures \ref{F:adi_cooling_temp} and \ref{F:adi_dens_press}, but for only Rd1k10.
	            There is extremely rapid cooling past the shock front due to effective radiation transport. 
	            As before, the lowest $b_i$ 
			particles interact with the embryo's atmosphere, maintaining high densities and pressures for extended time periods.}
	\label{F:rad1_k10_hist}
\end{figure}

Particle histories are displayed in Figure~\ref{F:rad1_k10_hist}.  
The dust and chondrules provide a large supply of radiators, but do not make the shock front optically thick, allowing efficient heat loss by radiation.  
As a result, we see cooling rates in excess of $10^4~\rm K~hr^{-1}$ through the crystallization range (taken here to be between $\sim 1400$~K and 1800~K).  
Peak temperatures reach $\sim 2000$~K, before rapidly dropping. 
The gas density peaks at $13\times10^{-9} \rm~g~cm^{-3}$ for low $b_i$ particles, but only $8\times10^{-9}\rm~g~cm^{-3}$ for the particles that are more distant from the atmosphere.  
Gas pressure similarly peaks at 0.5 mbar for low $b_i$ and 0.3 mbar for larger $b_i$.

\subsubsection{Rad 0.1, $\kappa_c=$10 (Rd0.1k10) }   

\begin{figure}[hbt]
	\centering
	\subfloat{{\includegraphics[width=\textwidth]
		{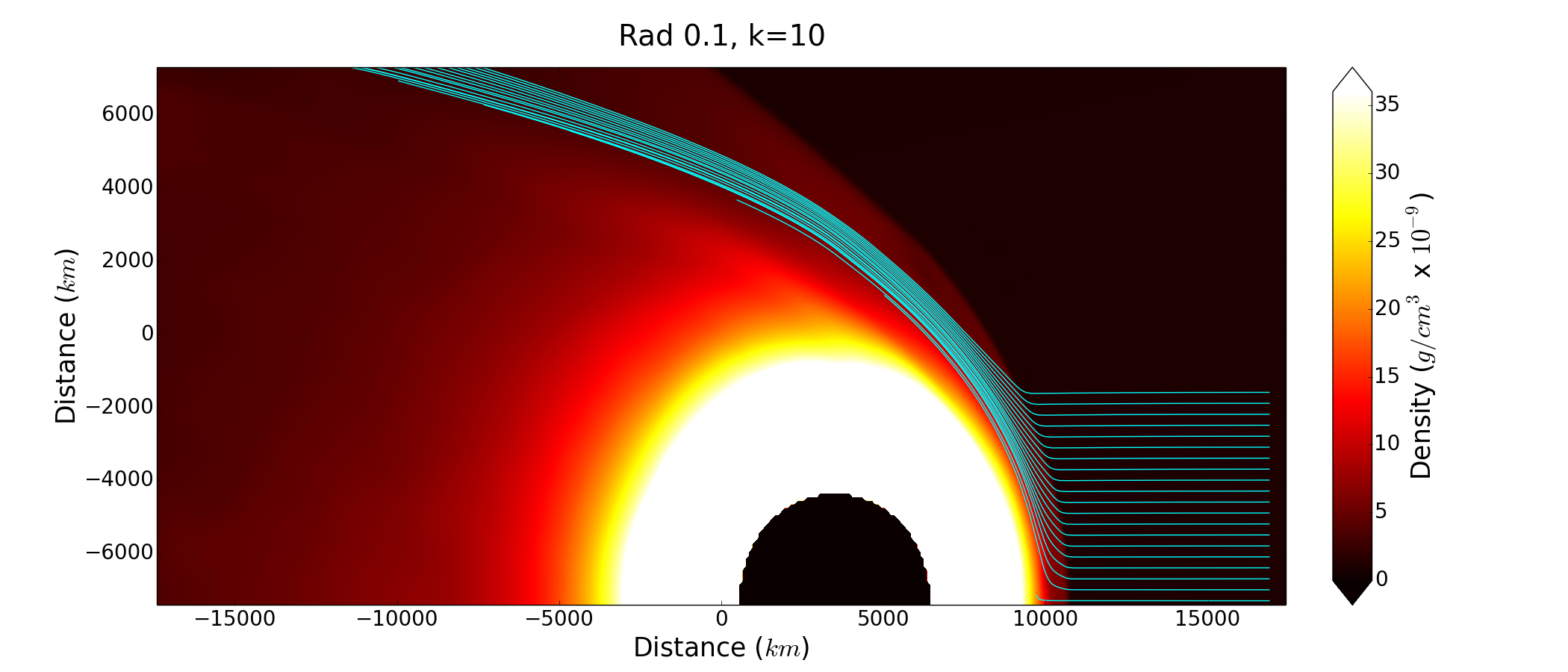} }}\\
	\subfloat{{\includegraphics[width=\textwidth]
		{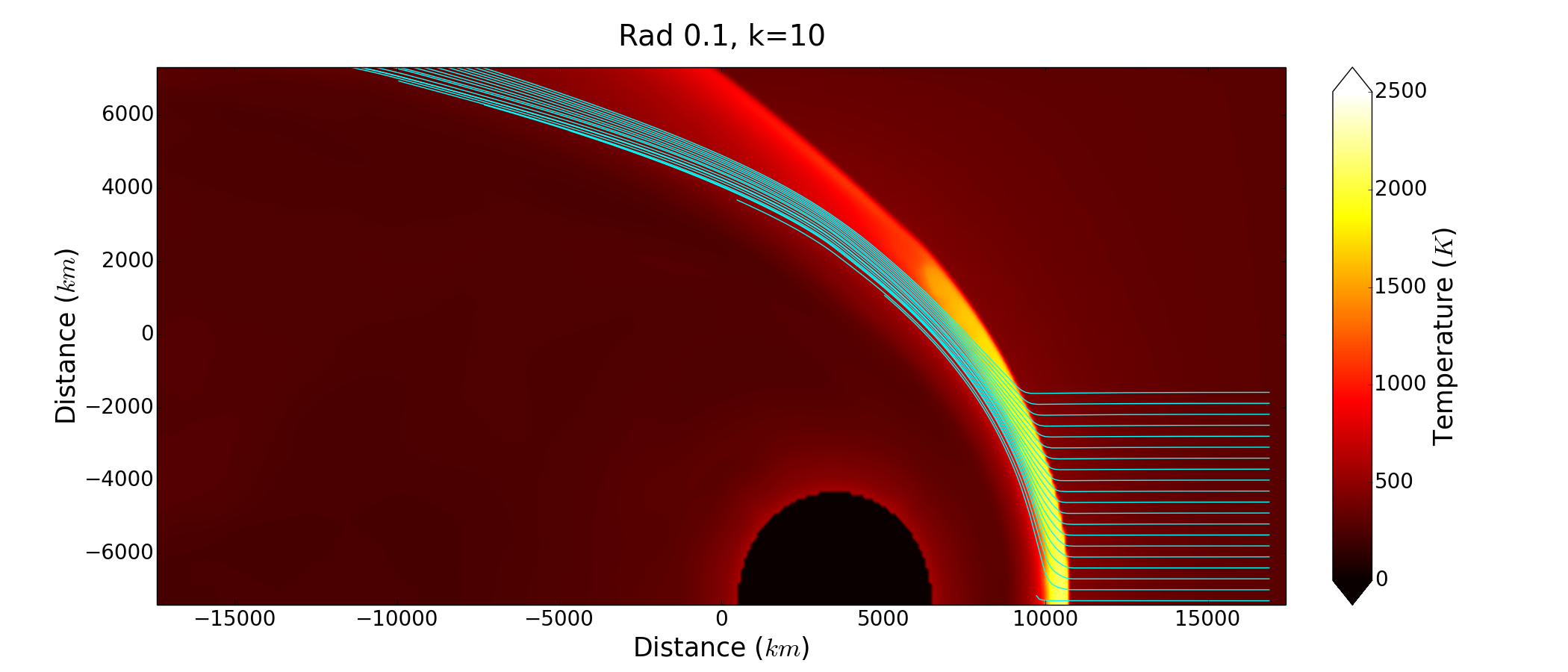} }}\\
	\caption{Similar to Figure \ref{F:rad1_k10_traject}, but for Rd0.1k10. A broader high temperature region is present due to less efficient cooling (compared with Rd1k10).
			There is still a noticeable radiative precursor, but remains less than about 500 K. }
	\label{F:rad0.1_k10_traject}
\end{figure}

Lowering the dust opacity by a factor of 10 reduces the number of effective radiators, but also decreases the optical depth.  
The net result is a lower radiative cooling efficiency for the gas, although chondrule precursors may still cool rapidly if thermal coupling with the gas is weak. 
The shock structure is shown in Figure~\ref{F:rad0.1_k10_traject}.  
The radiative precursor is small ($\sim 100$ K) due to inefficient radiation absorption.
The low radiative efficiency also allows the high-temperature shock region to be broader than seen in the Rad 1 case.

This case shows a markedly slower cooling of the gas than in the Rad 1 run, resulting in cooling rates of 4000-6000 K hr$^{-1}$ for the chondrule melts through the crystallization range (assuming efficient coupling between the gas reservoir and the chondrule).  Peak temperatures are only slightly lower, falling between 1800-2100 K according to $b_i$. 
Density climbs rapidly to $5\times10^{-9} \rm ~g~cm^{-3}$ upon entering the shock front, then rises more slowly as particles interact with the atmosphere. 
Low $b_i$ particles achieve  $15-20\times10^{-9} \rm ~g~cm^{-3}$ and may become trapped by the upper atmosphere.  
Pressure peaks at 0.5 mbar for particles very close to the atmosphere, again falling off with increasing $b_i$.

\begin{figure}[t]
	\centering
	\subfloat{{\includegraphics[width=0.48\textwidth]
		{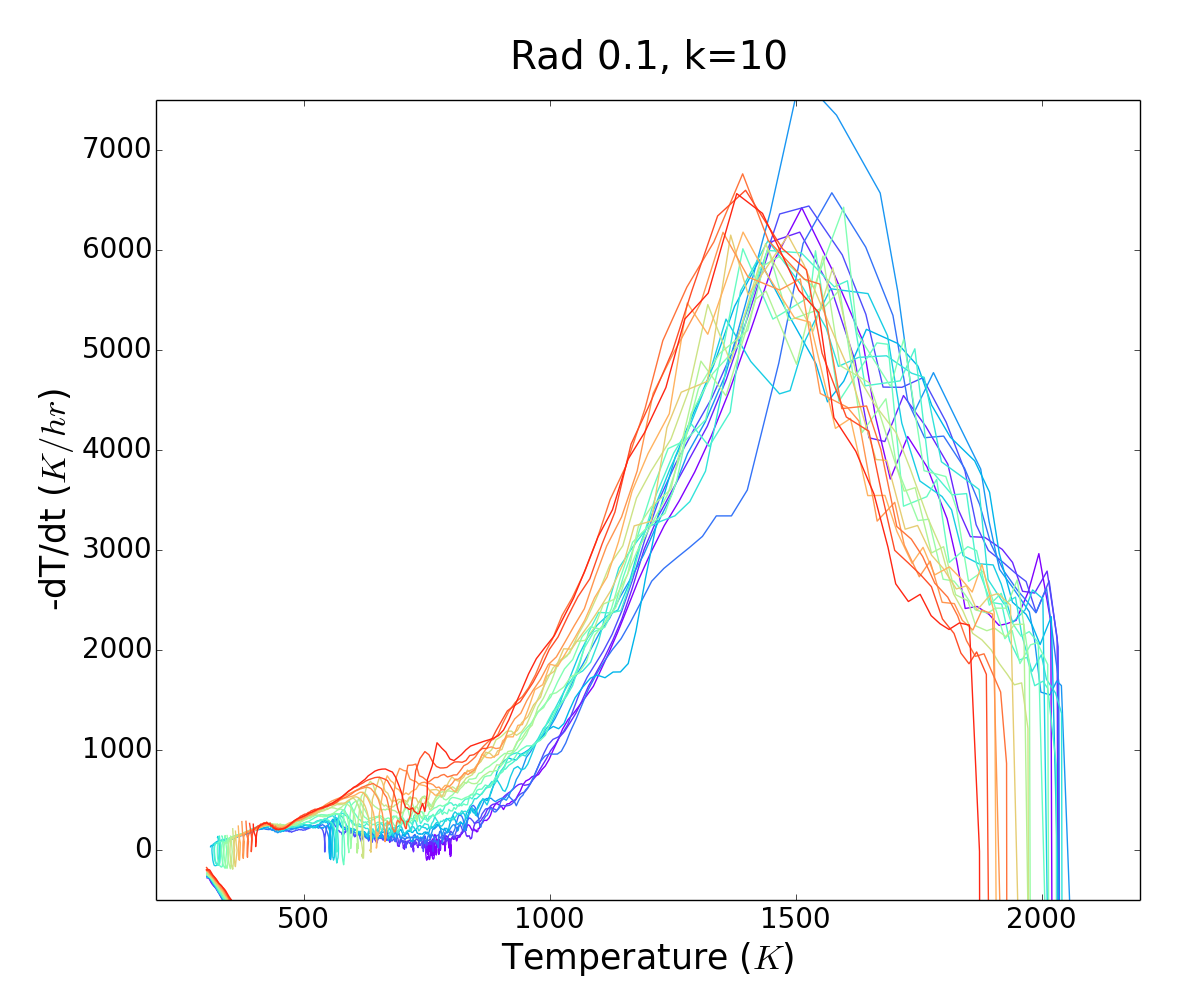} }}
	\subfloat{{\includegraphics[width=0.48\textwidth]
		{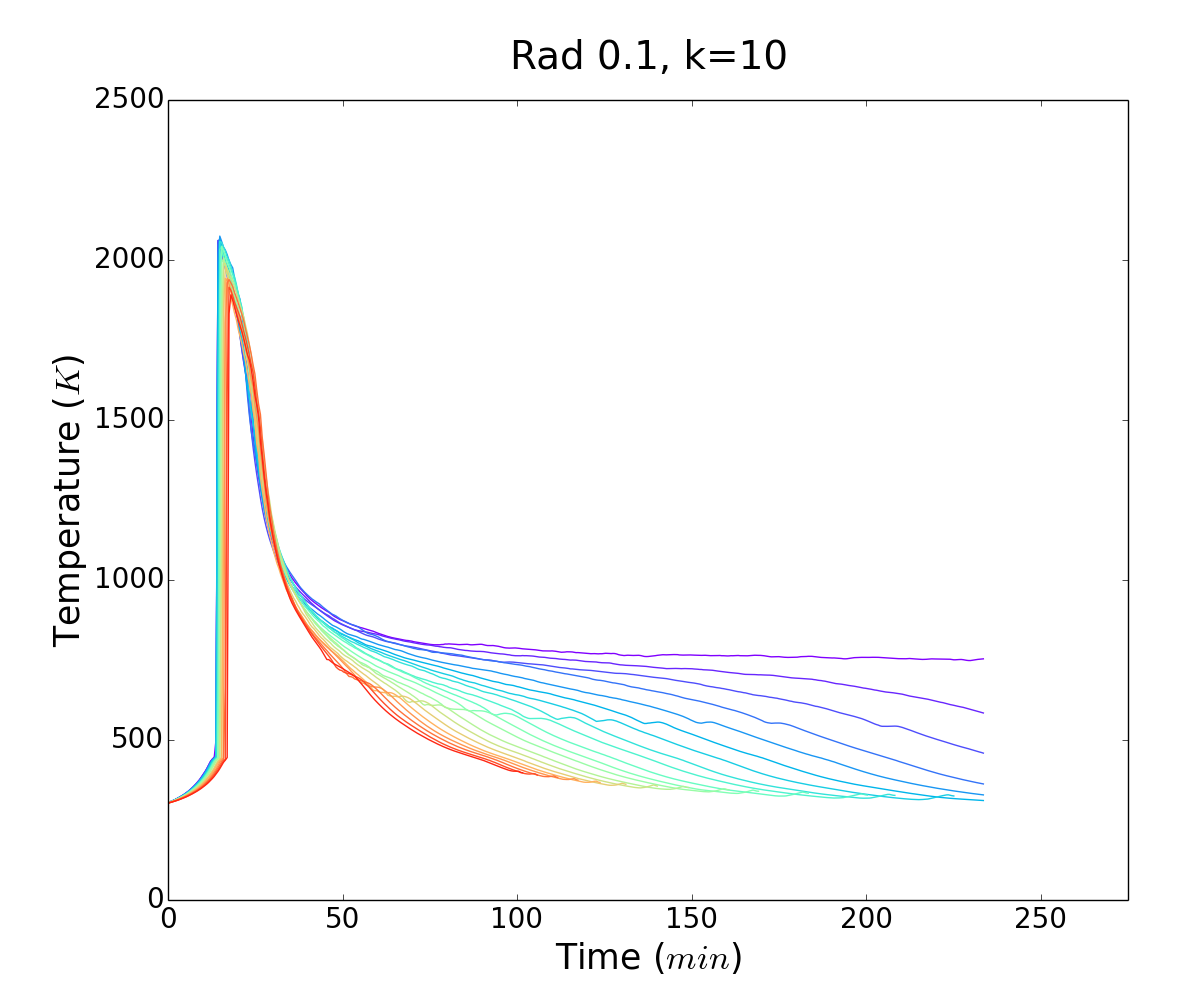} }}\\
	\subfloat{{\includegraphics[width=0.48\textwidth]
		{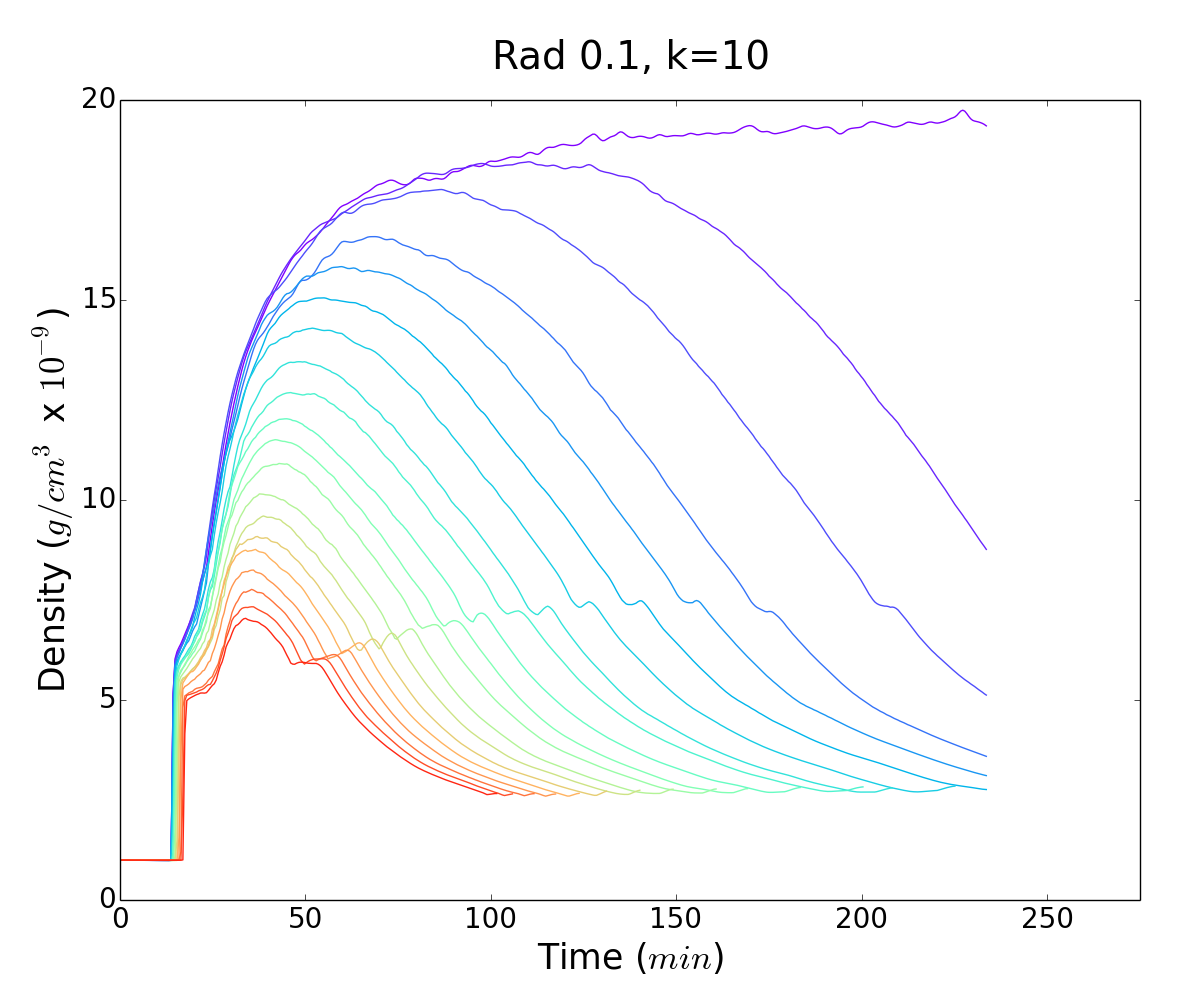} }}
	\subfloat{{\includegraphics[width=0.48\textwidth]
		{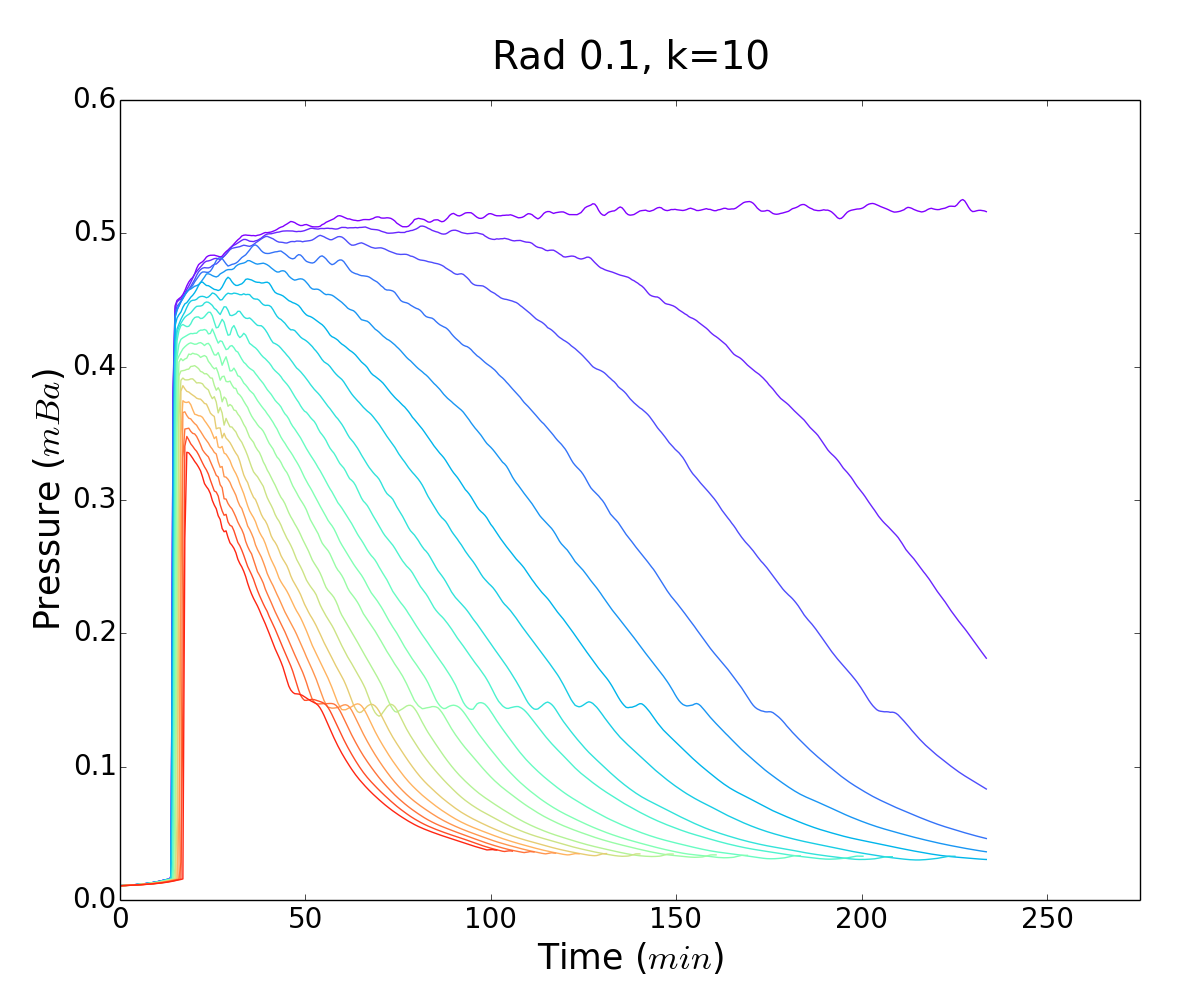} }}\\
	\caption{Similar to Figure \ref{F:rad1_k10_hist}, but for Rd0.1k10.  The cooling rates have dropped below $10^4$ K hr$^{-1}$, though they are still quite high
			 for most chondrule formation.  Moreover, these cooling rates trace the evolution of the gas, and not necessarily the potential chondrule temperature.  
			 As before, the density history shows a slow rise after the particles pass through the shock and skirt the upper 
			 atmosphere of the embryo.  Pressures remain similar to those seen in Rd1k10. }
	\label{F:rad0.1_k10_hist}
\end{figure}

While cooling rates are reduced, we stress that the gas is optically thin.  
As a result, if gas densities are not high enough to enforce thermal coupling between the chondrule precursors and the reservoir, then the particles will be colder than that shown in Figure~\ref{F:rad0.1_k10_hist}. 
The bow shock conditions explored here are near the limit where thermal coupling becomes plausible (section \ref{sec:shocks_brief}).

\subsubsection{Rad 0.01, $\kappa_c=10$ (Rd0.01k10)}   

\begin{figure}[hbt]
	\centering
	\subfloat{{\includegraphics[width=\textwidth]
		{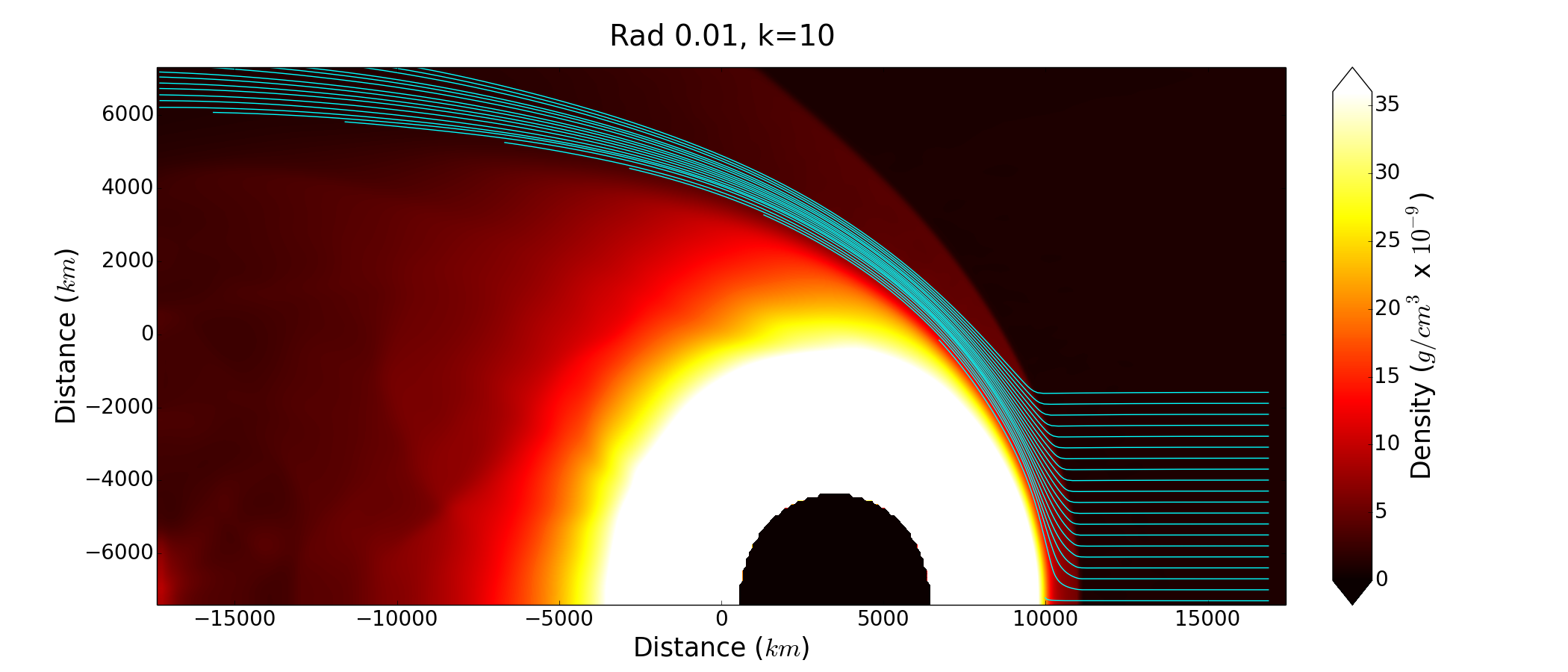} }}\\
	\subfloat{{\includegraphics[width=\textwidth]
		{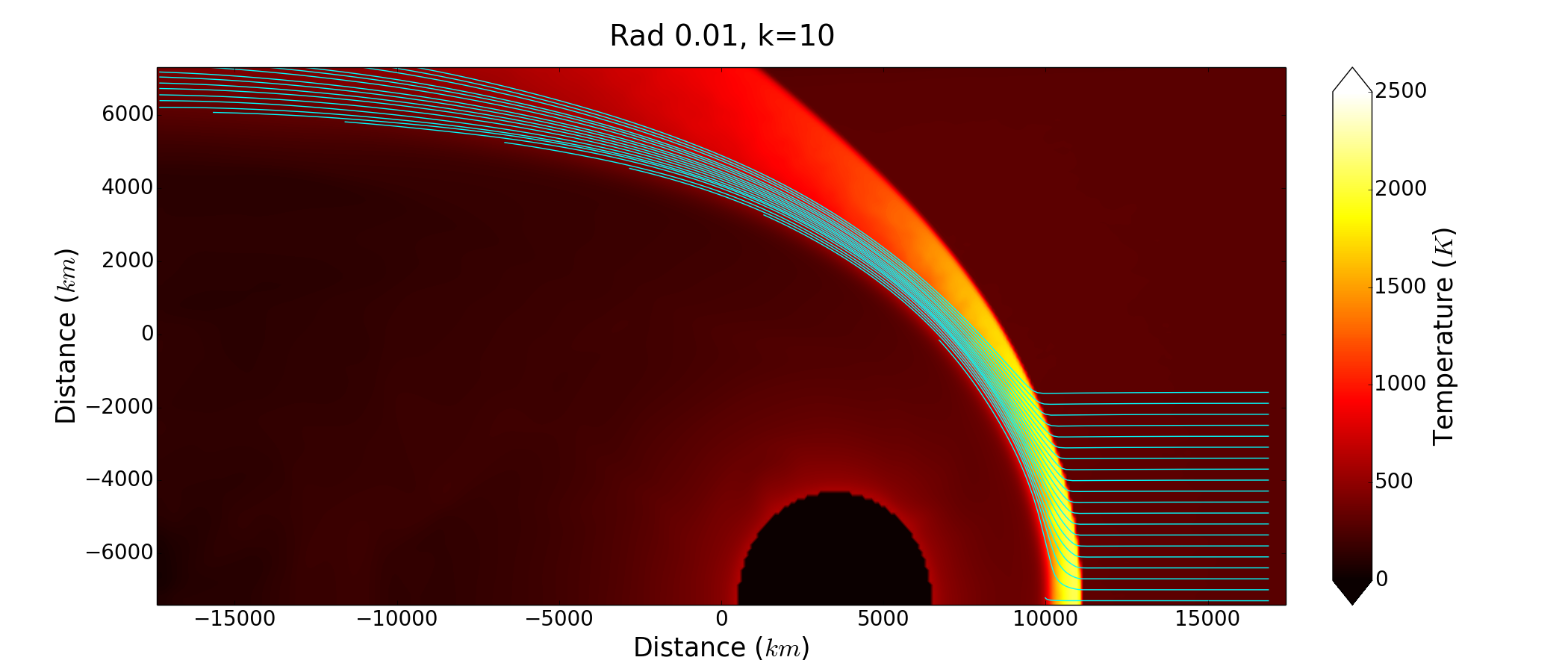} }}\\
	\caption{Similar to Figure \ref{F:rad1_k10_traject}, but for Rd0.01k10. 
	 Further broadening of the high-temperature zone in the shock front indicates a further reduction in radiative 
			efficiency from Rd0.1k10. Nonetheless, there is still sufficient cooling to maintain a noticeable difference between the bow shock morphologies of Rd0.01k10 and the adiabatic simulations.  Almost no radiative precursor is visible due to the low optical depths through the bow shock and the pre-shock regions. .}
	\label{F:rad0.01_k10_traject}
\end{figure}

By dropping the dust opacity further to 100 times lower than Rd1k10, we reduce radiative efficiency further (both absorption and emission).
  The shock front retains high temperatures for more extended regions, and there is almost no sign of radiative heating in the pre-shock zone (Fig.~\ref{F:rad0.01_k10_traject}).

\begin{figure}[t]
	\centering
	\subfloat{{\includegraphics[width=0.48\textwidth]
		{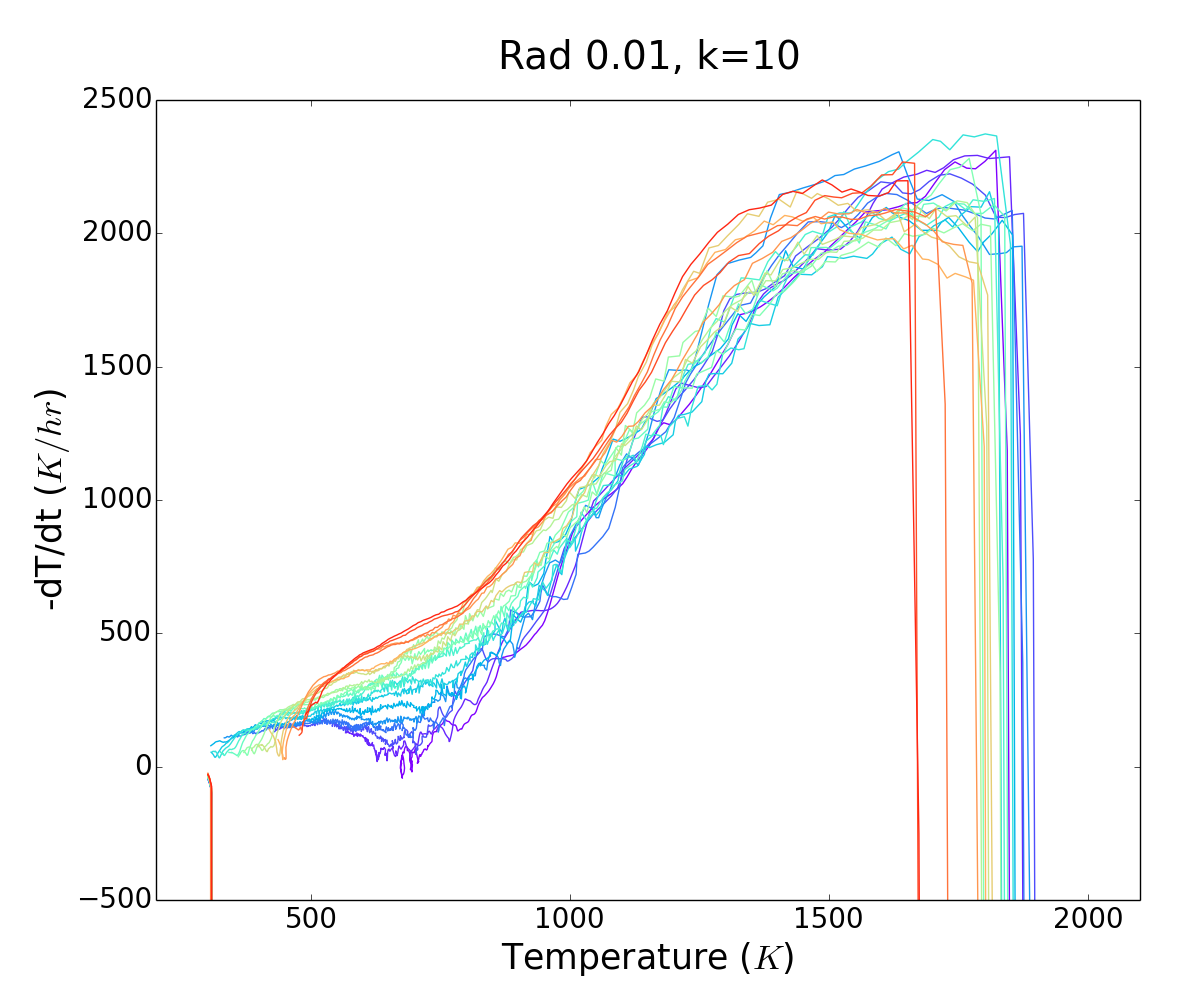} }}
	\subfloat{{\includegraphics[width=0.48\textwidth]
		{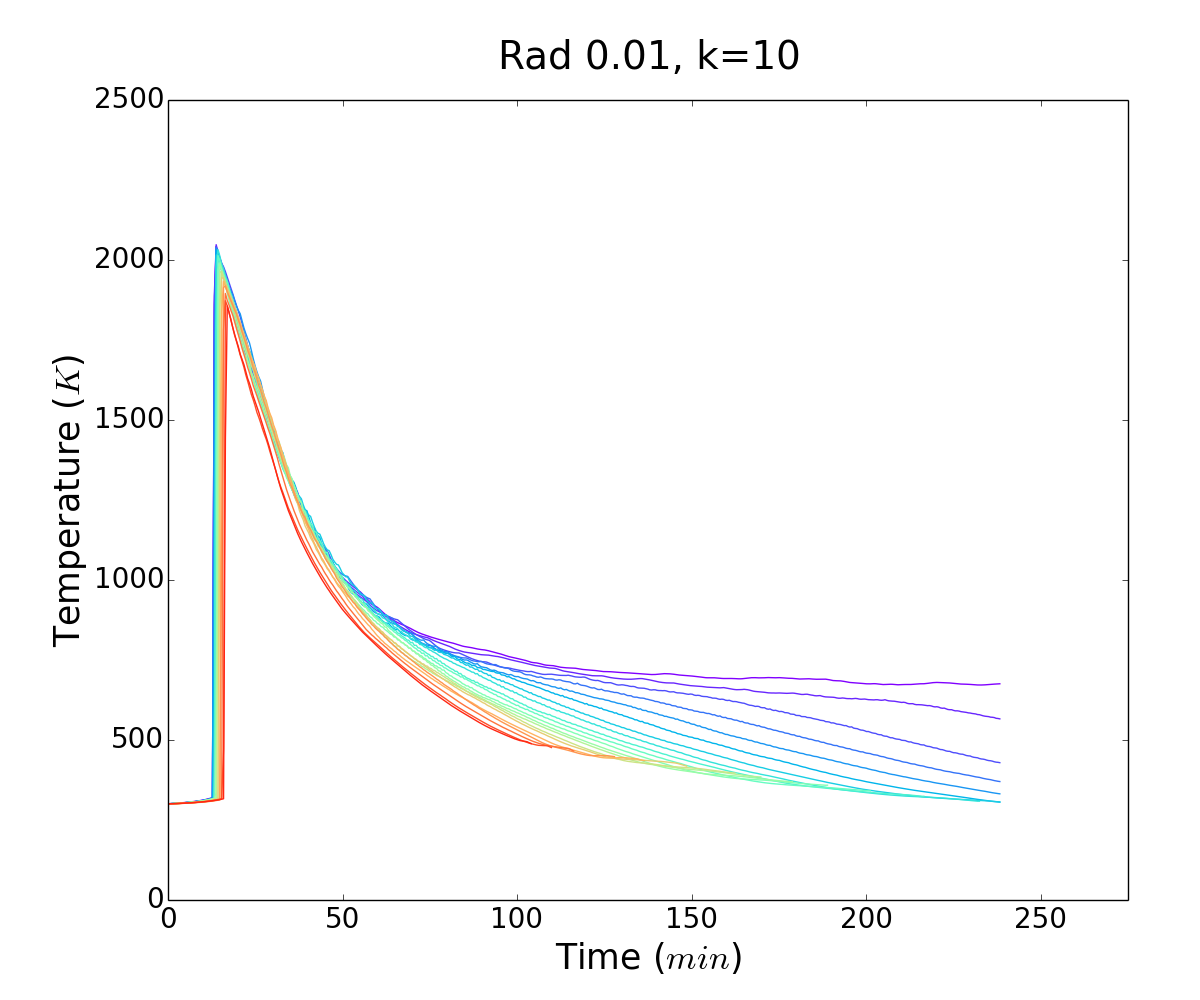} }}\\
	\subfloat{{\includegraphics[width=0.48\textwidth]
		{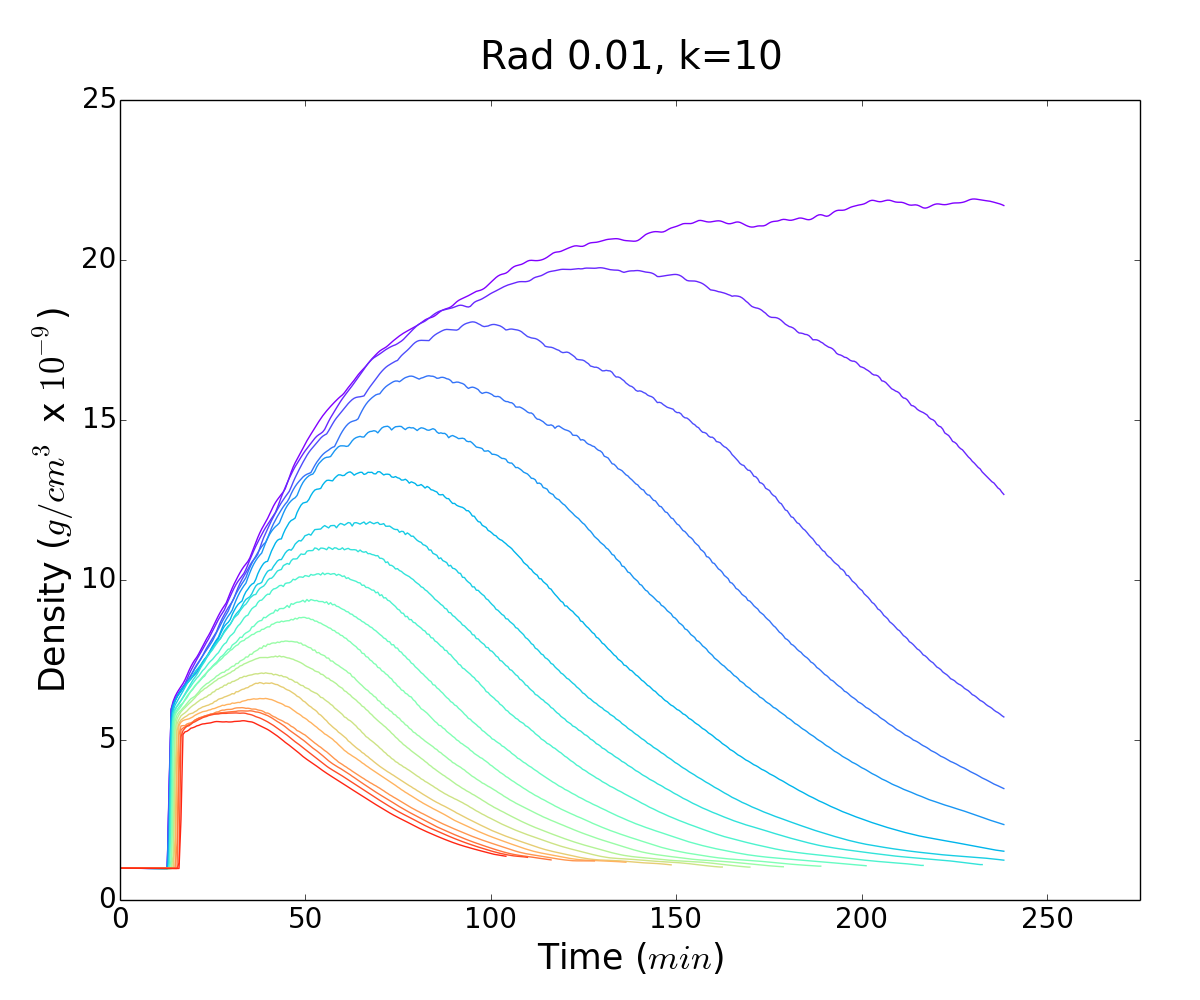} }}
	\subfloat{{\includegraphics[width=0.48\textwidth]
		{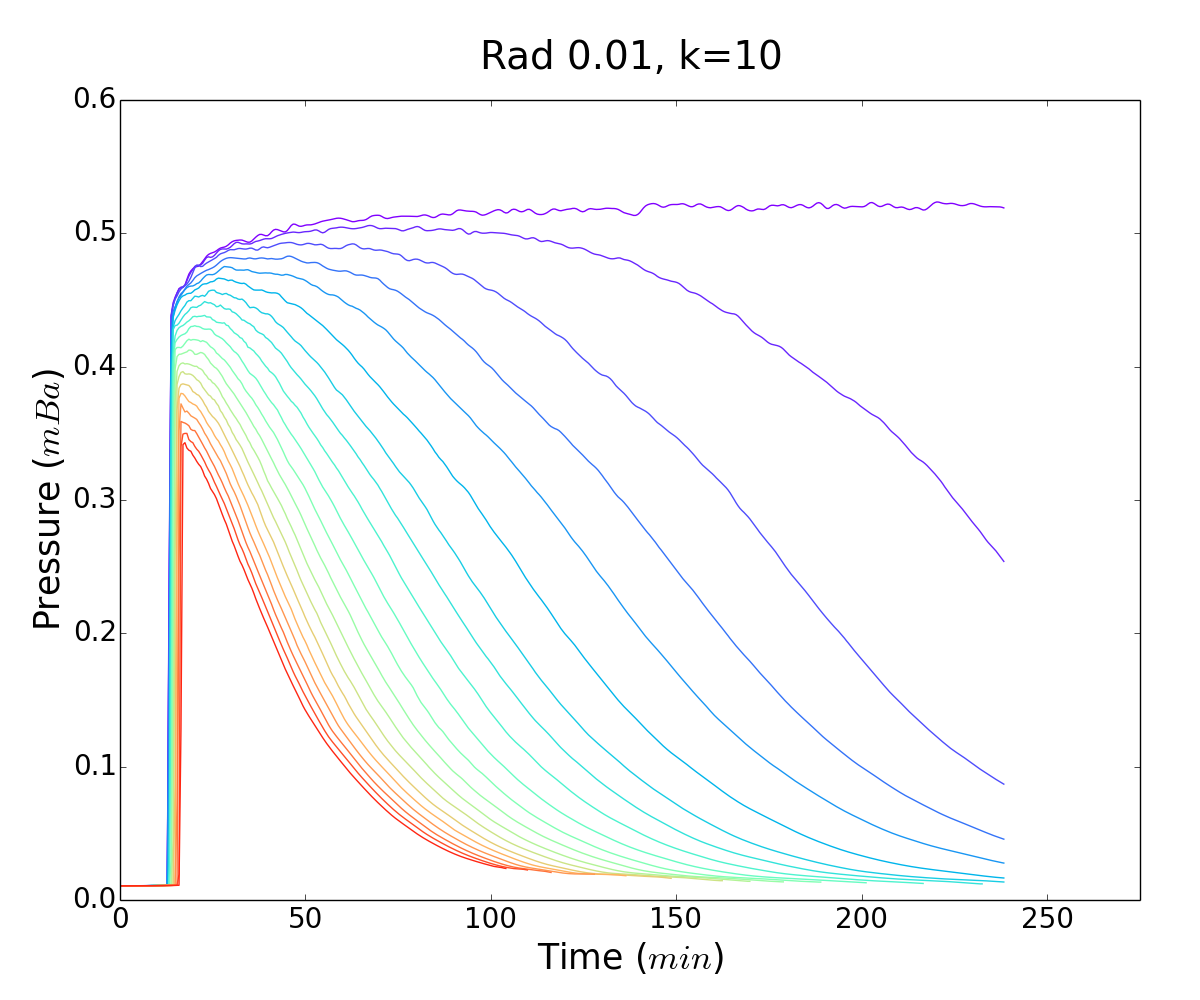} }}\\
	\caption{Similar to Figure \ref{F:rad1_k10_hist}, but for Rd0.01k10.  The cooling rates are plausibly in the chondrule-forming regime; however, very efficient (potentially unrealistic) thermal coupling between the gas and solids would be required.  Density and pressure
			 histories are again very similar to the Rd1k10 case. }
	\label{F:rad0.01_k10_hist}
\end{figure}

The cooling rates through the crystallization range are about  $2000~\rm K~hr^{-1}$.  
Peak temperatures are between 1800-2000 K for the considered $b_i$. 
 Density histories are similar to the Rad 0.1 case, with a sharp rise to $5\times10^{-9}\rm~g~cm^{-3}$ upon entering the shock followed by a slower rise to $6-20\times10^{-9} \rm~g~cm^{-3}$, depending on $b_i$.  This secondary, slow rise in density is more pronounced than seen in the other simulations.  
 The pressure histories are nearly identical to those in Rad 0.1.  
 The thermal histories for the super-particles are shown in Figure~\ref{F:rad0.01_k10_hist}.  
 While the apparently low cooling rates are more favorable for chondrule production, it must again be noted that this will only be true should there be efficient coupling between the gas and the chondrule precursors.

\subsubsection{Rad 0.01, $\kappa_c=100$ and the need for Rad 30, $\kappa_c=10$ (Rd0.01k100 and Rd30k10)\label{S:Rad_30}}  

\begin{figure}[hbt]
	\centering
	\subfloat{{\includegraphics[width=\textwidth]
		{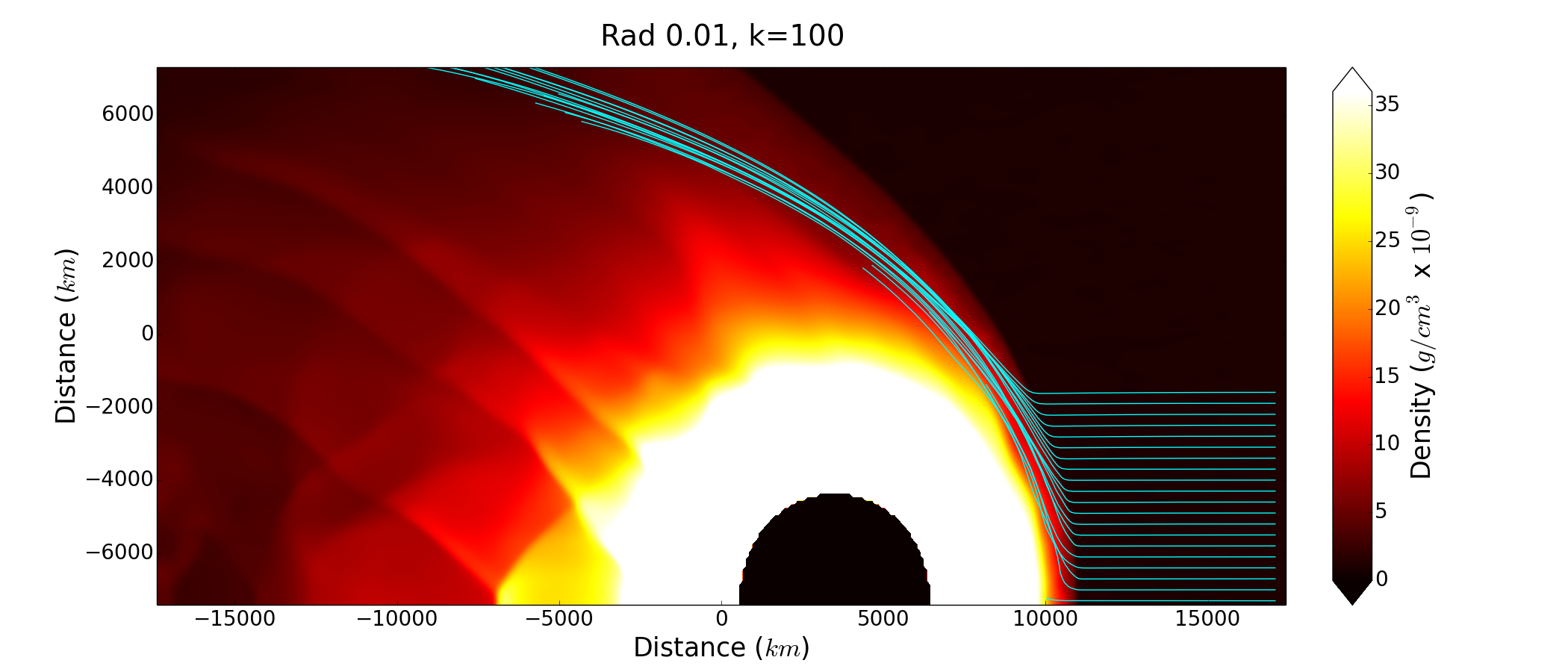} }}\\
	\subfloat{{\includegraphics[width=\textwidth]
		{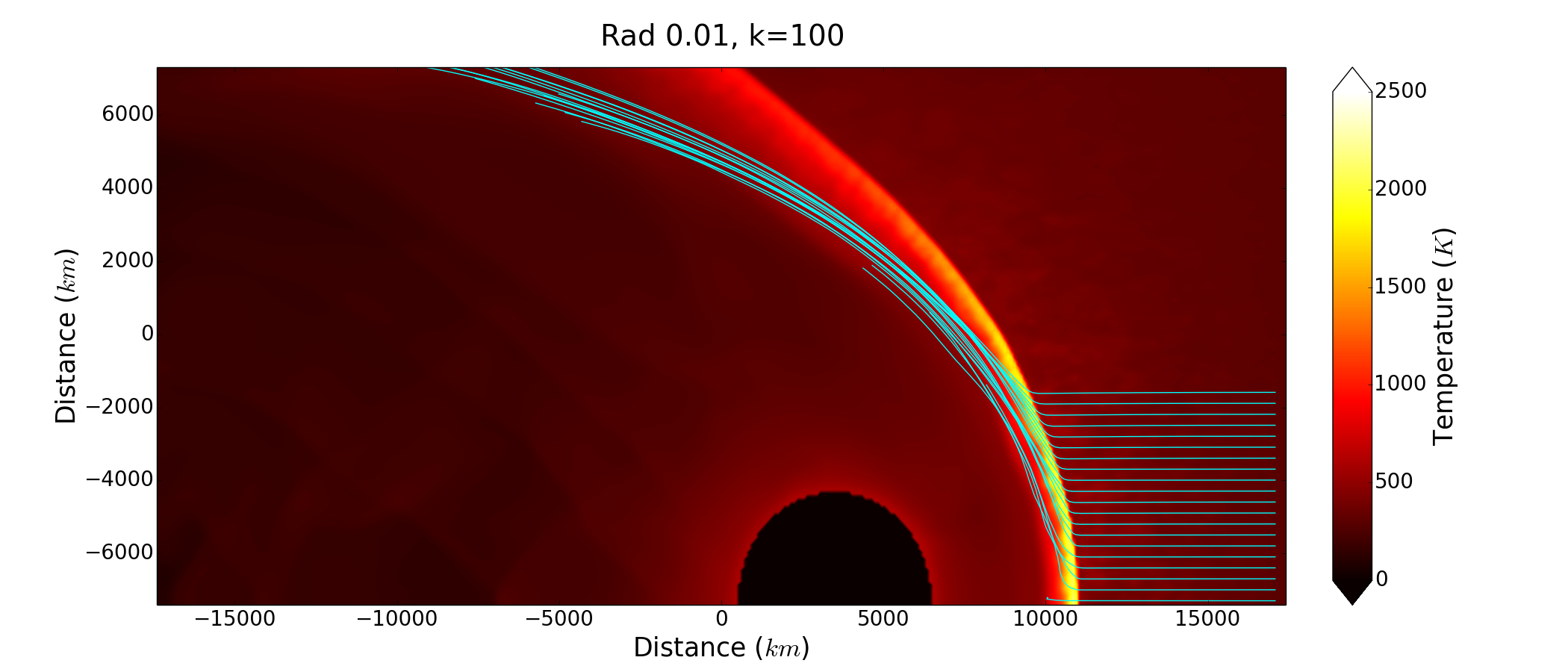} }}\\
	\caption{Similar to Figure \ref{F:rad1_k10_traject}, but for Rd0.01k100.
			The higher chondrule opacity increases the radiative cooling efficiency, as the gas is not optically thick.  
			The radiative precursor is again noticeable.
			 The mottled appearance is due to the discrete 
			nature of the simulated particles.}
	\label{F:rad0.01_k100_traject}
\end{figure}

\begin{figure}[t]
	\centering
	\subfloat{{\includegraphics[width=0.48\textwidth]
		{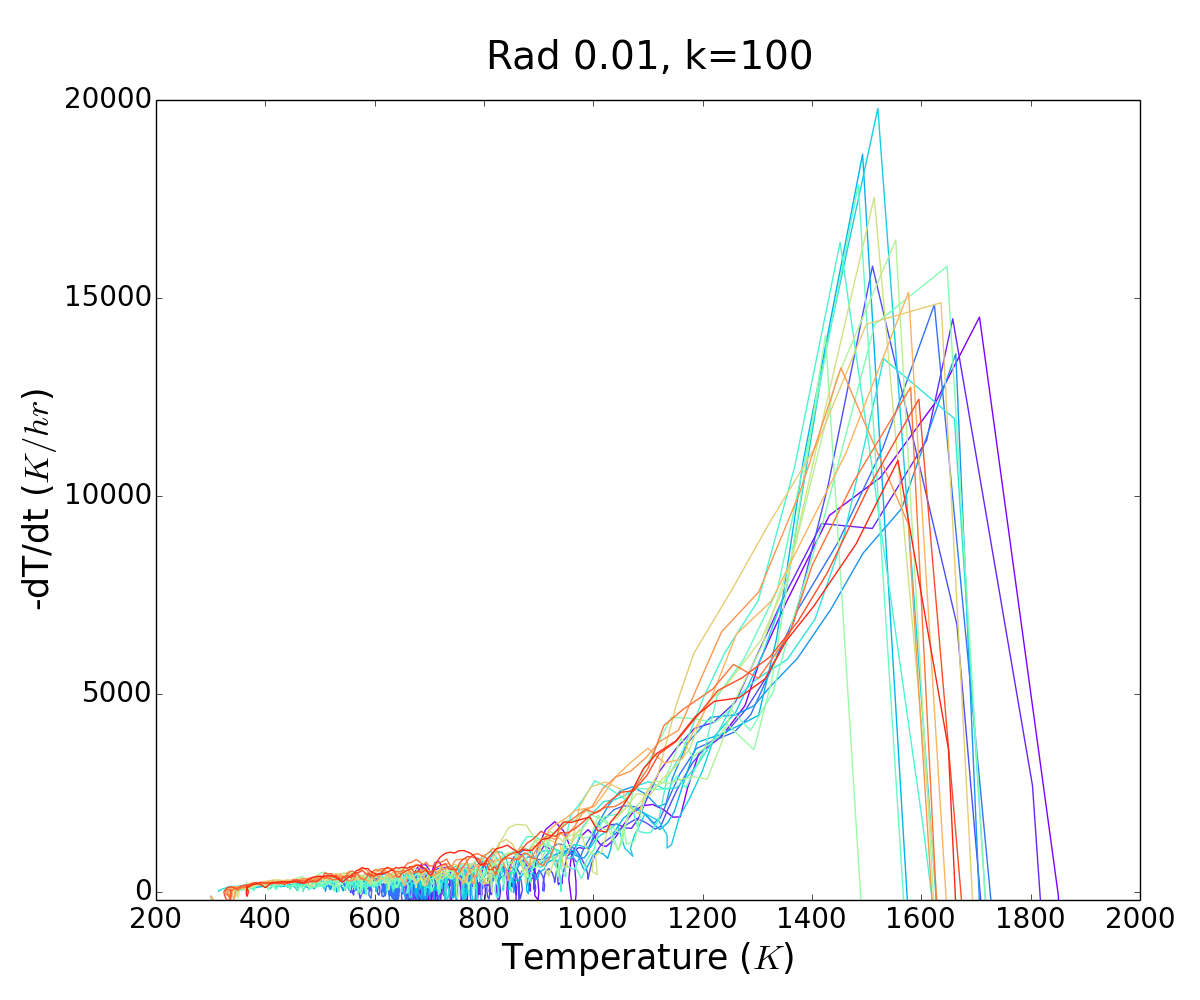} }}
	\subfloat{{\includegraphics[width=0.48\textwidth]
		{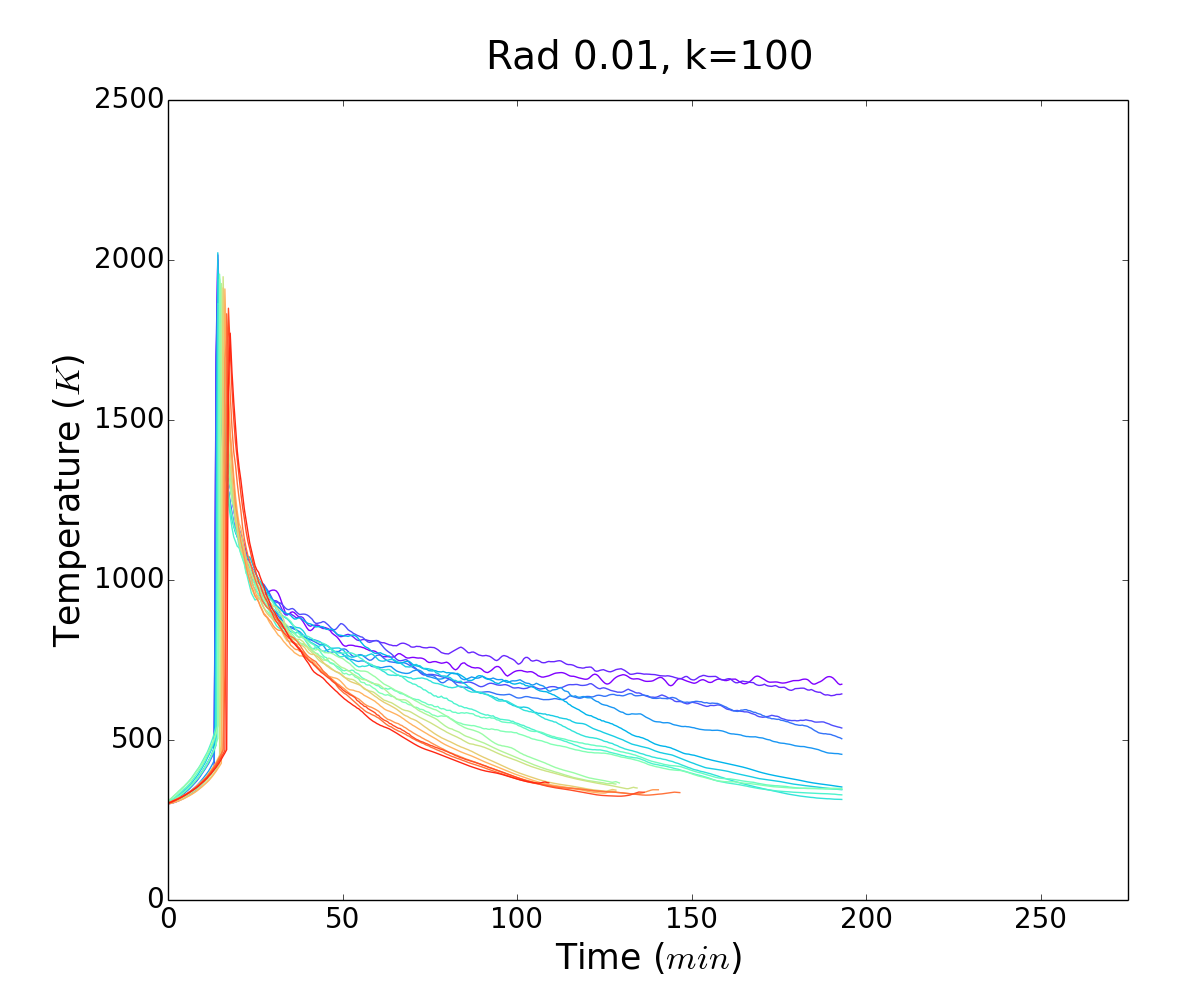} }}\\
	\subfloat{{\includegraphics[width=0.48\textwidth]
		{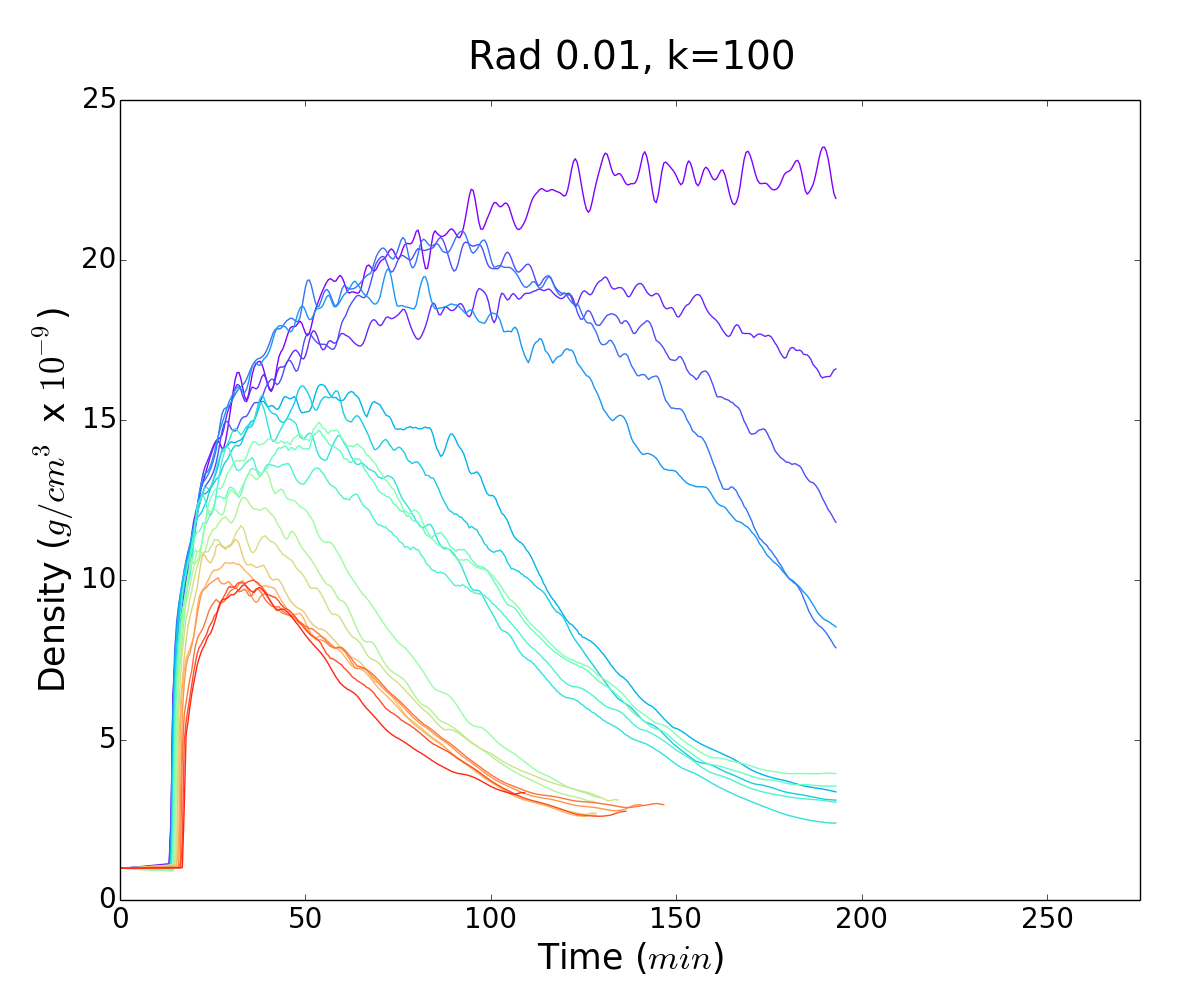} }}
	\subfloat{{\includegraphics[width=0.48\textwidth]
		{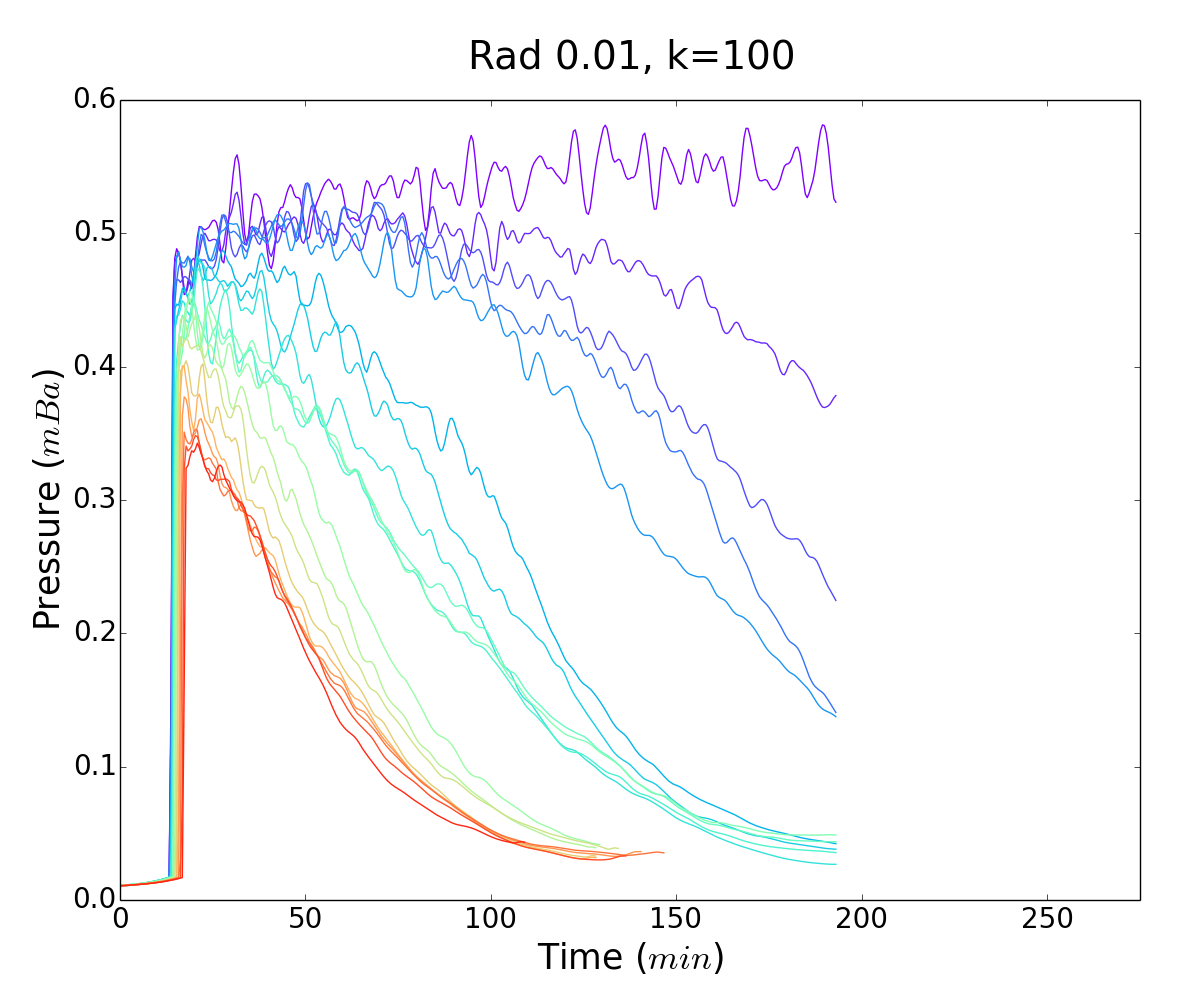} }}\\
	\caption{Similar to Figure \ref{F:rad1_k10_hist}, but for Rd0.01k100. 
	       Cooling rates are similar to those seen in Rd1k10, as are the density and pressure profiles. The noisy appearance in the curves is due to the super-particle discreteness effects, which are pronounced with the increased chondrule opacity.}
	\label{F:rad0.01_k100_hist}
\end{figure}

The run Rd0.01k100 explores a 10-fold increase in chondrule opacity (concentration $C=10$), as well as the same 100-fold decrease in dust opacity used in Rd0.01k10.  
The physical situation is a densely populated region of  chondrule precursors, with little surrounding dust. 
The bow shock density and temperature structures are shown in Figure~\ref{F:rad0.01_k100_traject}, which has a more mottled appearance than the other simulations. 
This is also seen in the corresponding particle histories (Fig.~\ref{F:rad0.01_k100_hist}).
The noise arises from the discrete nature of our simulated particles and the strong dependence of the opacity on those particles. 
The $10^6$ particles that we use sample the simulation volume well, but still do not reach a near-continuous distribution of physical chondrule precursors.  

The high-temperature region of the shock maintains a large range of impact radii, but becomes very thin due to efficient radiative cooling. 
There is a small radiative precursor, followed by a very rapid rise and fall in temperature (Fig.~\ref{F:rad0.01_k10_traject}).

The cooling rates again become greater than $10^4~\rm K~hr^{-1}$ through the crystallization range.  
Peak temperatures lie within 1800-2000 K for the traced $b_i$.  
Density histories peak at $\sim18-22\times10^{-9} \rm ~g~cm^{-3}$ for those particles that interact strongly with the atmosphere, and $\sim10-17\times10^{-9} \rm ~g~cm^{-3}$ for those of increasing $b_i$.  Pressure histories look very similar to previous cases, peaking at 0.35-0.5 mbar depending on $b_i$.

As discussed in section \ref{sec:bow_shocks}, the bow shock structure leads to a geometrically thin shock.  
For $C=1$, the estimated corresponding optical depth through the high-temperature region is $\tau\sim 0.01$.  
Even with $C=10$, the optical depth through the shock's bow (taken normal to any given part of the shock) remains optically thin. 
Concentrations $C\sim100$ would still only lead to $\tau\sim1$, allowing very efficient radiation down and upstream.  
Higher concentrations of chondrule precursors could, in principle, create an environment that is optically thick, approaching the adiabatic limit.
However, this situation may be inconsistent with bow shock processing for at least two reasons:  (1) The heat capacity of the rock may become important to the total energy budget, and could affect the efficacy of precursor melting.  (2) The solids will have a mass that will become comparable to or exceed that of the gas (recall the assumed ratio for chondrule precursors is 0.00375 under nominal conditions).  
This could cause shielding of gas-drag effects, with feedback potentially altering the shock structure as well. 
We do not attempt to model this here.

Instead of having a very high concentration of chondrules, another way to approach the optically thick limit is to have an abundance of small grains.  
This would, for the most part, keep the aerodynamic features of the bow shock model, while not requiring a significant increase in total solid mass.
 We have run several radiative test simulations (not shown) with high dust opacities to determine which conditions would lead to bow shock structures that are similar to those seen for the adiabatic limits (collectively referred to as Rd30k10 here). 
 We find that to reach this adiabatic limit, the dust enhancement needs to be about 30 times the nominal value, i.e., a thirty-fold increase in the fine-grained dust opacity. 
 This is roughly consistent with our expectations as outlined in section \ref{sec:bow_shocks}.  
A dusty environment is distinctly different from bow shocks in a low opacity environment \citep[see][]{boley_etal_apj_2013}, the latter of which allows the gas to remain essentially adiabatic but will also allow chondrule melts to cool rapidly unless thermal coupling to the gas is more efficient than expected. 
In the high dust scenario,  the solids and the gas will remain thermally coupled, making the adiabatic limit a reasonable test scenario for cooling rates.  
Furthermore, if the high-dust model for the bow shock environment is plausible for the formation environments, then the chondrule sorting mechanism cannot be due to disk aerodynamics alone.
For example, models such as turbulent concentration rely on preferentially creating a high chondrule environment \citep{cuzzi_etal_2001}.  
If bow shocks represent the chondrule-forming mechanism, they require high chondrule concentration and high abundances of small-grain dust.


\subsection{Wind Speeds} 								

An embryo is expected to cause a diverse range of shock conditions, due to both the embryo's evolving eccentricity and to the different phases of an embryo's orbit (see Fig.~\ref{F:v_rel}).
As such, we also explore wind speeds of 5 and $6\rm~km~s^{-1}$.
Such speeds are also motivated as lower eccentricities are more probable than higher ones. 
The simulations are only evolved using adiabatic gas dynamics, which is justified by assuming the chondrules are in a dusty environment, as discussed above.

\begin{figure}[hbt]
	\centering
	\subfloat{{\includegraphics[width=\textwidth]
		{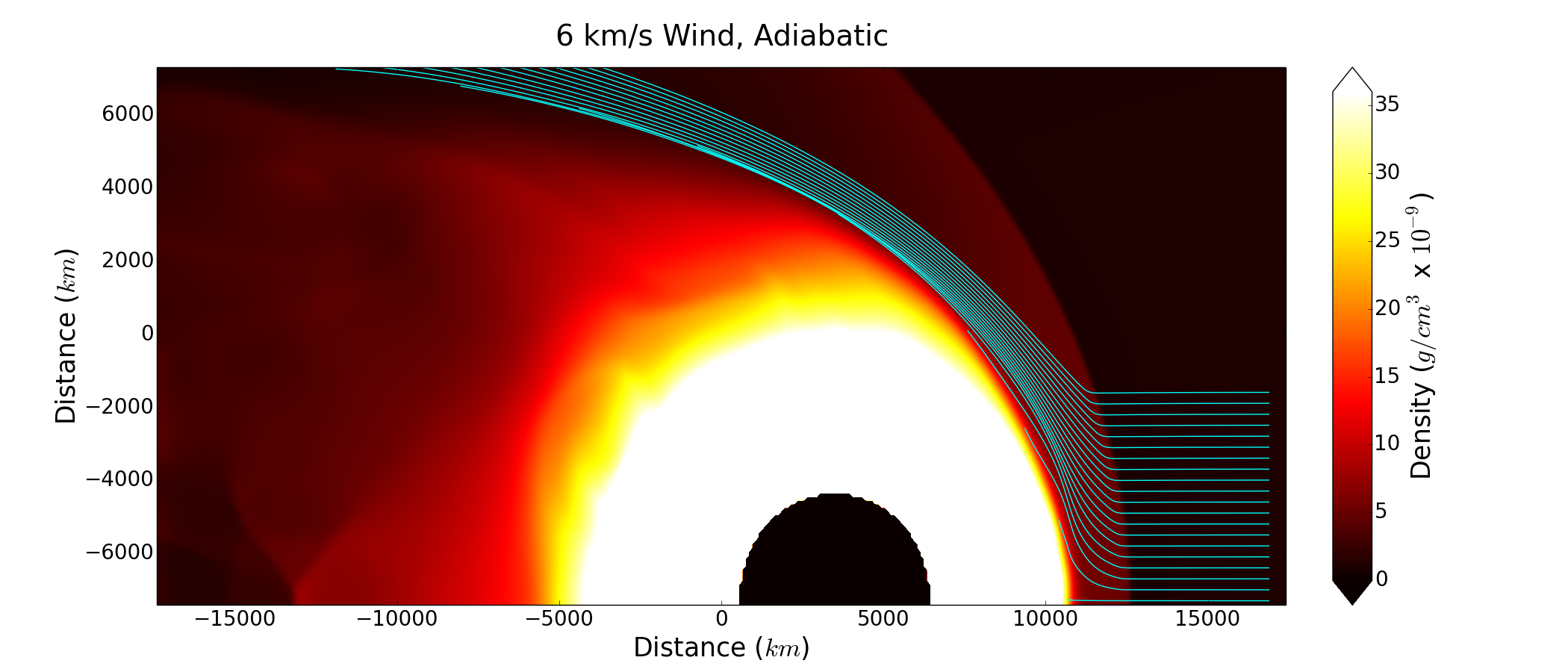} }}\\
	\subfloat{{\includegraphics[width=\textwidth]
		{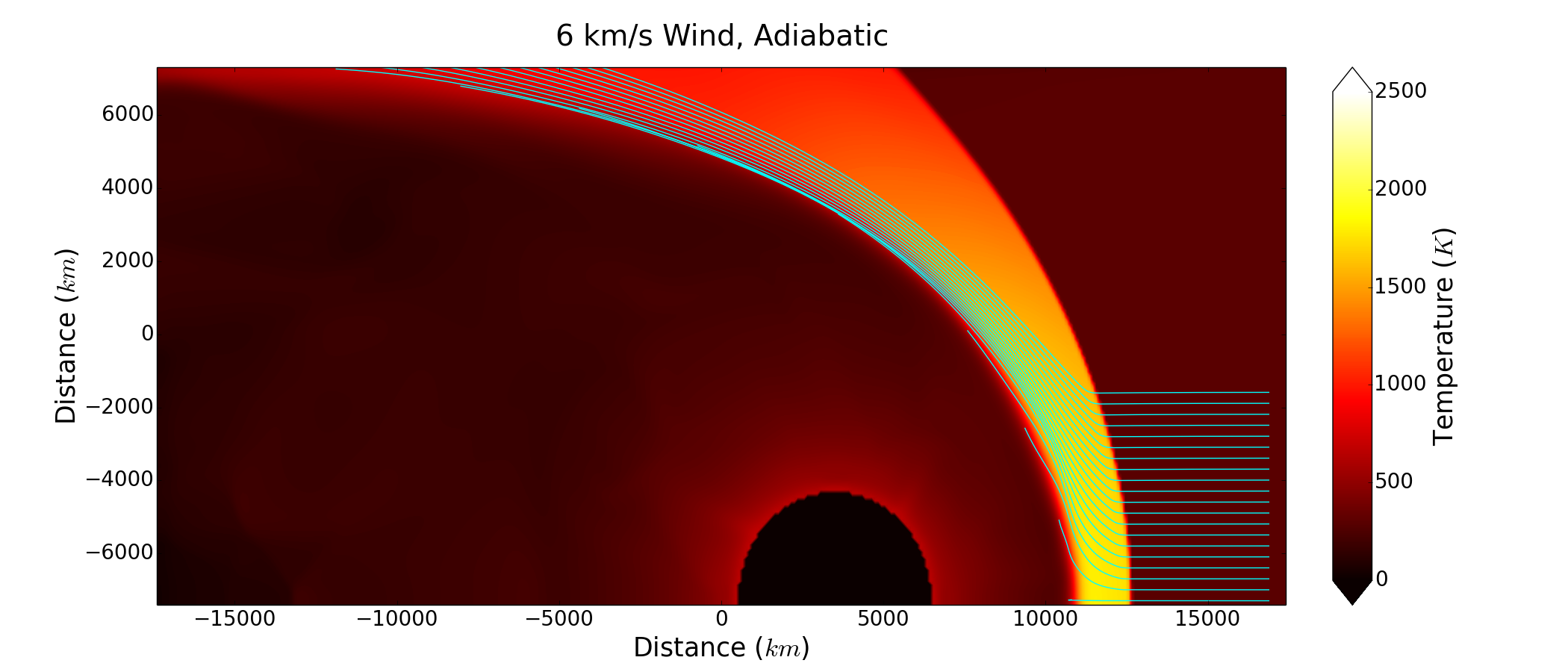} }}\\
	\caption{Similar to Figure \ref{F:rad1_k10_traject}, but for an adiabatic, 6 km s$^{-1}$ wind (AdiHi6). 
	          The slower wind speed widens the opening angle of the bow shock and also produces lower 
			peak temperatures.}
	\label{F:6k_traject}
\end{figure}

\begin{figure}[t]
	\centering
	\subfloat{{\includegraphics[width=0.48\textwidth]
		{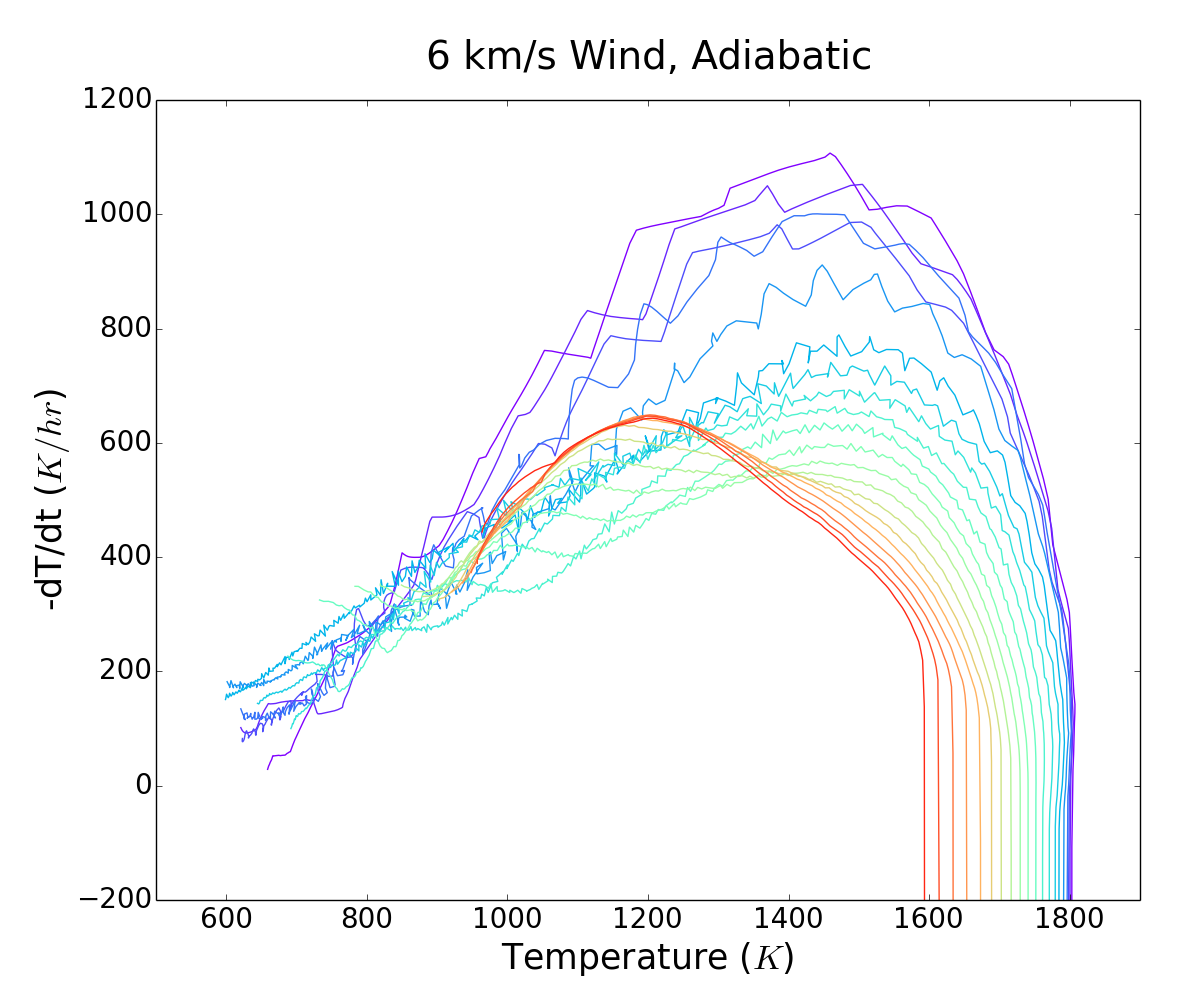} }}
	\subfloat{{\includegraphics[width=0.48\textwidth]
		{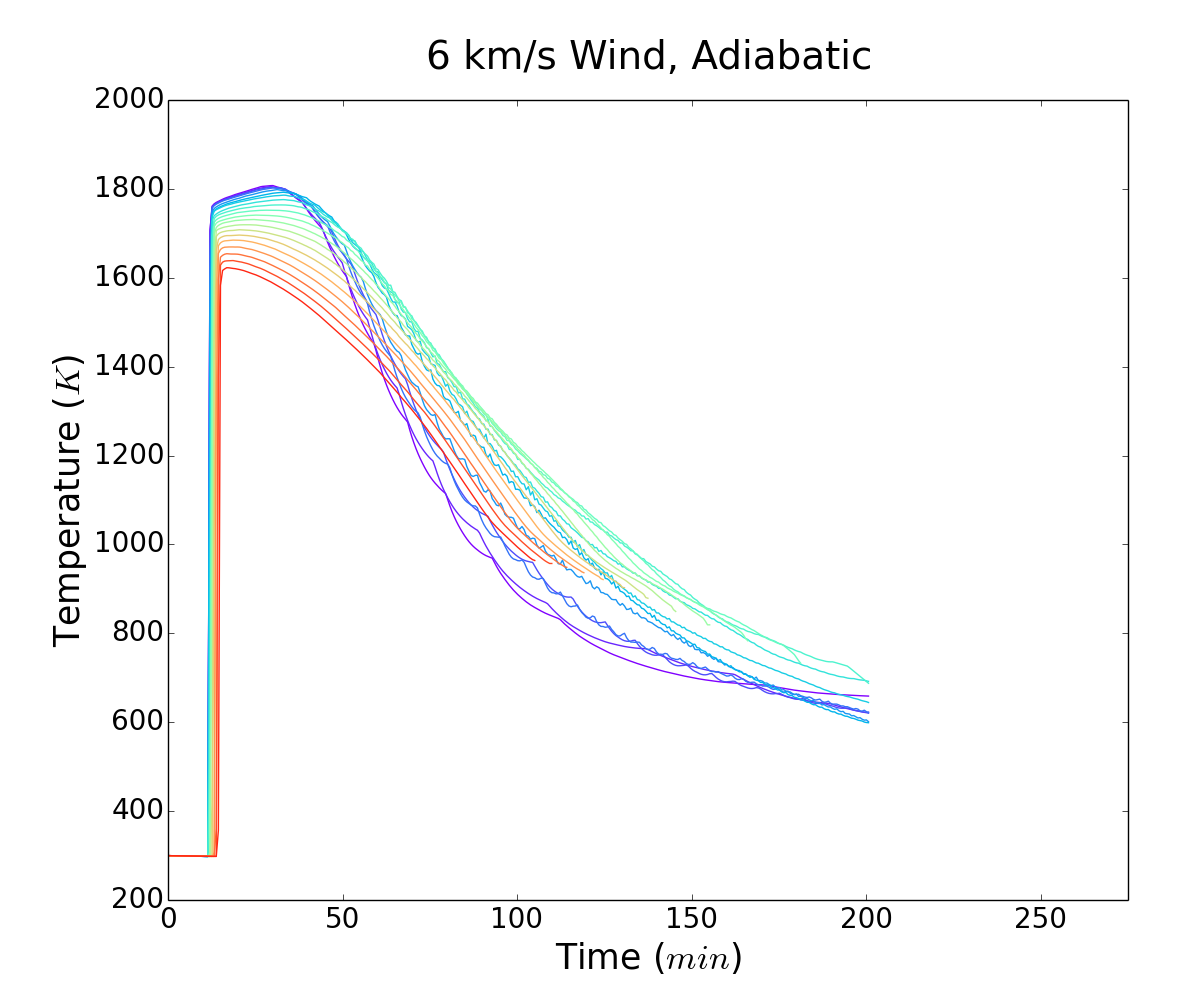} }}\\
	\subfloat{{\includegraphics[width=0.48\textwidth]
		{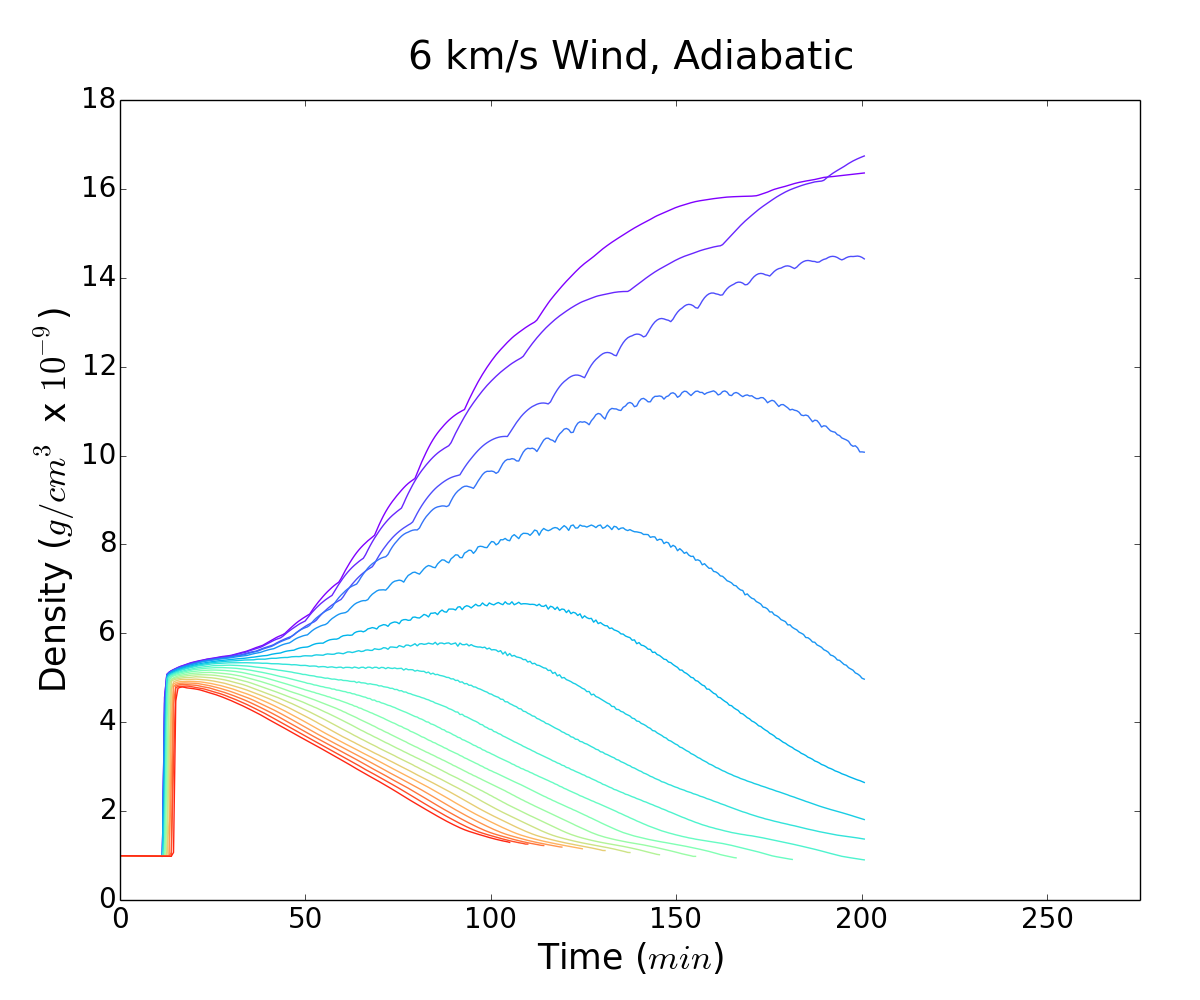} }}
	\subfloat{{\includegraphics[width=0.48\textwidth]
		{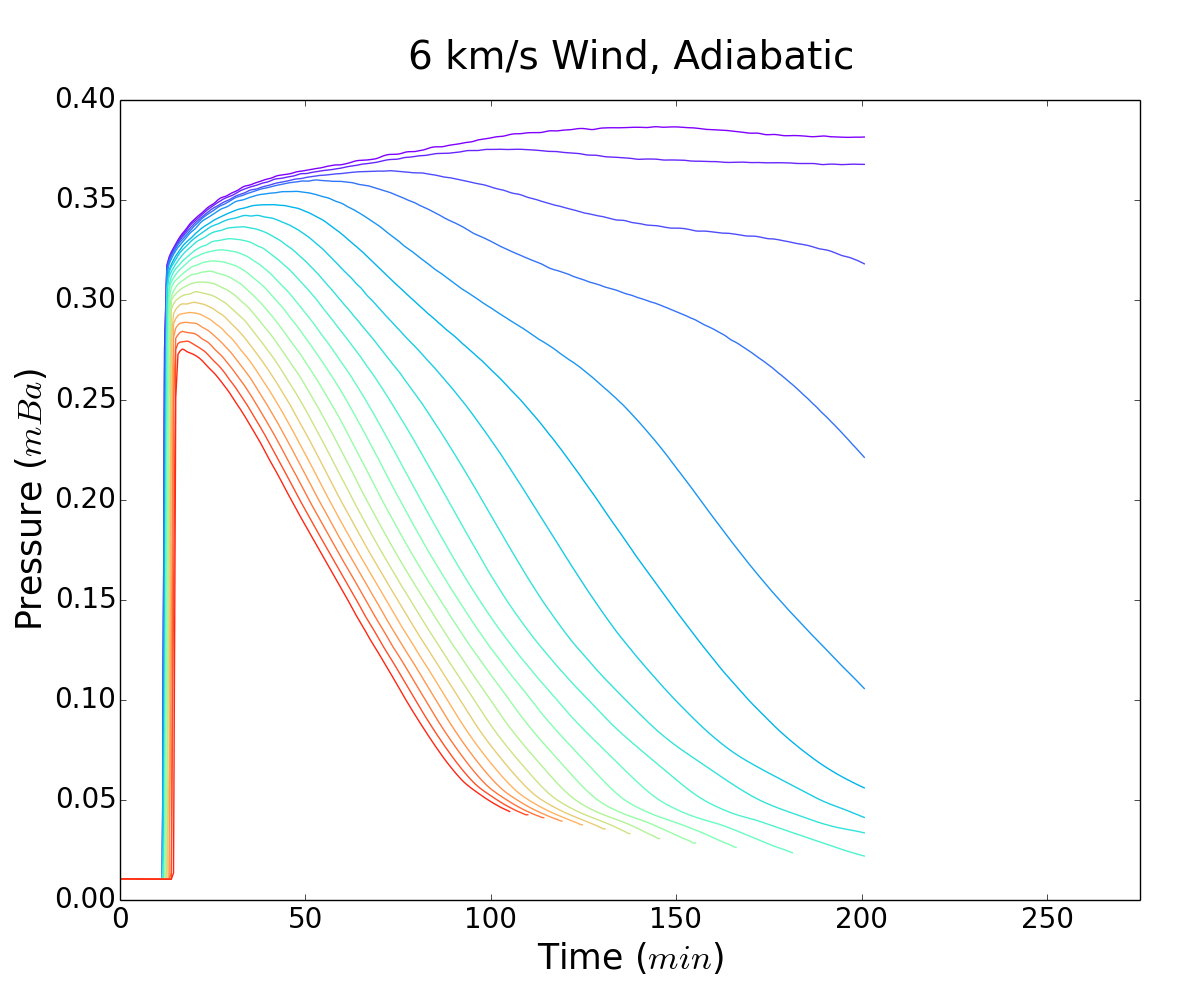} }}\\
	\caption{Similar to Figure \ref{F:rad1_k10_hist}, but for an adiabatic, 6 km s$^{-1}$ wind (AdiHi6). 
	The resulting cooling rates are largely consistent with furnace experiments. Bear in mind
			 that these rates correspond to an adiabatic simulation with perfect gas-solid thermal coupling, conditions that require a dusty environment to prevent chondrule melts from radiating away their energy efficiently.}
	\label{F:6k_hist}
\end{figure}

Figures~\ref{F:6k_traject} and \ref{F:6k_hist} show the shock structures, particle trajectories, and corresponding thermal profiles for particles going through the  $6\rm ~km~s^{-1}$ shock.   
The temperatures never peak above 1800 K, even for $b_i=0$.  
Strictly, there may be a brief period during which frictional heating from gas drag and chemical disequilibrium in H$_2$, which are not captured in these simulations,  allow higher peak temperatures (see  discussion in Boley et al.~2013). 
These shocks are therefore plausible for chondrule-forming conditions for a wide range of chondrules. 
In particular,  cooling rates are favourable for even prophyritic textures, with rates anywhere from 200-1100 K hr$^{-1}$ through crystallization. 

\begin{figure}[hbt]
	\centering
	\subfloat{{\includegraphics[width=\textwidth]
		{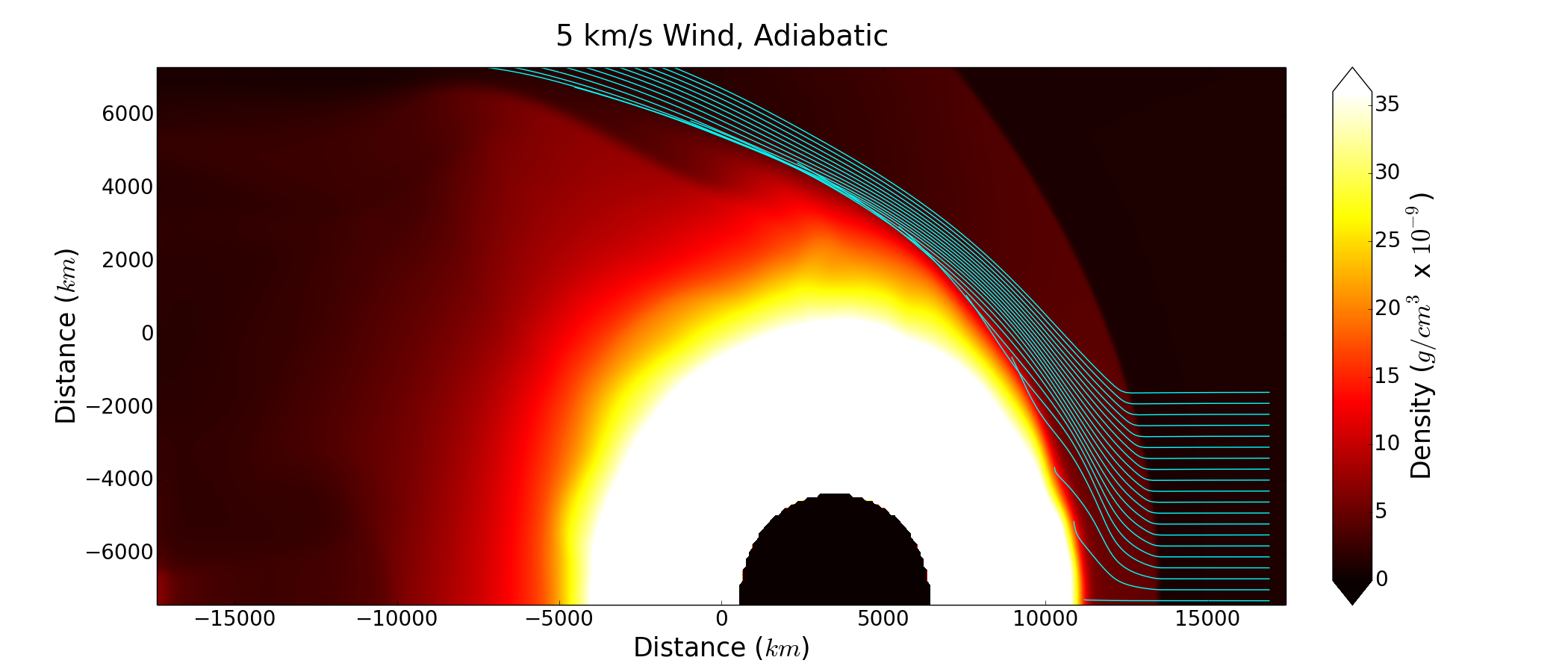} }}\\
	\subfloat{{\includegraphics[width=\textwidth]
		{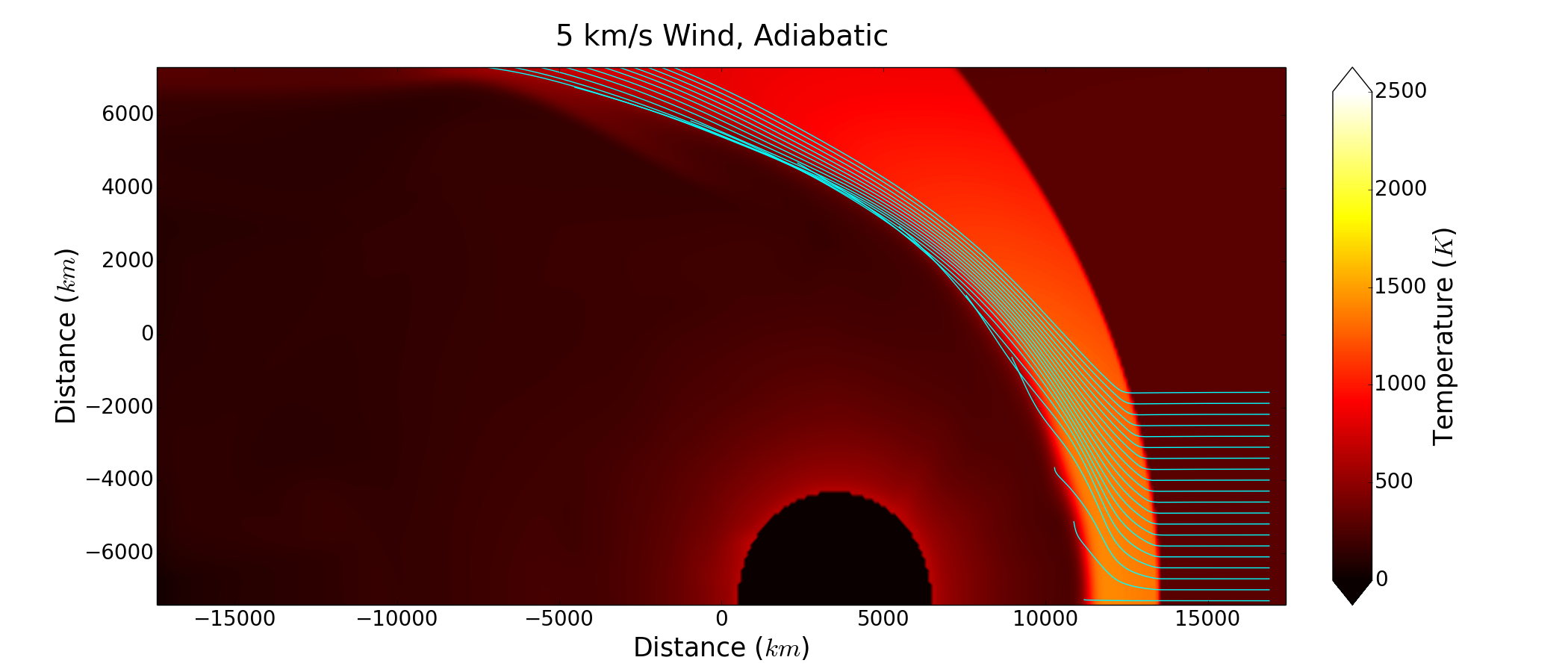} }}\\
	\caption{Similar to Figure \ref{F:6k_traject}, but for a 5 km s$^{-1}$ wind (AdiHi5).  The shock  is very broad, with low peak temperatures.}
	\label{F:5k_traject}
\end{figure}

\begin{figure}[t]
	\centering
	\subfloat{{\includegraphics[width=0.48\textwidth]
		{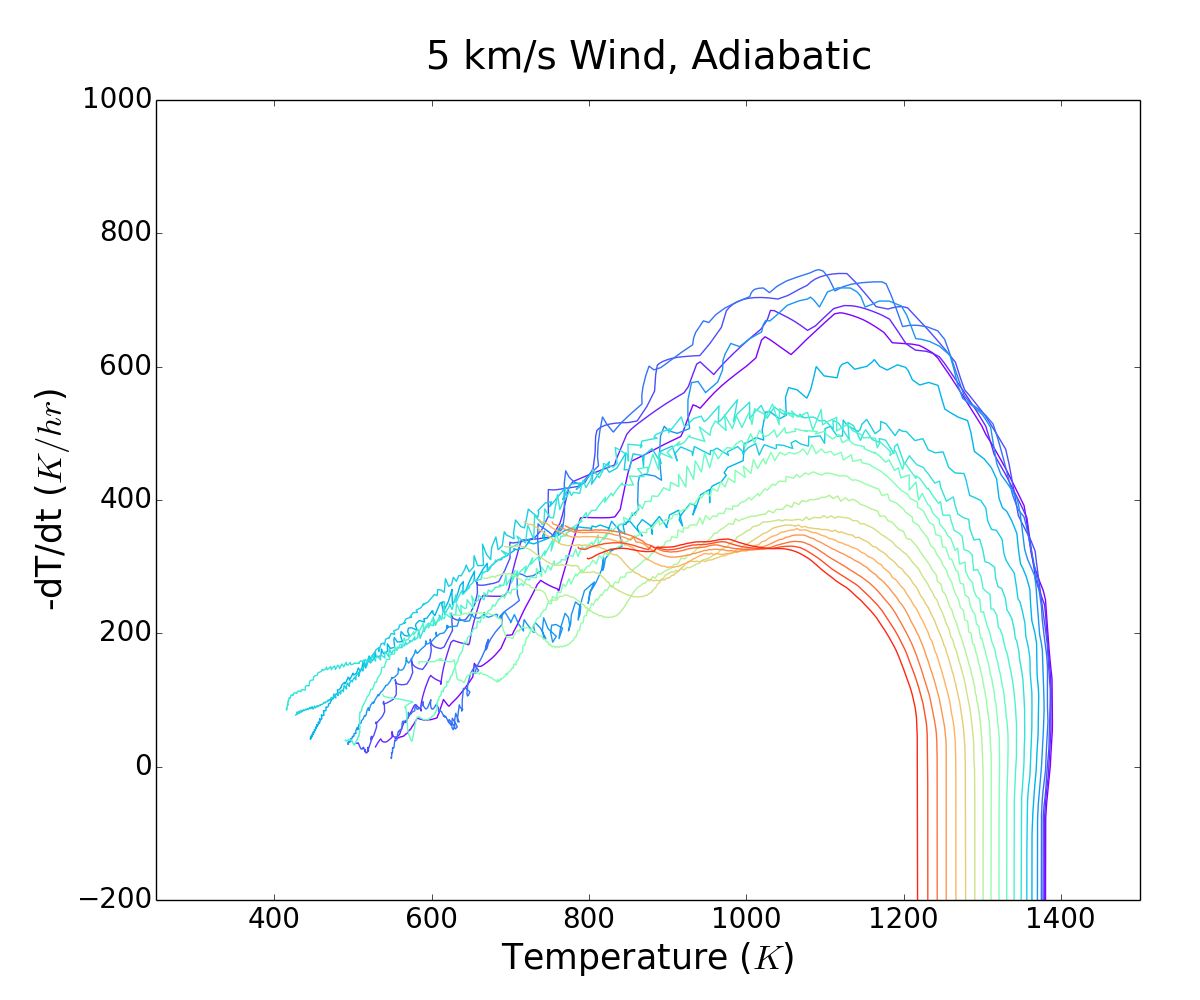} }}
	\subfloat{{\includegraphics[width=0.48\textwidth]
		{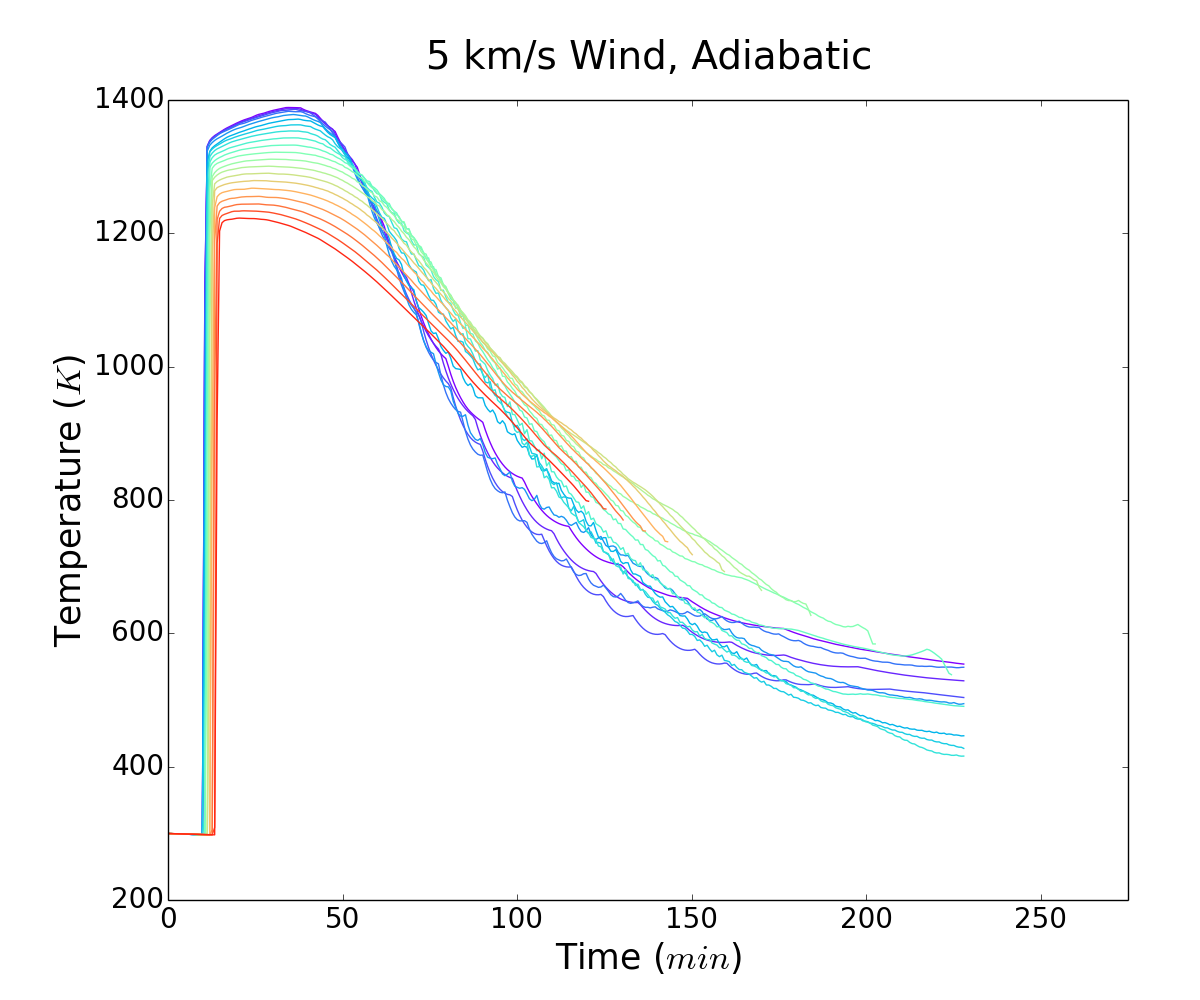} }}\\
	\subfloat{{\includegraphics[width=0.48\textwidth]
		{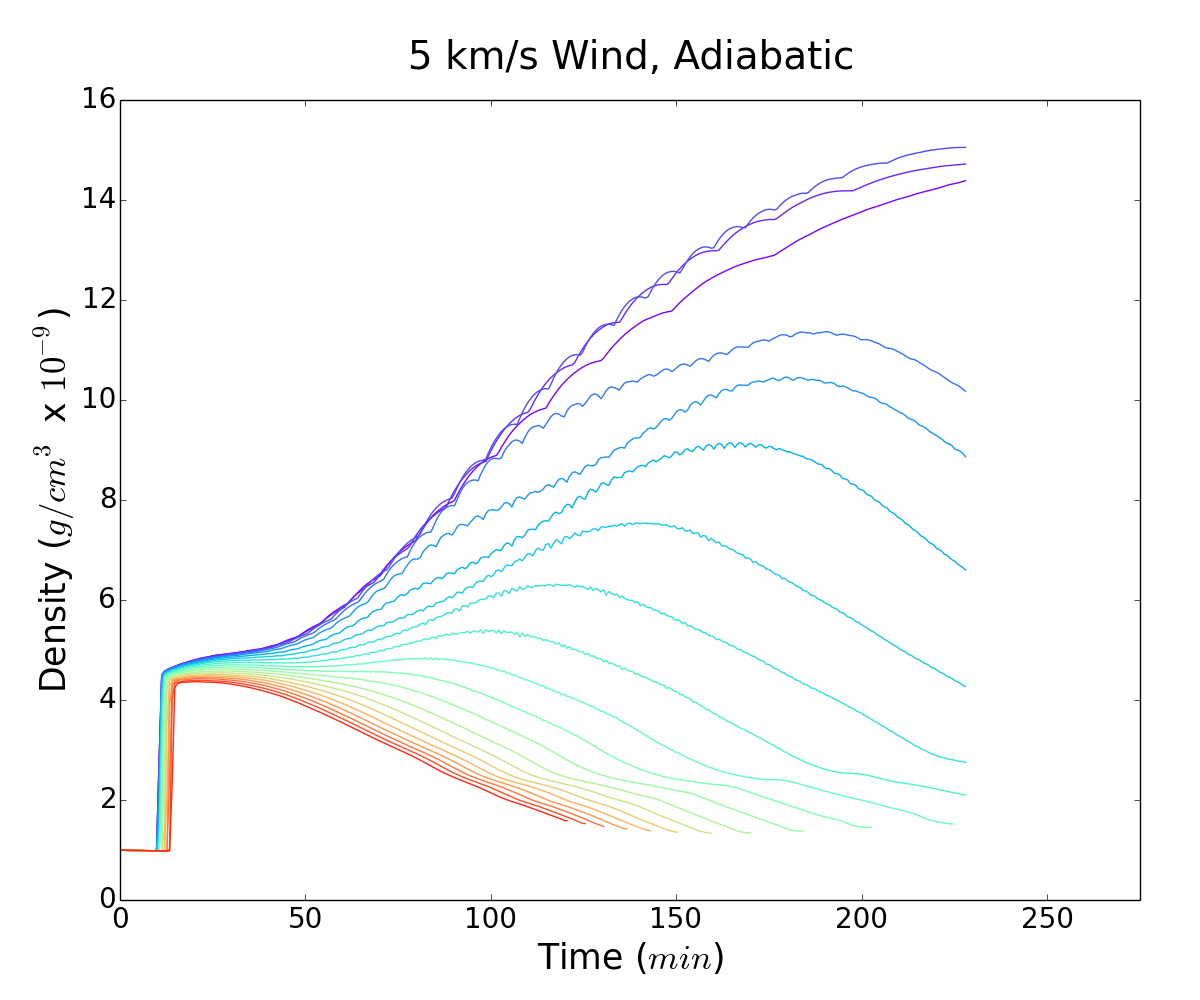} }}
	\subfloat{{\includegraphics[width=0.48\textwidth]
		{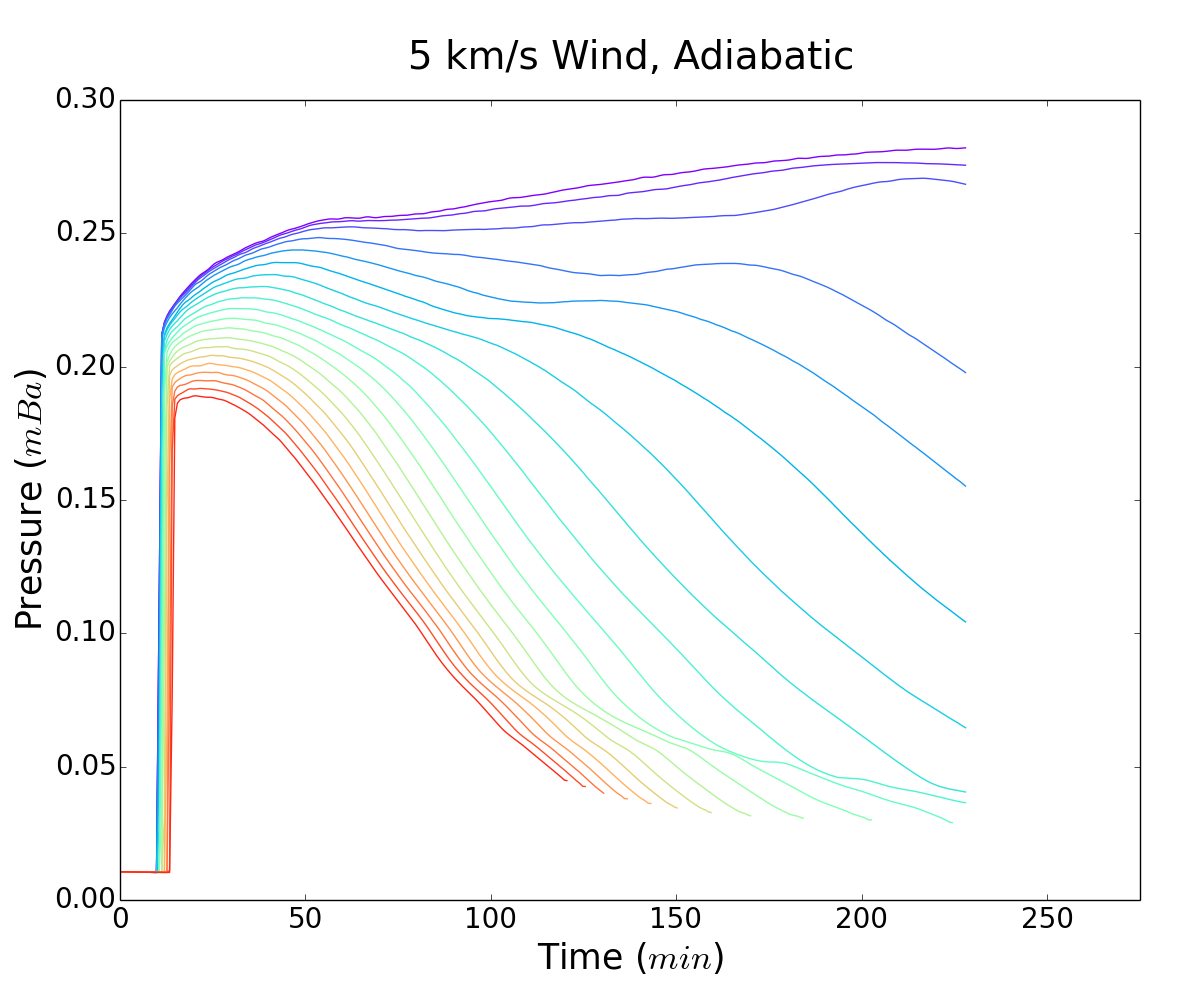} }}\\
	\caption{Similar to Figure \ref{F:6k_hist}, but for a 5 km s$^{-1}$ wind. Cooling rates have fallen even further, but the peak temperatures are no longer sufficient for the melting 
			chondrules.}
	\label{F:5k_hist}
\end{figure}

A 5 km s$^{-1}$ shock (Fig.~\ref{F:5k_traject} and \ref{F:5k_hist}), however, appears to be insufficient for melting chondrule precursors, with peak temperatures at $b_i=0$ of 1400 K.  
This is confirmation of the \cite{morris_desch_2010} results that wind speeds $\gtrsim 6$ km/s are needed to achieve temperatures that are high enough to melt chondrule precursors.


\section{Discussion}\label{S:discussion}

The presence of an atmosphere around an embryo affects the bow shock model in several ways:  
(1) Atmosphere gas prevents the formation of a highly rarefied, hot tail, as seen in atmosphere-free simulations \citep{boley_etal_apj_2013}.  
This also limits the occurrence of secondary shocks, which can cause re-heating events.  
(2) The effective cross section of the embryo is dramatically increased.  
This in turn creates a larger standoff distance, affecting the range of precursor sizes that can be deflected by the bow shock \citep[e.g.,][]{morris_etal_apj_2012,boley_etal_apj_2013}.  
It also creates a much larger deflection distance as the precursor is transported around the embryo. 
As a result, the cooling times for melts increases (at least in the adiabatic limit) because thermally processed material stays in contact with hot gas for a longer period of time. 
(3) Low $b_i$ particles that traverse through the entirety of the standoff region are not immediately lost to the embryo, and instead enter the embryo's upper atmosphere.  
Such solids will experience even higher densities and, in some cases, higher pressures than what are achieved in the shock. 
This exposure to the embryo's atmosphere may expose melts to high partial pressures of volatiles \citep{morris_etal_apj_2012},  which seems to be required to explain chondrule chemistry \citep{alexander_etal_sci_2008}. (4) As the gas and particles are redirected around the atmosphere, particle streamlines converge and even cross due to atmosphere undulations.  Because particles from different impact radii will experience different heating conditions,  stream convergence provides a mechanism for chondrules in different stages of melt to collide and form compound chondrules or to produce fragments.  
Both these effects can be observed in Figures~\ref{F:adi_traject_temp} and~\ref{F:adi_cooling_temp}.

\begin{figure}[h]
	\centering
	\includegraphics[width=0.7\textwidth]
		{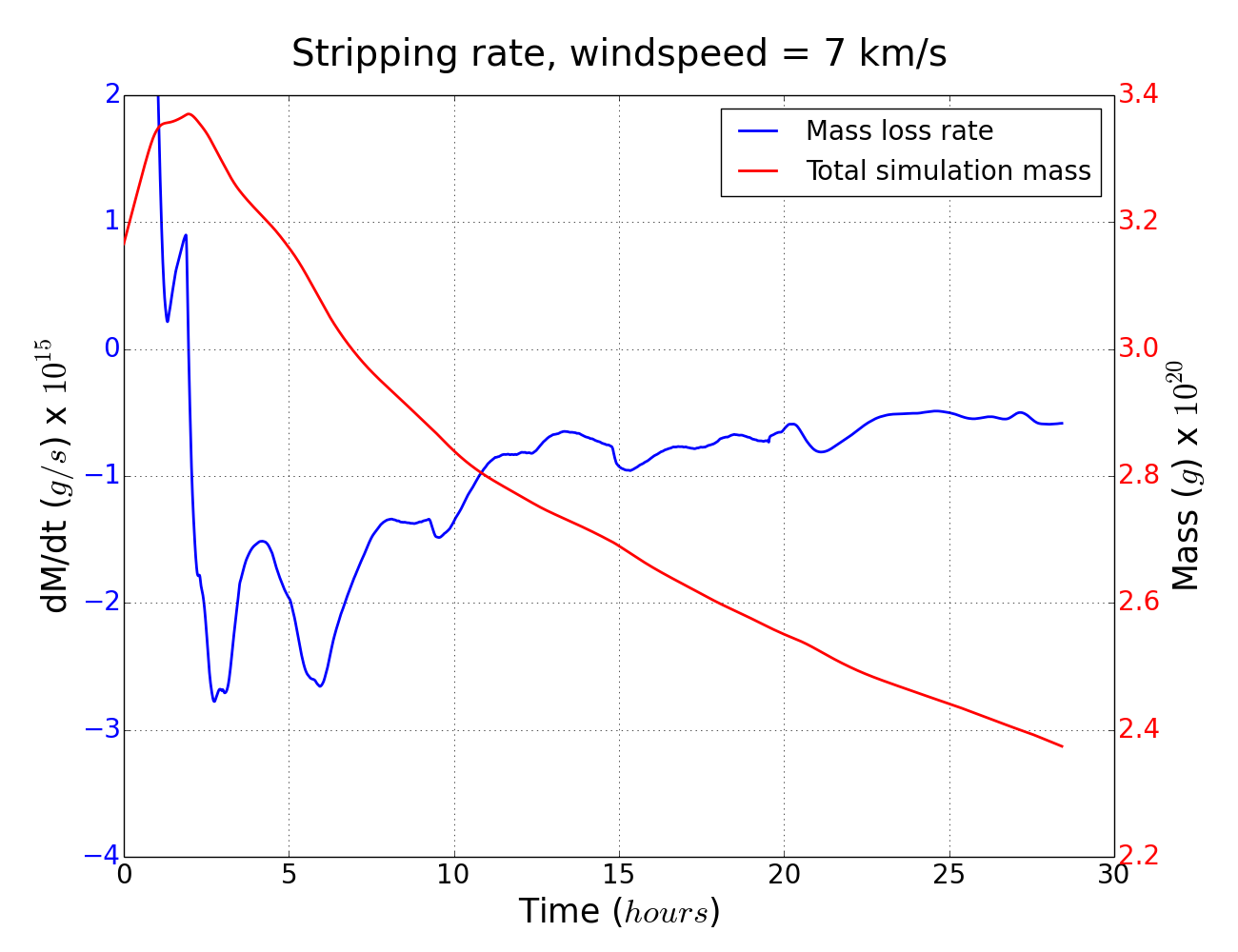}
	\caption{The red plot and the right-side scale indicate the total mass within the simulation space. The initial rise is due to
			the onset of the wind.  As the simulation progresses the total mass evolves toward a steady 
			decline.  The blue plot and left-side scale show the rate at which this total mass is changing.}
	\label{F:massloss}
\end{figure}

Nonetheless, the above effects are only important if an atmosphere can be retained, or at the very least, replenished at different phases during an embryo's orbit.  
Estimates for the mass loss rates suggest that even massive atmospheres should be short lived (section~\ref{SS:mass_rates}).  
For the conditions studied here, we estimated the stripping rate to be approximately $4\times 10^{15} C_w$~g s$^{-1}$.  The measured mass loss rate in our adiabatic, $7~\rm km~s^{-1}$ run settles to a value of $5\times 10^{14}\rm~g~s^{-1}$ (see Figure~\ref{F:massloss}).  
This suggests that $C_w\sim0.1$ and that the large atmosphere of $2.5\times10^{20}$ g would be stripped on the order of one week. 
 In section~\ref{SS:mass_rates} we showed that the accretion rate during low relative wind phases could exceed this stripping rate.
 Thus,  it is possible for the mass of the atmosphere to have periodic phases of growth and stripping.  
 Outgassing of the planet itself may somewhat offset stripping in high wind regions and increase the growth rate during quiescent portions of the orbit, but it would require massive amounts to greatly affect the bulk quantity of the atmosphere.  

Such stripping has potential cosmochemistry effects for the resulting planets.  
For example,  volatiles are expected to be outgassed during planetary accretion (particularly during the magma ocean stage), forming an atmosphere \citep{ringwood_1966}.  
 High relative winds, as explored here, provides a very efficient mechanism for stripping volatiles from embryos. 
The loss of volatiles by wind stripping could further affect early planetary evolution, such as removing a steam atmosphere that would otherwise prolong magma ocean stages \citep[e.g.,][see]{elkins-tanton_2012}.
 It has long been recognized that the Earth and other terrestrial planets are depleted in several volatile elements, as compared with solar abundances and CI chondrites \citep[e.g.,][]{ringwood_1966}. 
The stripping of outgassed atmospheres during the embryo stages would be an efficient process for driving this depletion.

\begin{figure}[h]
	\centering
	\includegraphics[width=0.7\textwidth]
		{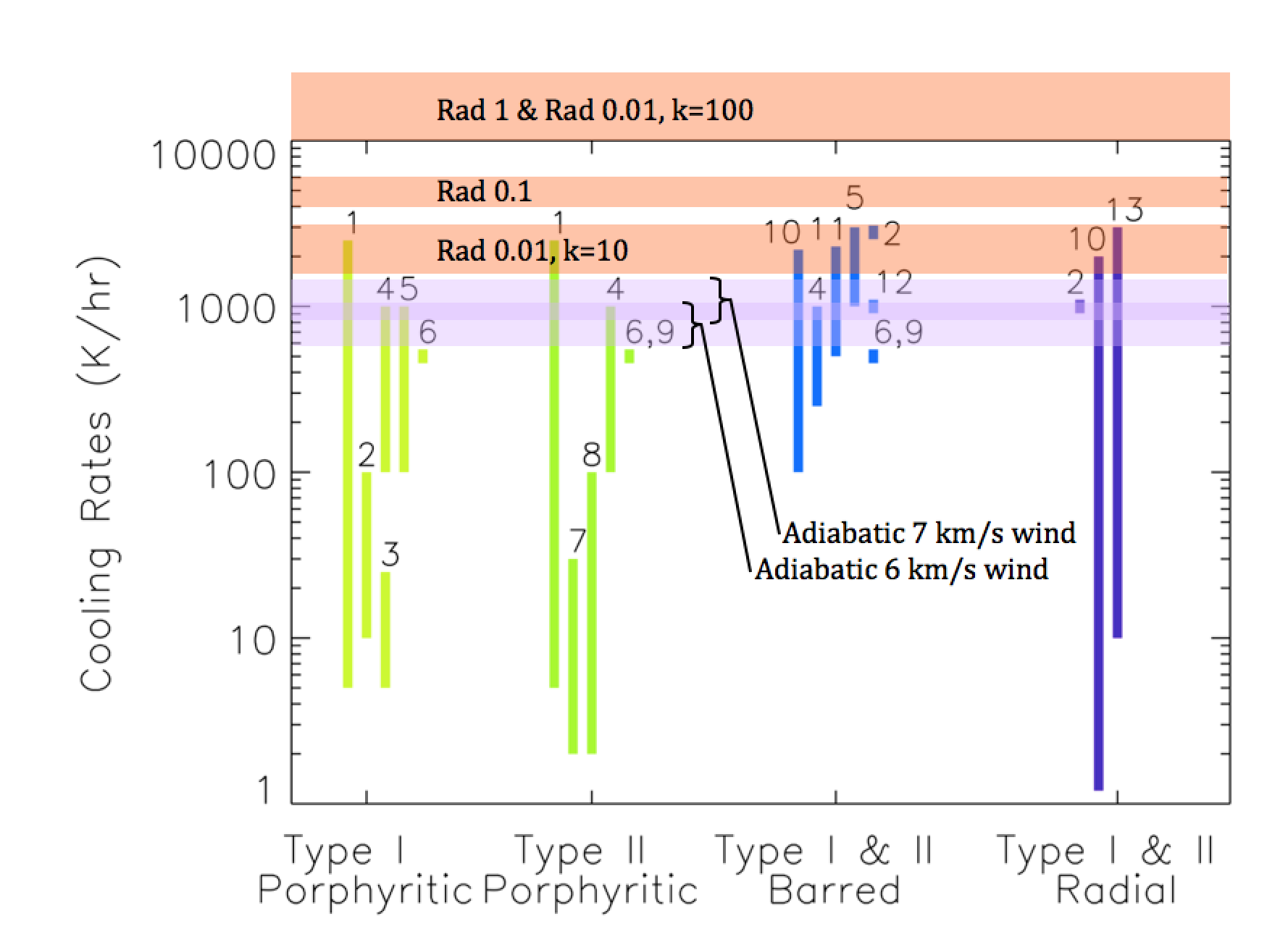}
	\caption{Cooling rates from our simulations compared with chondrule formation constraints. 
	The underlaying figure is from \cite{desch_etal_mps_2012}, which highlights the plausible cooling
			rates for different chondrule textures. Overlaid onto this figure are the cooling rates from the simulations presented here.}
	\label{F:desch_results}
\end{figure}

Assuming that at least some atmosphere can be effectively maintained through cycles of growth and stripping, we can compare our simulation results to data from laboratory furnace experiments and other inferred constraints (Fig.~\ref{F:desch_results}).    
For a gas density of $\rho_g = 10^{-9} \rm ~g~cm^{-3}$, a relative velocity of $V_{\rm rel} \gtrsim 6\rm~km~s^{-1}$ is required to reach chondrule-forming conditions (see section \ref{SS:thermal_hist_constraints}, which can potentially melt material out to impact radii of 2 $R_E$, assuming an atmosphere is present and that cooling is inefficient.

For all of the main radiative cases, the cooling rates were found to be too high to be consistent with most chondrules. Rad 1 resulted in cooing rates $>10^4~\rm K~hr^{-1}$,  Rad 0.1 in several $10^3$ K/hr, and Rad 0.01, $\kappa_c = 10$ dropped to $\sim 2000$ K/hr, which could potentially form chondrules with  pyroxene porphyritic, and barred textures. 
The Rad 0.01, $\kappa_c=100 \rm ~cm^2~g^{-1}$ ($C=10$) simulation again showed high cooling rates.
The adiabatic simulations provided the closest matches to inferred chondrule cooling rates (Fig.~\ref{F:desch_results}, although cooling rates were still in the range of $10^3 ~\rm K~hr^{-1}$. 
Additional radiative simulations with high concentrations of fine dust showed that the adiabatic limit could be reached if the dust concentration were as high as 30 times the standard nebula value (section \ref{sec:bow_shocks}), which could allow thermal coupling between the dust and the gas. 

The source of high concentrations of dust is not explored here.  However, there are potential candidates, such as impact-generated debris clouds.  
Such clouds are observed in the present day asteroid belt \citep{jewitt_etal_2011},
and would have been even more prevalent in the more crowded environment of the primordial belt.

Finally, we only focused on a single embryo size.  Generally, larger embryos give rise to larger bow shocks, which in turn increases the time a solid takes to traverse the shock for a given wind speed.
As long as the gas remains effectively adiabatic and the solids follow the evolution of the gas, then smaller embryos imply shorter cooling times, even for high concentrations of dust.  Nonetheless, additional embryo sizes should be explored with non-equilibrium modelling between the gas and the solids to determine whether smaller embryos (e.g., 1000 km) could create chondrule-producing shocks. 
Because smaller embryos will be more numerous than larger embryos, they could be more efficient at producing secondary chondrule heating events, particularly as atmospheres suppress the formation of strong tail shocks around a single embryo.

\section{Summary \& Conclusions\label{S:conclusions}}

In this study we used radiative hydrodynamics simulations with direct particle tracing to explore the thermal environment that chondrule precursors can experience while traversing planetoidal bow shocks. 
Plausible embryo atmospheres were included in our simulations, which allowed us to explore their effects on the bow shock structure and to determine the longevity of the atmospheres themselves when encountering strong relative winds. 
We investigated three different wind speeds under adiabatic conditions. 
We also presented in detail four cases with full radiative hydrodynamics, varying the opacities between cases. 
In addition, we ran a series of short radiative simulations that explored when small-grained dust would make the shock optically thick for the bow shock morphology to become comparable to the adiabatic case. 

The main findings of our study are: (1) Large atmospheres bring down cooling rates by increasing the effective size of the embryo.  This occurs because incoming particles are forced to spend more time in contact with hot gas. (2) Wind speeds of approximately 6-8 km/s produce the appropriate peak temperatures for chondrule melting.  
Wind speeds below this range do not achieve the required melting temperatures. 
(3) The best match of cooling rates comes from our adiabatic limit simulations.  
To realistically reach this limit, we require that chondrule precursors are embedded in fine-grained dust, at least 30 times the nominal solar nebula value.  
Increasing just the chondrule concentration does not appear to be consistent with the bow shock model.  
However, as small grains could be vaporized at high temperatures, the material dynamics of small grains in bow shocks requires further study. 
(4) Atmospheres are efficiently stripped from the embryos.  Atmospheres may only be maintained through cycles of accretion after periods of stripping at high relative wind speeds. This stripping may play a role in removing the volatile inventory of terrestrial worlds during planet building. 


\section{Acknowledgements}\label{S:ack}

CRM was supported, in part, by The University of British Columbia's Science Undergraduate Research Experience award.
ACB's contribution was funded, in part, by The University of British Columbia, the Canada Research Chairs program, and an NSERC Discovery grant.
MAM was supported by NASA Cosmochemistry grant NNX14AN58G and NASA Emerging Worlds grant NNX15AH62G.
This research was enabled in part by support provided by WestGrid and Compute Canada Calcul Canada.

We thank the anonymous referee for a helpful report 

\bibliographystyle{plainnat}
\bibliography{Chondrules_amsart}

\end{document}